\documentclass[11pt,a4paper]{article}
\pdfoutput=1 

\usepackage{jheppub} 
\usepackage[T1]{fontenc} 
\usepackage[italian,english]{babel}
\usepackage{hyperref}
\hypersetup{
	colorlinks=true,
	linkcolor=[rgb]{0.0,0.42,0.89},
	filecolor=magenta,      
	urlcolor=[rgb]{0.0,0.42,0.89},
	citecolor=[rgb]{0.64,0.0,0.0},
}

\usepackage{csquotes}  
\usepackage{ifpdf}
\usepackage{subfigure}
\usepackage{tikz}
\usepackage[compat=1.1.0]{tikz-feynman}
\usetikzlibrary{arrows,shapes}
\usetikzlibrary{trees}
\usetikzlibrary{matrix,arrows} 				
\usetikzlibrary{positioning}				
\usetikzlibrary{calc,through}				
\usetikzlibrary{decorations.pathreplacing}  
\usepackage{pgffor}							

\usetikzlibrary{decorations.pathmorphing}	
\usetikzlibrary{decorations.markings}
\tikzset{
	vector/.style={decorate, decoration={snake}, draw},
	fermion/.style={draw=black, postaction={decorate}}, 
	scalar/.style={dashed,draw=black, postaction={decorate}}}
\tikzstyle{block} = [draw, rectangle, 
minimum height=3em, minimum width=6em]

\usepackage{amssymb}
\usepackage{amsfonts}
\usepackage{amsmath}
\usepackage{rotating}
\usepackage{graphicx}
\usepackage{amsmath}
\usepackage{fancyhdr}
\usepackage{lineno}
\usepackage{babel}
\usepackage{graphics}
\usepackage{physics}
\usepackage{pstricks}
\usepackage{color}
\usepackage{multirow}
\usepackage{array}
\usepackage{diagbox}
\usepackage{cancel}
\usepackage{array,booktabs, makecell}
\usepackage{mathrsfs}
\usepackage{pifont}

\graphicspath{{./plots/}}

\newcolumntype{P}[1]{>{\centering\arraybackslash}p{#1}}
\newcolumntype{M}[1]{>{\centering\arraybackslash}m{#1}}
\newcommand{\nn}{\nonumber}
\newcommand{\lsim}{\mathrel{\mathop{\kern 0pt \rlap
			{\raise.2ex\hbox{$<$}}}
		\lower.9ex\hbox{\kern-.190em $\sim$}}}
\newcommand{\gsim}{\mathrel{\mathop{\kern 0pt \rlap
			{\raise.2ex\hbox{$>$}}}
		\lower.9ex\hbox{\kern-.190em $\sim$}}}

\newcommand{\be}{\begin{equation}}
	\newcommand{\ee}{\end{equation}}
\newcommand{\bea}{\begin{eqnarray}}
	\newcommand{\eea}{\end{eqnarray}}

\def\abs#1{\left| #1 \right|}

\newcommand{\fbi}{fb$^{-1}$ }
\newcommand{\abi}{ab$^{-1}$ }

\newcommand{\sarah}{\texttt{SARAH} }
\newcommand{\chep}{\texttt{CalcHEP} } 

\newcommand{\py}{\texttt{Pythia8} }

\newcommand{\mic}{\texttt{micrOMEGAs} }

\def\rd#1{\textcolor{red}{#1}}

\newcolumntype{P}[1]{>{\centering\arraybackslash}p{#1}}

\newcommand{\sgvann}{\langle \sigma_{\rm ann} v \rangle}
\newcommand{\bit}{\begin{itemize}}
\newcommand{\eit}{\end{itemize}}

\newcommand{\ww}      {{W^+ W^-}}
\newcommand{\mmu}      {{\mu^+ \mu^-}}

\newcommand{\ompl}      {\Omega_{\rm DM}^{\rm Planck} }
\newcommand{\omdm}      {\Omega_{\rm DM}}
\newcommand{\sgsi}      {\sigma_{\rm SI}}
\newcommand{\ssgsi}      {\sigma_{\rm SI}^{\rm scaled}}
\newcommand{\gm}{\gamma}

\newcommand{\wpm}{W^\pm}

\newcommand{\tpm}{T^\pm}

\newcommand{\beq}{\begin{equation}}
\newcommand{\eeq}{\end{equation}}
\newcommand{\ben}{\begin{enumerate}}
\newcommand{\een}{\end{enumerate}}

\newcommand{\sw}{s_{W}}

\newcommand{\br}{{\rm{Br}}}
\newcommand{\cdt}{{\rm{DCT}}}

\newcommand{\dm}{\Delta M}
\newcommand{\mh}{m_{h}}

\newcommand{\mtch}{M_{T^\pm}}
\newcommand{\mtz}{M_{T^0}}

\newcommand{\muf}      {\mu_{\rm f}}
\newcommand{\muc}      {{\mu_{\rm c}}}

\newcommand{\Dt}{\Delta}
\newcommand{\sg}{\sigma}

\newcommand{\lm}{\lambda}

\newcommand{\lmt}{\lambda_t}
\newcommand{\lmht}{\lambda_{ht}}
\newcommand{\drad}{c \tau \beta_T \gamma} 
\newcommand{\trad}{\rho_{\rm tr}} 
\newcommand{\pdct}{\mathcal{P}_{\rm DCT}} 
\newcommand{\pdctf}{\mathcal{P}_{\rm DCT} (\theta,\rho_{\rm d})} 

\newcommand{\lf}{\left(}
\newcommand{\ri}{\right)}

\newcommand{\met}      {{\rm MET}}

\newcommand{\cm}{{\,{\rm cm}}}

\newcommand{\iab}{{\,{\rm ab}^{-1}}}

\newcommand{\mev}{{\;{\rm MeV}}}
\newcommand{\gev}{{\;{\rm GeV}}}
\newcommand{\tev}{{\;{\rm TeV}}}

\newcommand{\tot}{{\rm tot}}
\newcommand{\trc}{{\rm Tr}}
\newcommand{\lumtot}{\mathcal{L}_{\rm tot}}
\newcommand{\dom}{\Delta\Omega}

\newcommand{\cmark}{\ding{51}}%
\newcommand{\xmark}{\ding{55}}%

\title{Probing Inert Triplet Model at a multi-TeV muon collider 
via vector boson fusion with forward muon tagging}

\author[a,b]{Priyotosh Bandyopadhyay,}
\author[a]{Snehashis Parashar,}
\author[a]{Chandrima Sen}
\author[c]{and Jeonghyeon Song}

\affiliation[a]{Indian Institute of Technology Hyderabad, Kandi,  Sangareddy-502284, Telengana, India}
\affiliation[b]{Korea Institute for Advanced Study, Seoul, 02455, Republic of Korea}
\affiliation[c]{Department of Physics, Konkuk University, Seoul 05029, Republic of Korea}

\emailAdd{bpriyo@phy.iith.ac.in} 
\emailAdd{ph20resch11006@iith.ac.in} 
\emailAdd{ph19resch11014@iith.ac.in}
\emailAdd{jhsong@konkuk.ac.kr}

\preprint{ IITH-PH-0002/23, 
KIAS-P23063}

\begin{document}

\abstract{ 
This study investigates the potential of a multi-TeV Muon Collider (MuC) for probing the Inert Triplet Model (ITM), which introduces a triplet scalar field with hypercharge $Y=0$ to the Standard Model. The ITM stands out as a compelling Beyond the Standard Model scenario, featuring a neutral triplet $T^0$ and charged triplets $T^\pm$. Notably, $T^0$ is posited as a dark matter (DM) candidate, being odd under a $Z_2$ symmetry. Rigorous evaluations against theoretical, collider, and DM experimental constraints corner the triplet scalar mass to a narrow TeV-scale region, within which three benchmark points are identified, with $T^\pm$ masses of 1.21 TeV, 1.68 TeV, and 3.86 TeV, for the collider study. The ITM's unique $TTVV$ four-point vertex, differing from fermionic DM models, facilitates efficient pair production through Vector Boson Fusion (VBF). This characteristic positions the MuC as an ideal platform for exploring the ITM, particularly due to the enhanced VBF cross-sections at high collision energies. To address the challenge of the soft decay products of $T^\pm$ resulting from the narrow mass gap between $T^\pm$ and $T^0$, we propose using Disappearing Charged Tracks (DCTs) from $T^\pm$ and Forward muons as key signatures. We provide event counts for these signatures at MuC energies of 6 TeV and 10 TeV, with respective luminosities of 4 ab$^{-1}$ and 10 ab$^{-1}$. Despite the challenge of beam-induced backgrounds contaminating the signal, we demonstrate that our proposed final states enable the MuC to achieve a $5\sigma$ discovery for the identified benchmark points, particularly highlighting the effectiveness of the final state with one DCT and one Forward muon. 
}

\maketitle
\flushbottom

\newpage

\section{Introduction}
\label{sec:Intro}

The landmark discovery of a Higgs boson with a mass of $125\gev$ at the LHC~\cite{ATLAS:2012yve,CMS:2012qbp}
has reinforced the foundations of the Standard Model (SM).
The significance of this Higgs boson is particularly profound due to its connections to several unresolved fundamental questions, including the nature of Dark Matter (DM)~\cite{Navarro:1995iw,Bertone:2004pz}, neutrino masses, the metastability of the SM vacuum~\cite{Degrassi:2012ry}, and the naturalness problem~\cite{Dimopoulos:1995mi,Chan:1997bi,Craig:2015pha}.
Given that these questions call for a new particle physics theory Beyond the SM (BSM),
the concept of an extended Higgs sector incorporating a DM candidate is both rational and intriguing.
However, despite intensive efforts to unveil potential extensions in the Higgs sector, 
no clues of new signals have emerged as of yet:
the properties of the observed Higgs boson align seamlessly with SM expectations,
and endeavors to detect additional scalar bosons have yielded no significant findings.
This situation strongly motivates the pursuit of a BSM theory that has eluded experimental detection up to this point.

In this context, the Inert Triplet Model (ITM) emerges as a compelling theoretical extension. 
The ITM introduces an additional $SU(2)_L$ triplet scalar field with hypercharge $Y=0$,
accommodating a neutral triplet $T^0$ and a pair of charged triplets $\tpm$. 
To incorporate a DM candidate $T^0$, the model adopts a discrete $Z_2$ symmetry,
under which the triplet scalars flip sign but the SM fields remain intact.
The theoretical appeal of the ITM is significant, offering potential solutions to a suite of core issues
such as the DM problem~\cite{Araki:2010nak,Araki:2011hm,YaserAyazi:2014jby,Jangid:2020qgo}, 
neutrino mass~\cite{Kajiyama:2013zla,Dong:2013ioa,Okada:2015vwh,Lu:2016dbc,Lu:2016ucn},
leptogenesis~\cite{Arina:2011cu,Lu:2016dbc,Lu:2016ucn},
strong first-order electroweak phase transition~\cite{Kazemi:2021bzj},
the metastability of the SM Higgs vacuum~\cite{Jangid:2020qgo}, and
the muon $g-2$ anomaly~\cite{DeJesus:2020yqx}. However, the Higgs portal quartic coupling ($\lambda_{ht}$), a large value of which is necessary for the first order phase transition \cite{Bandyopadhyay:2021ipw}, is constrained by the requirement of perturbativity up to the Planck scale,  along with the self quartic coupling ($\lambda_t$) \cite{Jangid:2020qgo, Bandyopadhyay:2021ipw}. Considering all these theoretical constraints along with DM relic and direct dark matter cross-section data, the inert triplet DM mass obtains a TeV-scale upper bound. Consequently, the detection of the new triplets in current high-energy collider experiments is challenging
due to several factors: the absence of tree-level couplings to fermions, 
the requirement for TeV-scale triplet masses to align with DM experimental data, and an extremely compressed mass spectrum.

A crucial question arises: how can we detect the signals of this theoretically robust yet experimentally elusive model 
at high-energy colliders? 
 One promising answer lies in the Disappearing Charged Track (DCT) phenomenon~\cite{Belyaev:2016lok,Curtin:2017bxr,Mahbubani:2017gjh,Fukuda:2017jmk,Lopez-Honorez:2017ora,Calibbi:2018fqf,Curtin:2018mvb,Han:2018wus,Alimena:2019zri,Saito:2019rtg,Bharucha:2018pfu,Belanger:2018sti,Filimonova:2018qdc,Jana:2019tdm,Chiang:2020rcv,Belyaev:2020wok,Calibbi:2021fld}.
This phenomenon is primarily due to the extremely small mass difference, $\mtch-\mtz \simeq 166\mev$~\cite{Cirelli:2005uq}, 
which results in a proper decay length of approximately $5.7$ cm for $\tpm$. 
As $\tpm$ is electrically charged, its trajectory can be tracked within the inner detector layers.
However, it vanishes before reaching the calorimeters, as its decay products remain elusive: 
in the dominant decay mode $\tpm \to \pi^\pm T^0$, 
the pions are too soft to reconstruct, and $T^0$ leaves no trace in any detector component.

Building on this understanding, the HL-LHC has been explored for its potential in probing the ITM using the DCT phenomena
in the Drell-Yan (DY) production accompanied by an additional photon~\cite{Chiang:2020rcv}.
However, there is a limitation: the detectable mass range is restricted to $\mtz \lsim 520\gev$, 
even when considering an integrated luminosity of $\lumtot=3\iab$~\cite{ATLAS:2017oal}. 
This range is less than ideal, as it fails to cover even 10\% of the observed DM relic density~\cite{Planck:2018vyg}.
Looking towards the future, projections for a 100 TeV proton-proton collider with a luminosity of $30\iab$ 
indicate a potential for detecting triplet scalar masses up to $3\tev$, if pileup effects are efficiently mitigated~\cite{Chiang:2020rcv}. 
However, reaching such a high luminosity remains a long-term goal, underscoring the need for alternative research avenues.
 
A promising alternative is a multi-TeV muon collider (MuC)~\cite{Palmer:1996gs,Ankenbrandt:1999cta}. 
The MuC's unique ability to achieve multi-TeV center-of-mass (c.m.) energies in a compact setup, combined with the cleaner collision environment of muon interactions, positions it as a powerful tool for BSM searches \cite{Capdevilla:2020qel, Bandyopadhyay:2021pld, Sen:2021fha, Asadi:2021gah, Huang:2021nkl}.
Unlike hadron colliders like the LHC, where the hard scattering process depends on parton-parton collisions within protons, the MuC utilizes collisions between two fundamental particles, thus fully harnessing the beam energy.
This feature significantly enhances the efficiency of the MuC in exploring heavy BSM particles, including DM searches~\cite{Choi:1999kn,Han:2020uak,Han:2021udl,Jueid:2021avn,Han:2022edd,Black:2022qlg, Belfkir:2023vpo, Jueid:2023zxx}.
Consequently, the feasibility and potential of the MuC have been increasingly recognized, especially since the European strategy endorsed MuC research, culminating in the establishment of the Muon Collider Collaboration in 2020~\cite{Schulte:2021eyr}. 
Furthermore, recent advancements in addressing critical challenges, such as cooling muon beams~\cite{Antonelli:2013mmk,Antonelli:2015nla} and reducing beam-induced backgrounds (BIB)~\cite{Collamati:2021sbv,Ally:2022rgk}, have significantly enhanced the prospects of the MuC program.

There exist extensive studies on DM signatures employing DCTs at the MuC, focusing on
the scalar/fermion $SU(2)_L$ multiplet, Higgsino, and Wino DM candidates~\cite{Bottaro:2021snn, Bottaro:2022one,Han:2020uak,Han:2022ubw,Capdevilla:2021fmj, Asadi:2023csb}. 
These explorations have been centered around DY pair production, 
using various triggers such as mono-photon, mono-$V$ ($V=Z,\wpm$), $VV'$, or mono-muon, 
within a pseudorapidity coverage of $\abs{\eta} \leq 2.5$~\cite{MuonCollider:2022ded}.
However, a dedicated study of ITM signatures at the MuC remains uncharted, 
which we aim to address in this paper. 
Our initial step involves updating the allowed parameter space for the ITM, incorporating the latest DM experimental constraints from LUX-ZEPLIN (LZ)~\cite{LZ:2022ufs} and the Fermi-LAT~\cite{MAGIC:2016xys} and HESS~\cite{HESS:2022ygk} experiments, as well as theoretical constraints from Planck scale perturbativity \cite{Jangid:2020qgo, Bandyopadhyay:2021ipw}. We will show that the masses of triplet scalar bosons are confined to an upper bound of $\sim 4$ TeV.

In our investigation of the  phenomenological signatures of ITM at the MuC, 
we propose two innovative approaches.
First, we will focus on Vector Boson Fusion (VBF) processes, 
which are distinct from DY processes in that their cross-sections increase with c.m.~energy.
This characteristic makes VBF processes exceptionally well-suited for exploiting the high collision energy of the MuC~\cite{Costantini:2020stv, Bandyopadhyay:2020otm}.
Second, we utilize Forward ($\abs{\eta} > 2.5$) muons~ \cite{Ruhdorfer:2023uea, Forslund:2023reu} as innovative triggers in neutral-current VBF processes, moving beyond the conventional pseudorapidity coverage limit of $|\eta|<2.5$. 
This typical limit, primarily imposed due to tungsten nozzles designed to shield the detector by absorbing the majority of soft BIB particles~\cite{Collamati:2021sbv, Ally:2022rgk}, is not an actual barrier
for high-energy Forward muons, as they can effectively penetrate the nozzles.
Recent initiatives within the MuC community to integrate Forward muon detectors into the collider's design further reinforce this approach~\cite{Accettura:2023ked,Ruhdorfer:2023uea, Forslund:2023reu}.

Our new strategy---employing VBF pair production of triplet scalars with Forward muon triggering---is particularly robust in probing ITM, owing to
the presence of $TTVV$-type vertices (where $V \equiv W^\pm/ Z$ and $T=T^0,\tpm$).
These vertices, absent for fermionic DM like the Higgsino and Wino, 
notably enhance the VBF cross-sections, 
thereby enabling a more effective analysis of the DCT signals. 
Utilizing the comprehensive DCT reconstruction efficiencies in Ref.~\cite{Capdevilla:2021fmj}, 
we will demonstrate that final states featuring a DCT and a Forward muon at a 10 TeV MuC 
can effectively probe the viable parameter space of the ITM. 
Our findings will highlight the MuC as an ideal platform for investigating the ITM, 
marking a significant contribution to the field.

The structure of this paper unfolds in the following manner: 
We begin with a concise review of the ITM in Sec.~\ref{sec:model}, 
where we also discuss the constraints from theoretical requirements and collider experiments. 
Following this, Sec.~\ref{sec:DM:constraints} details the identification of the parameter space
which is consistent with various DM measurements. 
This section also introduces three benchmark points for further study at the MuC.
In Sec.~\ref{sec:golden:channels}, our focus shifts to the VBF pair production of the triplet scalars at the MuC. 
Here, we identify four distinct final states that involve DCTs and Forward muons. 
Section \ref{sec:characteristics:signal} delves deeper into the signal characteristics. 
Particular attention is given to the long-lived $\tpm$, 
along with the reconstruction efficiencies of DCTs and the detection feasibility of Forward muons.
Our results are presented in Sec.~\ref{sec:results}, 
where we report the number of events for each final state at two distinct MuC energies, 6 TeV and 10 TeV. 
Furthermore, in Sec.~\ref{sec:projection}, we discuss the luminosity projections required for $5\sigma$ discoveries over a range of triplet scalar masses. 
The paper concludes with Sec.~\ref{sec:conclusions}, 
where we wrap up our study with key conclusions and insights.

\section{Inert Triplet Model}\label{sec:model}

\subsection{Brief review of ITM}
\label{sec:review}

In the ITM, the SM Higgs sector is extended by introducing an $SU(2)_L$ triplet scalar field $\mathcal{T}$, 
which carries a hypercharge of $Y=0$ in the $Q=T_3 + Y/2$ convention.\footnote{The ITM variant with $Y=2$ is significantly constrained by DM direct detection experiments due to the notably enhanced spin-independent cross-section via $Z$-mediated scattering processes~\cite{Araki:2011hm}. Consequently, we have excluded this scenario from our analysis.}
The SM scalar doublet $\Phi$ (with a hypercharge of $Y=1$)
and the new scalar field $\mathcal{T}$ are described as:
\begin{equation}	
	\Phi = \begin{pmatrix}
		\phi^+ \\ \phi^0
	\end{pmatrix}, \quad \mathcal{T} = \frac{1}{2} \begin{pmatrix}
		T^0 & \sqrt{2}T^+ \\ \sqrt{2}T^- & -T^0 
	\end{pmatrix}. \label{eq:tripdef}
\end{equation}
Thus, this model introduces three new scalar bosons: a neutral $CP$-even scalar $T^0$ and charged scalars $T^\pm$.
To incorporate a DM candidate, we adopt a discrete $Z_2$ symmetry,
under which all SM particles are designated as even
while the newly introduced triplet field is classified as odd.  
This classification ensures that the new scalar bosons do not directly couple to SM fermions, 
thereby earning the mode its name as the \emph{Inert} Triplet Model. 
In addition, the neutral component of $\mathcal{T}$ 
does not acquire a non-zero vacuum expectation value. 
Within this ITM, therefore, 
 it is solely the Higgs doublet $\Phi$ that is responsible for electroweak symmetry breaking,
as $\Phi$ acquires a non-zero VEV given by $\phi^0 = \left( v+h^0 \right) /\sqrt{2}$ (where $v \approx 246\gev$).

In our proposed model, the Lagrangian for  the scalar fields is specified as:
\begin{equation}
\label{eq:Lagrangian}
\mathcal{L} = \left| D_\mu \Phi \right|^2 + \text{Tr} \left[ \left( D_\mu \mathcal{T} \right)^\dagger \left( D^\mu \mathcal{T} \right) \right] - V(\Phi, \mathcal{T}),
\end{equation}
where the covariant derivative of the triplet scalar field is  defined by
\begin{equation}
D_\mu \mathcal{T} = \partial_\mu \mathcal{T} - i \frac{g_2}{2} \left[\tau^a W^a_\mu , \mathcal{T}\right].
\end{equation}
Then, the gauge interaction Lagrangian manifests as:
\begin{align}
\label{eq:gauge}
\mathcal{L}_{\text{gauge}} &= {i g_2}\left( \sin\theta_W A_\mu + \cos\theta_W Z_\mu \right)  ( T^- \overset{\leftrightarrow}{\partial^\mu} T^+ ) \\[2pt]
&\quad -ig_2  W^-_\mu  ( T^0 \overset{\leftrightarrow}{\partial^\mu}  T^+ )
+ ig_2  W^+_\mu ( T^0\overset{\leftrightarrow}{\partial^\mu}  T^- ) \nonumber \\[2pt]
&\quad - g_2^2 T^0 \left( \sin\theta_W A_\mu + \cos\theta_W Z_\mu \right) ( T^- W^{+ \mu} + T^+ W^{- \mu} ) \nonumber \\[2pt]
&\quad + g_2^2 T^+ T^-  \left( \sin\theta_W A_\mu + \cos\theta_W Z_\mu \right)^2 + g_2^2 W^{+}_\mu W^{- \mu} ( T^0 T^0 + T^+ T^- ) \nonumber \\[2pt]
&\quad -\frac{g_2^2}{2} \left( W^+_\mu W^{+ \mu} T^- T^- + W^-_\mu W^{- \mu} T^+ T^+  \right) , \nonumber
\end{align}
where the derivative is represented by $(  f \overset{\leftrightarrow}{\partial^\mu} f' ) = f\partial^\mu f' -(\partial^\mu f) f'$, $g_2$ is  the $SU(2)_L$ gauge coupling, and $\theta_W$ is the electroweak mixing angle.

The scalar potential $V(\Phi, \mathcal{T})$, conforming to the $Z_2$ symmetry, is given by
\begin{align} \label{eq:itm_pot}
	V(\Phi, \mathcal{T}) \ &= \mu_{h}^2 \, \Phi^\dagger \Phi + \mu_{T}^2\, \trc \lf \mathcal{T}^\dagger \mathcal{T}\ri
	+\lambda_{h} \left|\Phi^\dagger \Phi \right|^2 
	+\lmt \lf \trc \left| \mathcal{T}^\dagger \mathcal{T} \right| \ri^2 +
	\lmht \lf \Phi^\dagger \Phi\ri \trc \lf \mathcal{T}^\dagger \mathcal{T}\ri .
\end{align} 
Consequently, the ITM is characterized by three parameters:
\begin{equation}
\label{eq:model:parameters}
\left\{ \mu_T, \; \lmt, \; \lmht \right\}.
\end{equation}
For the DM and collider phenomenologies of the ITM, the parameters $\mu_T$ and $\lmht$ are of primary focus.
While $\lmt$ does not directly impact the phenomenologies,
it plays a significant role in modulating the behavior of $\lmht$ under renormalization group equations. 
The theoretical stability of the model up to the Planck scale requires $\lmt$ to be 
within the range of [0.01, 0.8]~\cite{Jangid:2020qgo}. 
In this study, we have chosen to set $\lmt$ at a value of 0.2.  
This decision aligns with both perturbativity constraints and the theoretical stability requirements outlined in the literature. 

The scalar potential in \autoref{eq:itm_pot} determines the tree-level mass spectrum:
\begin{align}
\label{eq:mass}
	\mh^2 &= 2 \lambda_h v^2, \\
	\mtz^2 &= \mtch^2 = \frac{1}{2} \lmht v^2 + \mu^{2}_T . \nonumber
\end{align}
At the tree level, $T^\pm$ and $T^0$ exhibit mass degeneracy. 
However, a small mass deviation emerges due to loop corrections through the interactions with SM gauge bosons. 
At the one-loop level, the mass difference is given by~\cite{Cirelli:2005uq}:
\begin{align}
	\Delta M &\equiv \mtch-\mtz\\
	\nn
	&= \frac{\alpha_2}{4 \pi} M_{T^0} \Bigg[\mathcal{F} \left(\frac{m_W}{M_{T^0}}\right) +(s_W^2 -1) \mathcal{F} \left(\frac{m_Z}{M_{T^0}}\right)\Bigg],
\end{align}
where the form factor $\mathcal{F}(x)$ is defined as
\bea
 \mathcal{F}(x)= -\frac{x}{4}\left[2x^3 \ln x + (x^2 -4)^{\frac{3}{2}} \ln \frac{x^2 -2 -x \sqrt{x^2 -4}}{2}  \right] .
 \eea

\begin{figure}[t]
	\centering
	\includegraphics[width=0.65\linewidth]{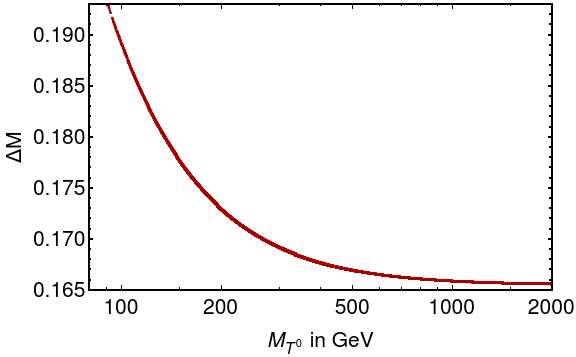}
	\caption{The mass difference between $\tpm$ and $T^0$ as a function of $\mtz$. 
	This mass difference is computed at the one-loop level, induced by the interactions with SM gauge bosons.}
	\label{fig:mass_split}
\end{figure}

In \autoref{fig:mass_split},
we present the mass splitting $\Dt M$ as a function of  $M_{T^0}$.
It  is evident that for a higher mass scale ($M_{T^0} \gsim 800$\,GeV), 
$\Delta M$ stabilizes around 166 MeV. 
As $\mtz$ decreases, the mass difference experiences a slight increase but consistently remains below 200 MeV.
Notably, even with the inclusion of two-loop corrections, 
the mass difference undergoes only minor changes, in the order of a few MeV~\cite{Ibe:2012sx}. 
As $T^0$ is the lightest stable particle with an odd $Z_2$ parity, it emerges as a viable DM candidate.

Let us now examine the decay modes of the charged triplet $\tpm$.
Given that the mass difference exceeds the mass of pion, 
$\tpm$ predominantly decays into $T^0 \pi^\pm$, $T^0 e^\pm \nu_e$, or $T^0 \mu^\pm \nu_\mu$.
The partial decay widths and the corresponding branching ratios are as follows~\cite{Cirelli:2005uq,Franceschini:2008pz}:
\begin{align}
\label{eq:width}
	\Gamma (T^\pm \to T^0 \pi^\pm ) &= \frac{2 G_F^2\, V_{ud}^2 \, \Delta M^3\, f_\pi^2}{\pi} \sqrt{1- \frac{m_\pi^2}{\Delta M^2}} ,
	&(&\br \simeq 97.7\%),\\[3pt] \nn
	\Gamma (T^\pm \to T^0 e^\pm \nu_e) &= \frac{2 G_F^2 \, \Delta M^5}{15 \pi^3} ,& (&\br \simeq 2\%),\\[3pt] \nn
	\Gamma (T^\pm \to T^0 \mu^\pm \nu_\mu) &= 0.12\,\, \Gamma (T^\pm \to T^0 e^\pm \nu_e) , &(&\br \simeq 0.25\%),
\end{align}
where $f_\pi \simeq 130.4\mev$ is the decay constant of the pion~\cite{Colangelo:2000zw}.
The overwhelmingly dominant decay channel for $\tpm$ is into $T^0 \pi^\pm$.

Another critical observation from \autoref{eq:width} is 
that the partial decay rate of the dominant $\tpm\to T^0 \pi^\pm$ is directly proportional to $\Delta M^3$. 
Consequently, the highly compressed mass spectrum leads to a notably small total decay width for $\tpm$, 
resulting in a substantial proper lifetime of $0.19\,$ns. 
This corresponds to a decay length for $T^\pm$ in its rest frame of approximately 5.7 cm~\cite{Han:2022ubw}. 
Such a decay length is sufficiently long to be explored by high-energy collider detectors, 
emphasizing its crucial significance in experimental observations.

\subsection{Viable parameter space from the theoretical and collider constraints}
\label{sec:viable:theory:collider}

Prior to investigating the implications from diverse DM observations, 
it is imperative to identify the viable parameter points 
that are consistent with theoretical requirements and collider constraints. 
To this end, we consider the following aspects:\footnote{We do not consider the constraints from the electroweak oblique parameters, 
$S$, $T$, and $U$~\cite{ParticleDataGroup:2022pth}, 
since the extremely small $\Dt M$ and the TeV-scale triplet masses 
result in negligible contributions from the triplet scalar bosons.
Note that in the ITM, $S=0$, $T \simeq (\dm)^2/(24 \pi \sw^2 m_W^2)$, 
and $U \simeq \dm/(3\pi \mtch)$~\cite{Chakrabarty:2021kmr}. 
}
\begin{description}
\item[Theoretical requirements]~
\begin{itemize}
	\item[--] All quartic couplings, $\lambda_h$, $\lmt$, and $\lmht$, must be perturbative, 
	i.e., $|\lm_{h,ht,t}| \leq 4\pi$.
\item[--] The unitarity condition imposes an additional requirement: $|\lmt| \leq \pi$.
\item[--] The scalar potential should be bounded-from-below, necessitating
\bea
\lambda_{h} > 0, \quad \quad \lmt > 0, \quad \quad 2\sqrt{\lambda_{h} \lmt} > |\lmht|.
\eea
\end{itemize}
Given that $\lambda_{h} = 0.129$, all three conditions collectively lead to
\bea
-1.27 < \lmht < 4\pi.
\eea
\item[Higgs precision data at the LHC]~\\
The charged triplets can significantly affect 
di-photon decay rate of the Higgs boson via loop corrections, resulting in the partial decay width given by
\begin{equation}
	\Gamma(h\to\gamma\gamma) = \frac{G_F \alpha^2 m_h^3}{128\sqrt{2}\pi^3}
	\abs{\sum_{f}N_c Q_f^2 \mathcal{R}_{1/2}(\xi_f) + \mathcal{R}_1(\xi_W) + \frac{\lmht v^2}{2M_{T^\pm}^2}\mathcal{R}_0(\xi_{T^\pm})}^2,
	\label{eq:hrr}
\end{equation}
where $N_c$ and $Q_f$ are the color factor and the charge of the fermions, respectively,
and $\xi_a = {4 M_a^2}/{m_h^2}$.
The form factors for spin-0, spin-1/2, and spin-1 are
\begin{align}
	\mathcal{R}_0(\xi) &= -\xi\left[1-\xi \, \mathcal{P}(\xi)\right], \\[3pt] \nn
	\mathcal{R}_{1/2}(\xi) &= 2\xi\left[1+(1-\xi)\mathcal{P}(\xi)\right] , \\[3pt]\nn
	\mathcal{R}_1(\xi) &= -\left[2+3\xi+3\xi(2-\xi)\mathcal{P}(\xi)\right] \label{eq:a1}, 
\end{align}
where
\begin{equation}
	\mathcal{P}(\xi) = 
	\begin{cases}
		\arcsin[2](\xi^{-\frac{1}{2}}) & \text{if } \xi \leq 1, \\[5pt]
		-\frac{1}{4}\left[\ln(\frac{1+\sqrt{1-\xi}}{1-\sqrt{1-\xi}} - i\pi)\right]^2 & \text{if } \xi > 1.
	\end{cases}
\end{equation}
The di-photon signal modifier $\mu_{\gamma \gamma}= \Gamma(h \to \gamma \gamma)/\Gamma(h \to \gamma \gamma)|_{\text{SM}}$ is constrained to be $\mu_{\gamma\gamma} =1.04_{-0.09}^{+0.10}$ at the $1\sigma$ limit
by the recent ATLAS measurement~\cite{ATLAS:2022tnm},
which limits $\lmht$ through \autoref{eq:hrr}.
\end{description}

\begin{figure}[h]
	\centering
	\includegraphics[width=0.6\linewidth]{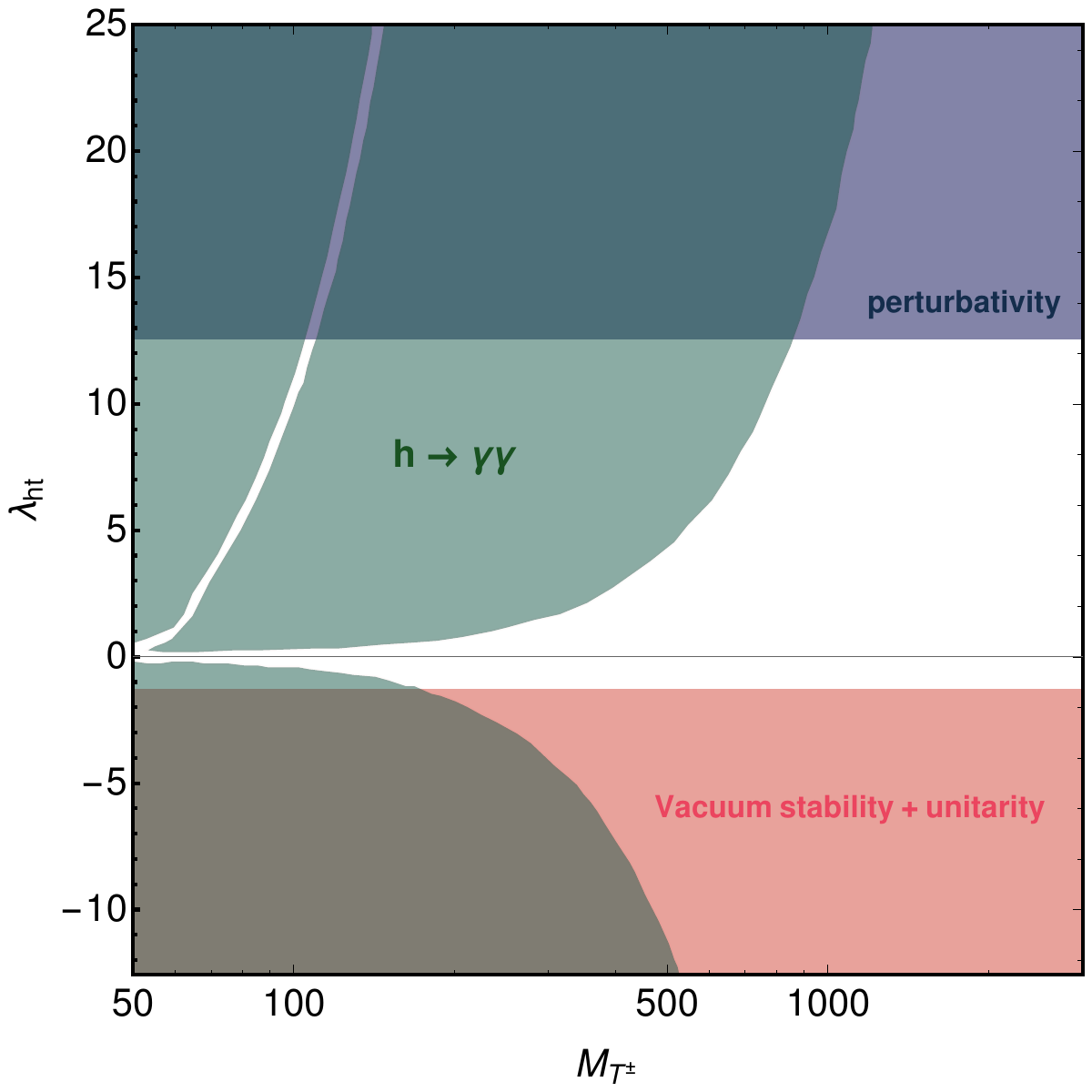}
	\caption{Excluded regions in the $(\mtch,\lmht)$ parameter space, indicated by the perturbativity constraint (blue), vacuum stability and unitarity requirements (pink), and the $1\sigma$ constraints from the Higgs boson decay into two photons (green).}
	\label{fig:bounds}
\end{figure}

In \autoref{fig:bounds}, we illustrate the excluded regions within the parameter space $(\mtch,\lmht)$, 
based on the constraints discussed earlier. 
The combined criteria for vacuum stability and unitarity exclude the region shown in pink, 
effectively ruling out a substantial portion of the negative $\lmht$ space. 
The perturbativity condition disqualifies the region where $\lmht > 4\pi$, shown in blue.
Additionally, constraints from the decay rate of $h \rightarrow \gamma\gamma$ eliminate a large area of the parameter space, especially for lighter $M_{T^0}$ values. When $M_{T^0} \lesssim 200$ GeV, the magnitude of $\lmht $ is constrained to be less than one. Conversely, for heavier $M_{T^0}$ values, such as those above 1 TeV, the parameter space with sizable $\lmht $ remains viable. In summary, the white region in \autoref{fig:bounds} represents the parameter space that meets all theoretical and collider-imposed requirements.

\section{Dark Matter constraints}
\label{sec:DM:constraints}

In this section, we reassess the constraints on the ITM parameter space in light of the latest DM observations, focusing on the higher masses of the viable parameter space delineated in the previous section. 
This reassessment is conducted via an extensive random parameter scan within the following ranges:
\bea
\label{eq:scan:range}
\mu_T \in [500.0, 5000.0]\gev, \quad 
\lmht \in [0.0,2.0], \quad \lmt = 0.2.
\eea

We reiterate that $\lmt$ does not exert a significant influence on the DM sector. The upper limit on $\lmht$ is chosen with consideration of bounds from Planck-scale perturbativity of the model~\cite{Jangid:2020qgo, Bandyopadhyay:2021ipw}. The scan over the DM masses terminates at 5 TeV, motivated by the kinematical feasibility of pair production of the scalars at a future 10 TeV Muon Collider \cite{Accettura:2023ked}. 

Our analysis incorporates constraints from three key domains: relic density, direct detection, and indirect detection. 
For the DM calculations, we initially generated the \chep model files~\cite{Belyaev:2012qa} 
by incorporating the new Lagrangian into \sarah~\cite{Staub:2013tta}. 
Subsequently, we utilized \mic version 5.3.41~\cite{Belanger:2001fz}. 
In the subsequent subsections, we will provide detailed accounts of these constraints. 

\subsection{DM relic abundance}
\label{subsec:relic}

The triplet DM particles, initially in thermal equilibrium with the SM particles in the early universe, 
froze out when their annihilation rate fell below the Hubble expansion rate. 
The latest result for the DM relic density, 
as reported by the Planck Collaboration based on data from the Planck 2018 release~\cite{Planck:2018vyg}, is
\begin{equation}
\label{eq:Omega:Planck}
\ompl h^2 = 0.1198 \pm 0.0012,
\end{equation}
where $\Omega_{\rm DM}$ denotes the ratio of the DM relic density to the critical density $\rho_c = 3 H^2/(8\pi G_N)$. 
Here, $H$ represents the Hubble constant, expressed as $H = 100h$~km/s/Mpc. 
Roughly, $\Omega_{\rm DM} h^2$ is inversely proportional to $\langle \sigma_{\rm ann} v \rangle$,
the thermal average of the DM annihilation cross-section multiplied by their relative velocity. We will be denoting this observed relic as $\Omega_{\rm obs}$ for the rest of the article.

We now turn our attention to DM annihilation at the tree level,  within the ITM framework. 
The DM candidate $T^0$ has four primary annihilation channels: 
 $T^0 T^0 \to W^+ W^-,\, ZZ,\, h h, \, f\bar{f}$, where $f$ represents the charged SM fermions. 
The most dominant annihilation mode is $T^0 T^0 \to W^+ W^-$, due to $T^0$ being charged under $SU(2)_L$. 
This channel proceeds through three mechanisms:  the four-point vertex $T^0$-$T^0$-$W^+$-$W^-$, 
the $t$-channel exchange of $T^\pm$, and the $s$-channel exchange of the Higgs boson. 
The other annihilation channels into $ZZ,\,  hh, \,f\bar{f}$ contribute significantly less to the relic density, as they predominantly occur through $s$-channel Higgs mediation.

\begin{figure}[h]
	\centering
	\subfigure[]{\includegraphics[width=0.455\linewidth]{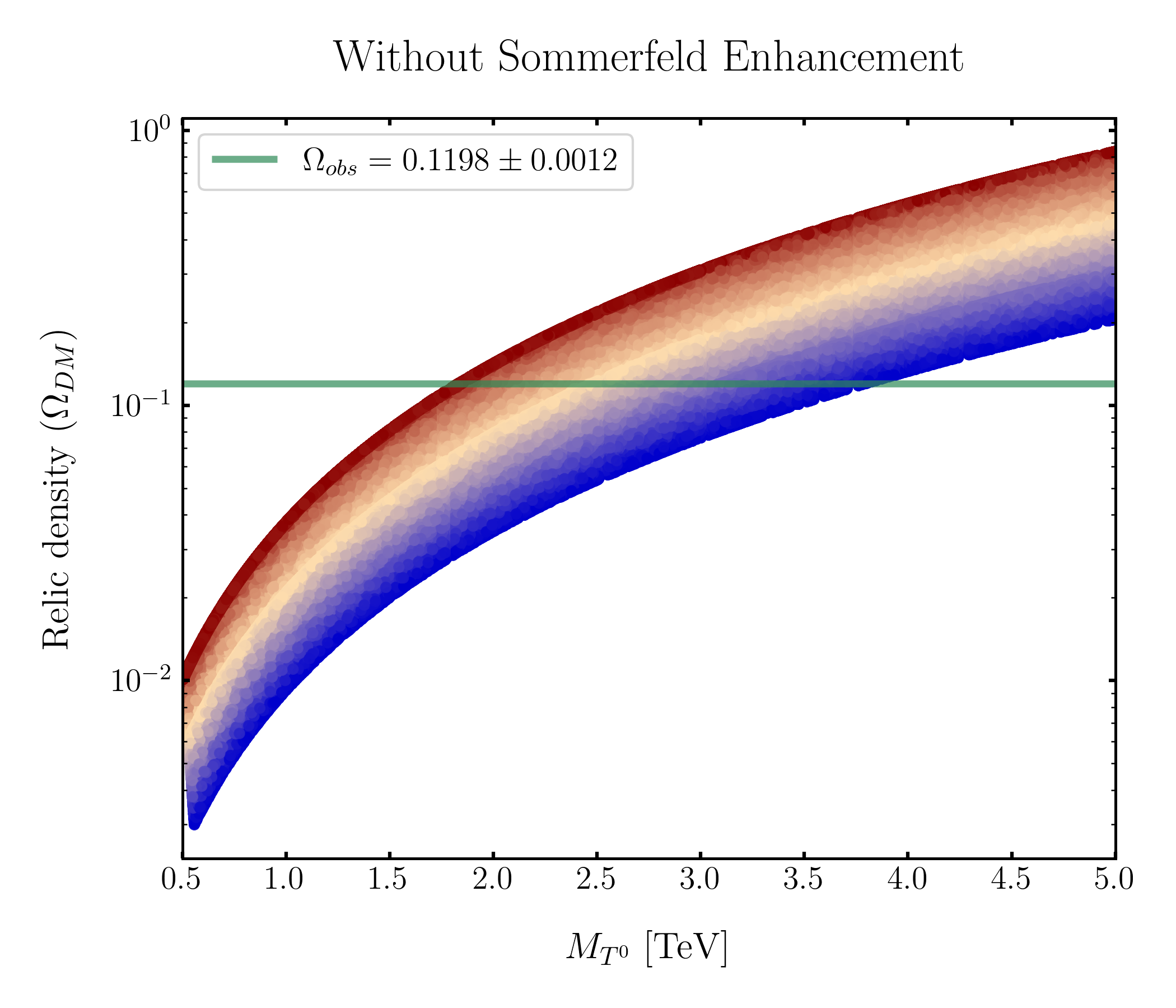}}
	\subfigure[]{\includegraphics[width=0.52\linewidth]{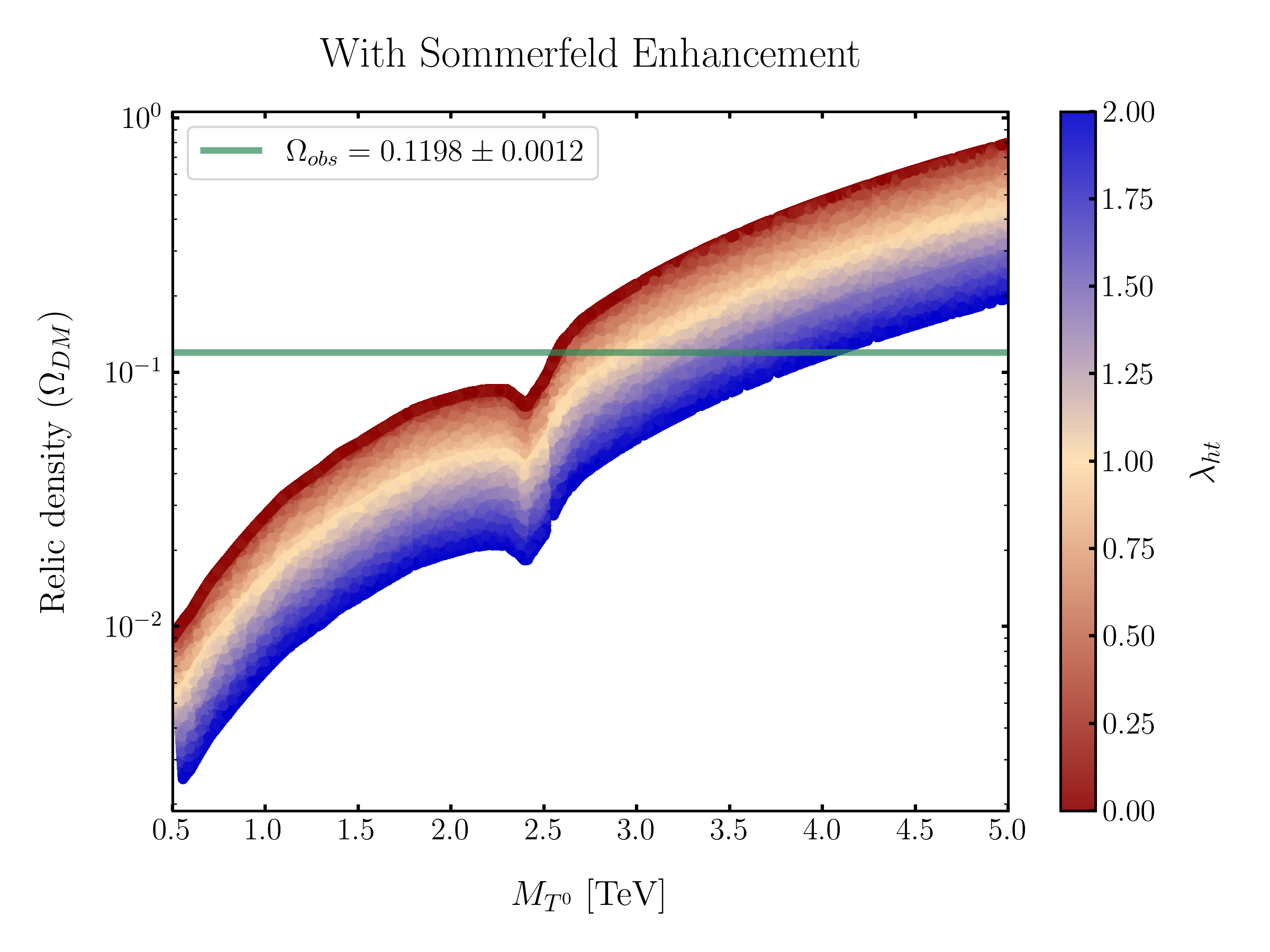}}
	\caption{Relic abundance ($\Omega_{\rm DM}$) plotted against the DM mass $\mtz$, color-coded by $\lmht$ (a) without Sommerfeld enhancement, (b) with Sommerfeld enhancement. 
	The horizontal dark green band represents the observed DM relic abundance by the Planck Collaboration~\cite{Planck:2018vyg}, specifically $\Omega_{\rm obs}= 0.1198 \pm 0.0012$.}
	\label{fig:rel}
\end{figure}

In the ITM, the additional contributions from coannihilation processes to the DM relic density
are sizable due to the small mass difference of $\Delta M \simeq 166\mev$~\cite{Griest:1990kh,Mizuta:1992qp}. 
Coannihilation occurs through the scatterings of $T^0 T^\pm$ and $T^+ T^-$. 
The former includes $T^0 T^\pm \to W^\pm Z, \, W^\pm \gamma, \, W^\pm h, \, f\bar{f}'$, 
while the latter comprises $T^+ T^- \to W^+W^-$, $ZZ$, $Z\gamma$, 
$\gamma\gamma$, $hZ$, $hh$, $h\gamma$, $f\bar{f}$, and $\nu\nu$. 
The most dominant coannihilation process is $T^0 T^\pm \to Z W^\pm$, 
which occurs via the $t$-channel exchange of $T^\pm$, 
the $s$-channel exchange of $W^\pm$, and a contact term involving $T^0$-$T^\pm$-$Z$-$W^\pm$.  In \autoref{fig:rel}(a) we present the DM relic $\Omega_{\rm DM}$ against the DM mass $\mtz$ for the tree-level annihilation and co-annihilation processes. A closer examination reveals notable correlations between the model parameters in accounting for the relic density. 
For a fixed $\lmht$, a higher $\mtz$ results in a larger relic density. 
This is attributed to smaller annihilation cross-sections for heavier triplet scalars, 
leading to a larger $\omdm$. 
For a fixed $\mtz$, the relic density decreases with increasing $\lmht$. 
This is because a higher $\lmht$ increases the Higgs boson couplings to $T^0 T^0$ and $T^+ T^-$, 
leading to larger annihilation cross-sections and thus a smaller relic density. 

However, as established in existing literature, these annihilations are enhanced by the so-called Sommerfeld Enhancement (SE), especially at higher masses where the SM gauge bosons can be treated as massless~\cite{Cirelli:2007xd, Hisano:2004ds}. A detailed study of the SE in context of this model can be found in contemporary literature~\cite{Katayose:2021mew}. To incorporate the SE into our \mic output, we obtain the ratios between the red and blue curves in the top left panel of the Figure 3 in Ref.~\cite{Cirelli:2007xd}, and multiply the factors to the relic density provided by \mic (the same procedure is followed by Ref.~\cite{Chiang:2020rcv}).

In \autoref{fig:rel}(b), we present the relic density ($\Omega_{\rm DM}$) values for the scanned parameter points, with $\mtz$ in the x-axis, keeping a color map over  $\lmht$, with the SE factors multiplied as mentioned before. The dark green horizontal band represents the DM relic abundance determined by Planck measurements,  as shown in \autoref{eq:Omega:Planck}. The results clearly demonstrate that the observed DM relic density can be accounted for within the ITM framework, within the ranges of $\mtz$ and $\lmht$ chosen for our scan.

In accordance with the established results for a scalar triplet DM, a sharp dip is observed at $\mtz \sim 2.4$ TeV, corresponding to the first zero-energy resonance peak of the Sommerfeld enhancement~\cite{Cirelli:2007xd, Hisano:2004ds, Chun:2015mka, Katayose:2021mew}. This effect keeps the model underabundant in relic for a higher range of masses that what is expected, if one considers only tree-level (co)-annihilation of the DM, which can be compared using \autoref{fig:rel}(a). From the \autoref{fig:rel}(b) we see that, for $\lmht = 0.0$, the lowest possible mass that satisfy the observed relic is $\mtz \approx 2.53$ TeV. With the increase in $\lmht$, this mass bound increases as the annihilation cross-sections are further enhanced in the Higgs portal. For an arbitrarily large $\lmht$, one can expect this relic-satisfying mass value to also increase arbitrarily. However, we restrict $\lmht$ to an upper limit of 2.0, in order to avoid any unwanted constraints from the perturbativity of the theory at some higher scale~\cite{Jangid:2020qgo, Bandyopadhyay:2021ipw}. With this upper bound on $\lmht$, the maximum possible mass that satisfies the observed relic for this model comes out to be $\sim 4.0$ TeV. Additionally, while an ideal DM model needs to always satisfy the observed relic, the points that are underabundant in relic are still allowed by the Planck data. For the prospect of early discovery of the triplet scenario at the colliders, it is crucial to consider the lower masses of the model as well, with some partial contribution to the total relic density of DM in our universe. Hence, we extract all the points in \autoref{fig:rel}(b) that either satisfy the correct relic or are underabundant, and proceed to the next subsection where we impose the constraints coming from the direct detection  experiments for Dark Matter.



\subsection{DM direct detection bounds}

Several DM direct detection experiments are currently in operation, 
aiming to detect signals resulting from the elastic scattering of DM with atomic nuclei. 
Most of these experiments are specifically designed 
to investigate the spin-independent interactions between DM particles and the target nuclei 
 with a high atomic number and mass.

In the ITM, the spin-independent scattering occurs through $t$-channel Higgs boson exchange. 
The cross-section $\sigma_{\rm SI}$ at the leading order is described by~\cite{Chao:2018xwz}:
\begin{equation}
	\label{eq:sigma:SI}
	\sigma_{\rm SI} \simeq \frac{\lmht^2 f_N^2}{4\pi \mh^4} \frac{M_N^4}{(M_N + M_{T^0})^2},
\end{equation}
where $f_N \approx 0.287\,(0.084)$ for $N=p\,(n)$ represents the nucleon form factor~\cite{Belanger:2013oya}, 
$\mh$ is the Higgs boson mass, and $M_N$ is the mass of the nucleon.

When comparing the model's prediction for $\sigma_{\rm SI}$ with the observed upper bounds on $\sigma_{\rm SI}$, it is essential to consider a scaling factor, ${\omdm}/{\Omega_{\rm obs}}$. If the model's predicted relic density is lower than the observed value $\Omega_{\rm obs}$, there is a corresponding reduction in the actual DM-nucleon scattering cross-section. Thus, we should account for this reduction factor.
Consequently, we introduce a scaled spin-independent cross-section, $\ssgsi$, defined as:
\bea
\label{eq:scale:SI:xsec}
\ssgsi  \equiv \frac{\omdm}{\Omega_{\rm obs}} \times  \sg_{\rm SI} .
\eea

\begin{figure}[h]
	\centering
	\includegraphics[width=0.7\linewidth]{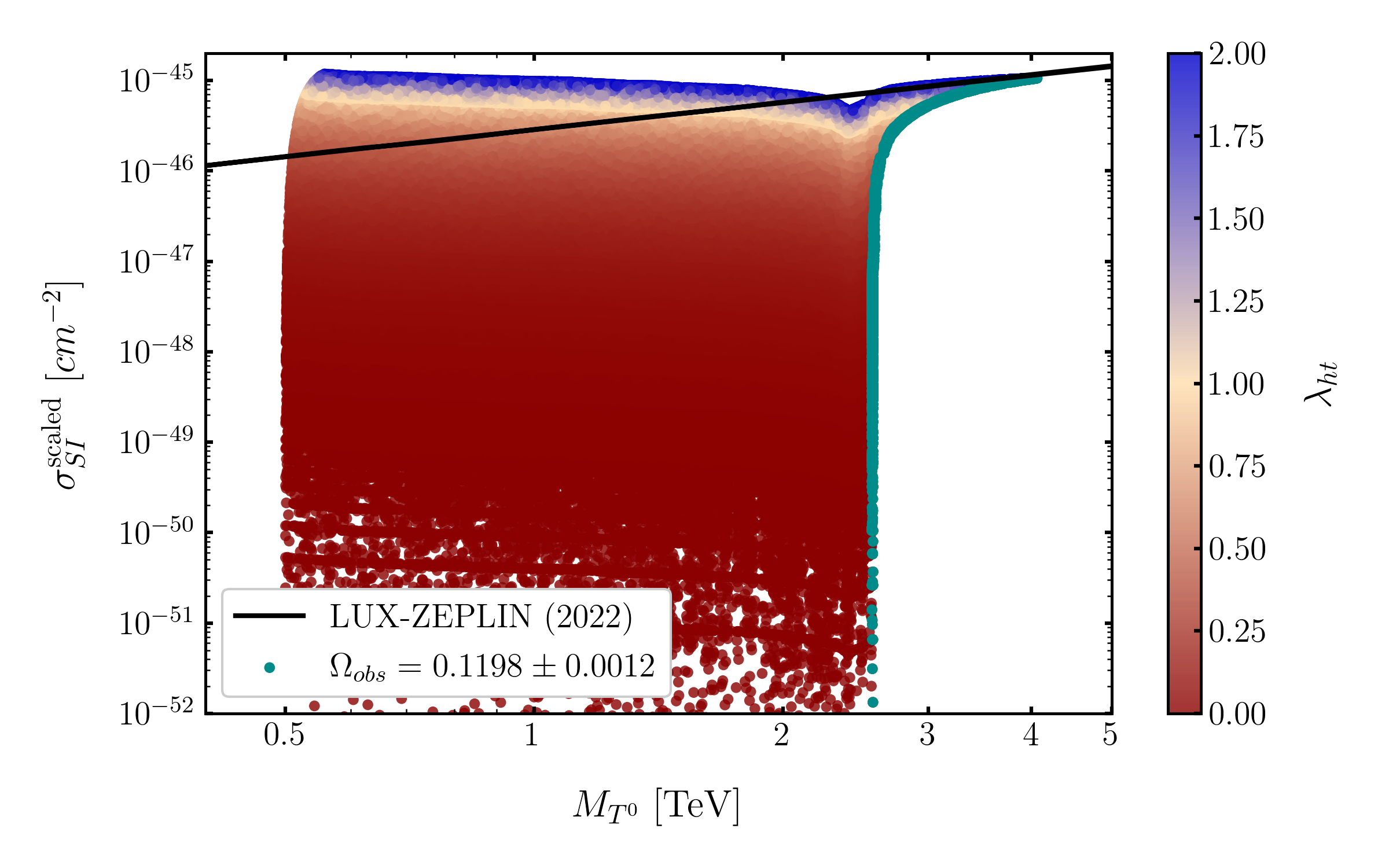}
	\caption{Scaled spin-independent DM-nucleon scattering cross-section of $T^0$ as a function of $\mtz$,
	over the parameter points avoiding the over-closure. 
	We include the upper bound on $\sgsi$ from the LUX-ZEPLIN experiment\cite{LZ:2022ufs} (black solid line). 
	The green data points represent parameter combinations that comply with the observed relic density $\Omega_{\rm obs} = 0.1198 \pm 0.00212$.}
	\label{fig:dd}
\end{figure}

In \autoref{fig:dd}, we depict $\ssgsi$ 
 as a function of  $\mtz$, 
based on our random parameter scan described in \autoref{eq:scan:range}. 
The color code corresponds to $\lmht$, a key factor in the spin-independent cross-section.  
Within the plot, we specifically showcase data points that satisfy $\omdm \leq \Omega_{\rm obs}$: 
the green data points correspond to the correct relic density. To constrain the parameter points, we include the upper bound on the $\ssgsi$ from the LUX-ZEPLIN (LZ) experiment\cite{LZ:2022ufs} (black solid line), which is the most stringent bound available till now. The LZ experiment excludes regions of high $\lmht$, with the upper limit extending as one moves up in the $\mtz$ values. To exemplify, for $\mtz \sim 2.5$ TeV, parameter points above $\lmht \sim 1.5$ are excluded, whereas $\lmht = 2.0$ is allowed for $\mtz \sim 4.0$ TeV. As a matter of fact, the points that provide the correct relic density are not excluded by the LZ bound. We extract all such allowed points, and move to the next subsection to discuss the constraints coming from the DM indirect detection experiments.

\subsection{DM indirect detection bounds}


From the previous subsection, we establish that, while a significant region of the parameter space is ruled out by the Planck data and the LZ experiment the region of higher mass and higher $\lmht$ still remains somewhat unbounded. Additional bounds can then be drawn from the DM indirect detection experiments, which look for the DM annihilation into gamma rays from the DM-rich regions of the universe, such as galactic centers and dwarf spheroidal galaxies. Dedicated experiments such as the Fermi-LAT~\cite{MAGIC:2016xys} and HESS~\cite{HESS:2022ygk} 
have jointly contributed  to the detection of gamma rays.
However, up to the present time, none of these experiments have observed significant excesses beyond the expected backgrounds. 
The absence of gamma ray excesses imposes stringent constraints on the parameter space of the ITM, 
which is the primary focus of this subsection. 



The gamma-ray flux from the annihilation of DM particles into the SM particles is expressed as: 
\begin{equation}
	\label{eq:gm:flux}
	\frac{d\phi_\gamma}{dE} = \frac{\sgvann}{8\pi \mtz^2} \sum_{\rm ch} \,\br_{\rm ch} \,\frac{dN_{\gamma}^{\rm ch} }{dE} J(\dom),
\end{equation}
where $J(\dom)$ is 
\bea
\label{eq:J:factor}
J(\dom) = \int_{\dom}  \int_{\rm LOS} \rho_{\rm DM}^2[ l (r,\theta)] d\Omega \, dl \, .
\eea
Here,
${dN_{\gamma}^{\rm ch}}/{dE}$ is the energy spectrum of gamma rays produced per DM annihilation in a given channel \enquote{ch}, $\br_{\rm ch}$ is the branching fraction of the channel,
and $ \rho_{\rm DM}[ l (r,\theta)]$ is the DM density profile along the line of sight (LOS)
over the region of interest with the solid angle $\dom$.
The gamma flux observations impose constraints on $\sgvann$ through \autoref{eq:gm:flux}, thus influencing the model parameters. 
These constraints also depend on the DM halo profiles, for which three widely considered profile functions are used in this work, namely, Einasto~\cite{Springel:2008by}, Navarro-Frenk-White (NFW)~\cite{Navarro:1996gj}, and Einasto 2~\cite{Cirelli:2010xx}. We utilize micrOMEGAs 5.3.41 to calculate the contribution of different DM annihilation channels to the $\sgvann$.

\begin{figure}[h]
	\centering
\includegraphics[width=0.7\linewidth]{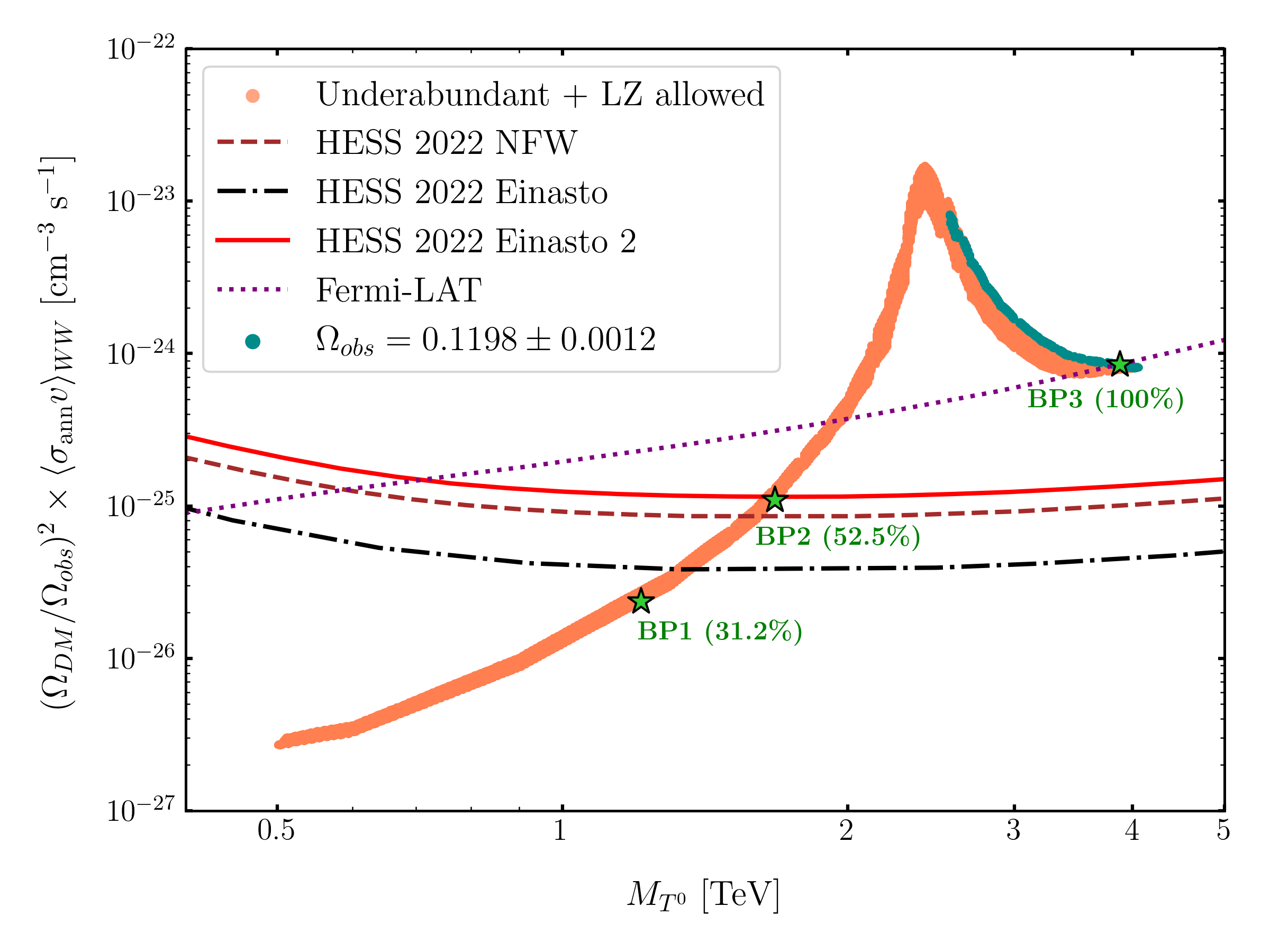}
	\caption{Thermally averaged DM annihilation cross-section of the dominant channel into $\ww$, 
	$\sgvann_{WW}$, 
	plotted against  $\mtz$
	for parameter points that fulfill both the DM relic density and direct detection measurements. The cross-sections are scaled with the square of the respective relic contributions, and the SE factors at $\beta = 10^{-3}$.
	The overlay includes bounds from the 2022 HESS indirect detection experiment~\cite{HESS:2022ygk}, 
	taking into account three different DM halo profiles. The bounds from the Fermi-LAT (2016) \cite{MAGIC:2016xys} are also shown in the purple dotted line. Dark green points represent regions where the relic density is satisfied, while orange points indicate regions of underabundance. The chosen benchmark points are marked in bright green stars, with the percentage contribution to the relic in parentheses.}
	\label{fig:ind}
\end{figure}

However, for the $SU(2)$ triplet DM, there are pronounced enhancements to the tree-level $\sgvann$ values due to the Sommerfeld effect. In consideration of the DM abundance in the current universe, the DM particles carry significantly less speed, with typical values of $\beta \sim 10^{-3}$~\cite{Cirelli:2007xd}. At such small $\beta$ values, the Sommerfeld enhancement factors are much higher compared to the epoch of the DM freeze-out~\cite{Katayose:2021mew}. Hence, we modify the tree-level $\sgvann$ values provided by \mic by multiplying the Sommerfeld enhancement factors for $\beta \sim 10^{-3}$, as extrapolated from the results in Refs.~\cite{Cirelli:2007xd, Hisano:2004ds}, for the DM annihilation in the dominant $W^+W^-$ channel. Taking into account the contribution of the points that are underabundant in relic, we scale these cross-sections with a factor of $(\Omega_{\rm DM}/\Omega_{obs})^2$~\cite{Katayose:2021mew}, and plot the final outcomes for the entire parameter space allowed by the Planck and LZ data, in \autoref{fig:ind}.

In \autoref{fig:ind},
we present $\sgvann_{WW}$ with respect to $\mtz$,
over the parameter points permitted by the 2022 LZ experiment and the relic density measurement of the Planck experiment. The points that satisfy the correct relic are marked in dark green. We include the upper limits from the HESS 2022 results \cite{HESS:2022ygk} in the three aforementioned DM halo profiles: NFW (brown dashed line), Einasto (black dot-dashed line), and Einasto 2 (red solid line). Additionally, with the purple dotted line, we also superimpose the bound from the Fermi-LAT (2016) experiment in the same channel. Based on these final set of bounds, we pick three benchmark points for the detailed analysis at a muon collider, which are marked with bright green stars on the \autoref{fig:ind}. The percentage contribution of the points to the total observed relic is written in parentheses underneath the stars, with further details in \autoref{tab:bp}.

\begin{table}[h]
	\renewcommand{\arraystretch}{1.4}
	\centering
	\begin{tabular}{|c|c|c|c|c|c|c|}
		\hline
		BP & $M_{T^0}$ [TeV] & $\lambda_{ht}$ & $\Omega_{\rm DM}$ & $\Omega_{\rm DM}/\Omega_{obs}$ & \makecell{HESS \\ Einasto 2 } & \makecell{Fermi-LAT }\\
		\hline
		BP1 & 1.21 & 0.026 & 0.037 & 0.312& \green{\cmark}& \green{\cmark}\\
		\hline
		BP2 & 1.68 & 0.0 & 0.063 & 0.525&\green{\cmark}& \green{\cmark}\\
		\hline
		BP3 & 3.86 & 1.861 & 0.119 & 1.000&\rd{\xmark}&\green{\cmark} \\
		\hline
	\end{tabular}
	\caption{Choices of benchmark points for the collider analysis.}
	\label{tab:bp}
\end{table}

The three benchmark points are chosen within the triplet mass range of 1.2-4.0 TeV, keeping in mind the possible pair/associated VBF production at the muon collider. The first two points are underabundant in DM relic while BP3 satisfies the 100\% relic, as can be seen from \autoref{tab:bp} and \autoref{fig:ind}. A point with $\lambda_{ht}=0$, which corresponds to the minimal DM scenario \cite{Cirelli:2005uq} is chosen as BP2. For BP1 and BP3, we chose non-zero $\lambda_{ht}$, keeping them within the perturbativity bound $\lambda^{\rm max}_{ht}=1.95$ for the Planck scale \cite{Bandyopadhyay:2021ipw, Jangid:2020qgo}. From \autoref{fig:ind}, it is evident that the underabundant benchmark points BP1 and BP2 both lie within the upper limits from the Fermi-LAT experiment, as well as from the HESS experiment, considering the Einasto 2 profile. However, while BP3 satisfies the correct relic, it is only allowed by the Fermi-LAT bound, and not by the HESS bound. Considering the fact that both of these are two independent experiments focusing on different DM-rich regions of the galaxy, we choose to move ahead with this set of benchmarks, strictly from a collider search perspective.


\section{Golden channels to probe the triplet scalars at a muon collider}
\label{sec:golden:channels}

In the previous section, 
we successfully constrained the masses of the ITM triplet scalars 
to three benchmark points with $\mtz=1.21,\, 1.68,\, 3.86\tev$.
This outcome was achieved by incorporating the theoretical requirements, collider experiment constraints, and various DM experiment results.  
The exploration of such a high-mass domain provides strong motivation 
for the utilization of a multi-TeV MuC~\cite{AlAli:2021let,Accettura:2023ked,Aime:2022flm}.

Regarding the choice of the c.m.~energy for the MuC, 
several options have been under consideration in the literature. 
The current focal point of the international MuC collaboration centers around a 3 TeV MuC, 
while ongoing discussions and activities also encompass higher energies up to 30 TeV~\cite{Han:2021udl,AlAli:2021let}. 
For our analysis, which focuses on the pair production of the ITM triplet scalars, 
we will assess the feasibility of both a 6 TeV and 10 TeV MuC setup, each designed with projected integrated luminosities of $4\iab$ and $10\iab$, respectively~\cite{AlAli:2021let}.

\subsection{Production modes of triplet scalars at a muon collider}\label{sec:prod}

\begin{figure}[h]
	\centering
	\includegraphics[width=0.8\linewidth]{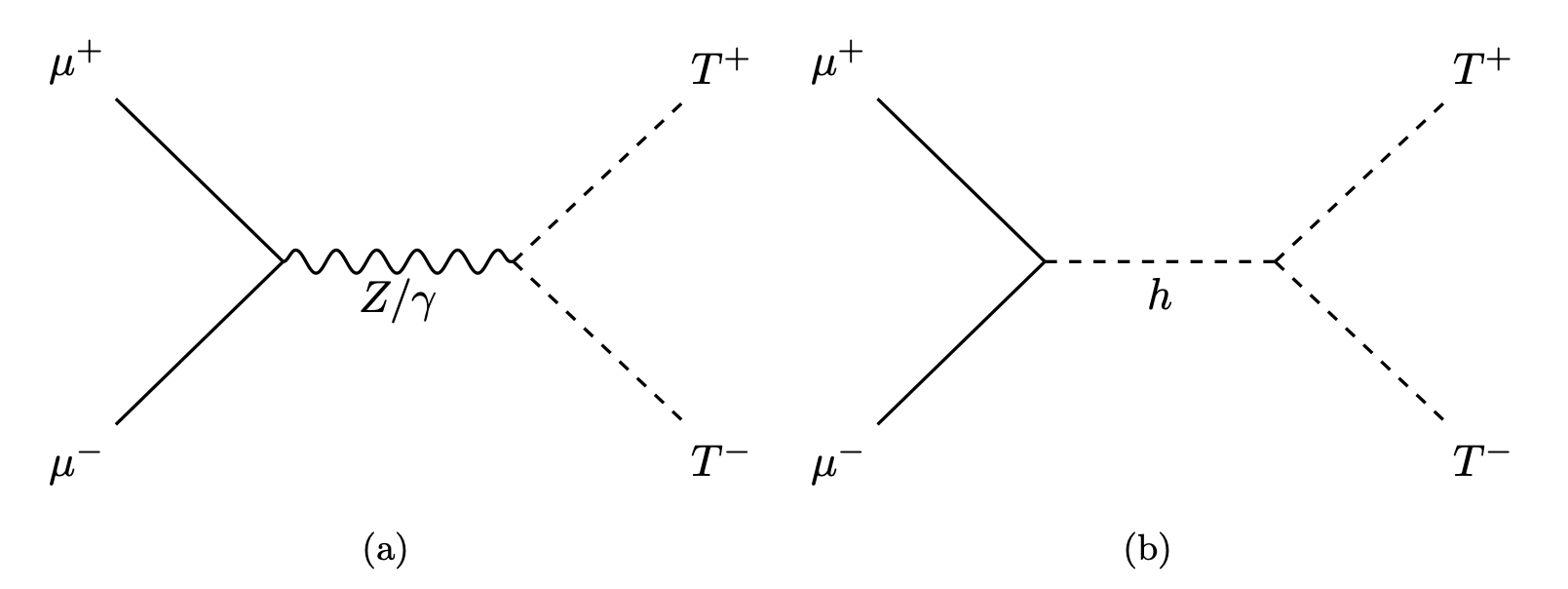}
	\caption{Feynman diagrams illustrating the Drell-Yan production of $T^+ T^-$ through $\mu^+\mu^-$ collisions. }
	\label{fig:dyprod}
\end{figure}

To identify promising channels for the ITM at a MuC, 
we explore various avenues for the pair production of triplet scalars.
We begin with the DY pair production of charged triplet scalars, $\mmu\to T^+ T^-$. 
The corresponding Feynman diagrams are shown in \autoref{fig:dyprod}: 
those mediated by gauge bosons of $\gamma$ and $Z$ in \autoref{fig:dyprod}(a) 
and the one mediated by the Higgs boson in \autoref{fig:dyprod}(b).\footnote{Notably, it is worth mentioning that due to the relatively small values of $\lambda_{ht}$ and the muon Yukawa coupling to the SM Higgs boson, the contribution from Higgs mediation proves negligible.}
However, these DY pair production modes do not exhibit high efficiency,
primarily due to the fact that the cross-section of an $s$-channel diagram decreases as $\sqrt{s}$ increases.
This reduction prevents us from fully harnessing the high 
collision energy of the MuC,
which was the initial motivation for probing the heavy triplet scalars.

More efficient processes are available at the MuC, 
the VBF processes. 
A standout advantage of VBF is its dramatically higher cross-section at high c.m.~energy,
compared to the DY cross-section~\cite{Costantini:2020stv}. 
The VBF cross-section exhibits growth according to $\log^2({s}/{m_V^2})$~\cite{Han:2021udl}, 
rendering VBF processes exceptionally well-suited for a high-energy MuC.
Moreover, there are multiple VBF production channels for two triplet scalars,
as follows:
\begin{align}
\label{eq:VBF:processes}
\mu^+ \mu^- \;&\to\; T^+ T^-  \nu \bar{\nu},
\\ \nn
\mu^+ \mu^-  \;&\to\;  T^0 \;T^0  \, \nu \bar{\nu},
\\ \nn
\mu^+ \mu^-  \;&\to\;  T^\pm T^0 \,  \mu^\mp \nu,
\\ \nn
\mu^+ \mu^-  \;&\to\;  T^+ T^- \,  \mu^+ \mu^-.
\end{align}
These diverse production channels significantly enhance the potential for probing the model.

\begin{figure}[h]
	\centering
	\includegraphics[width=0.32\linewidth]{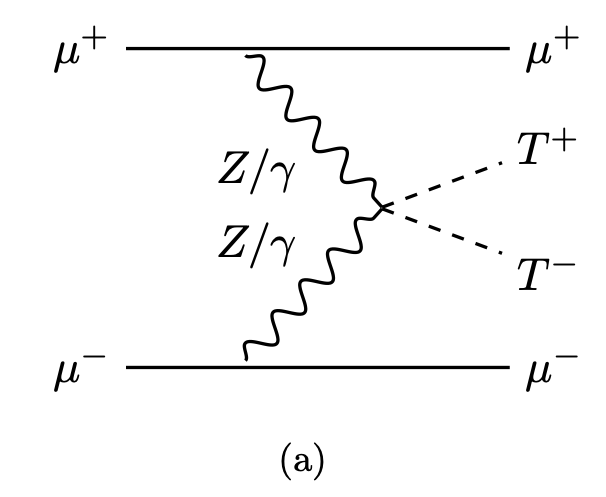}~
	\includegraphics[width=0.32\linewidth]{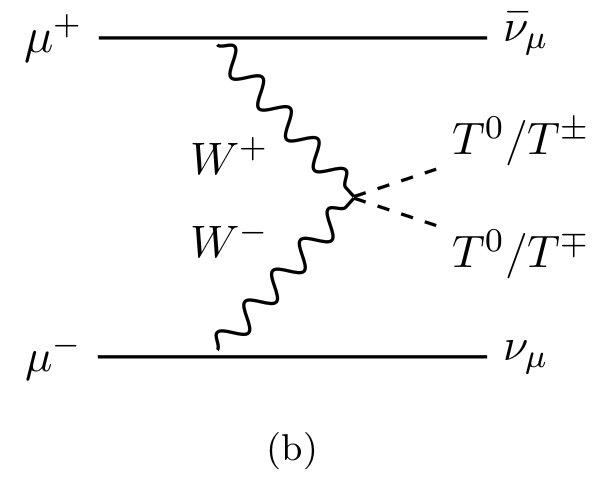}~
	\includegraphics[width=0.32\linewidth]{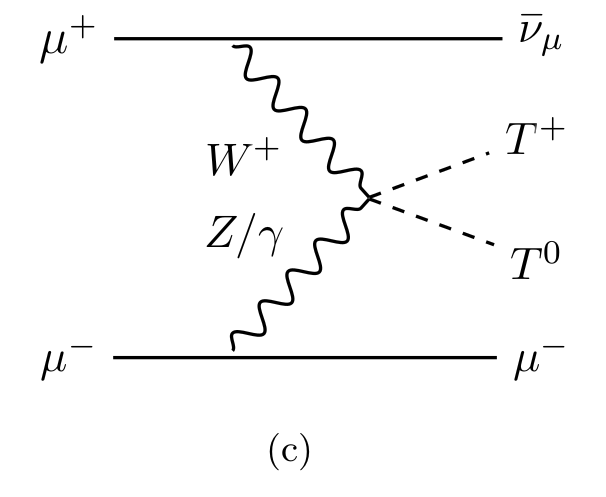}
	\caption{Representative Feynman diagrams illustrating the dominant contributions to the pair production of $T^\pm$ and $T^0$ through VBF.}
	\label{fig:Feynman:VBF}
\end{figure}

The VBF processes in \autoref{eq:VBF:processes} occur through several Feynman diagrams: 
the Higgs boson-mediated $s$-channel diagrams, the triplet-mediated $t$-channel diagrams, 
and the contact diagrams generated from the quartic couplings 
associated with the $V$-$V^{(\prime)}$-$T$-$T^{(\prime)}$ vertex 
(where $V^{(\prime)} = \gm,Z,\wpm$ and $T^{(\prime)} = T^0, \tpm$).
Notably, the major contributions are from the quartic vertices,
of which the Feynman diagrams are presented in \autoref{fig:Feynman:VBF}.
It is worth emphasizing that the $TTVV$ vertices are distinctive features of the ITM, 
setting it apart from other BSM models with fermion DM, such as Wino and Higgsino~\cite{Han:2020uak, Capdevilla:2021fmj}. 
The presence of these quartic vertices allows the triplet scalars to exhibit a high VBF production rate without encountering suppression from the triplet-mediated propagators. Although similar vertices are also found in Two-Higgs Doublet Models~\cite{Han:2021udl}, the coupling strength  of triplets significantly surpasses that of doublet scalars, resulting in substantially larger cross-sections at high collision energies.

\begin{figure}[h]
	\centering
	\subfigure[]{\includegraphics[width=0.48\linewidth]{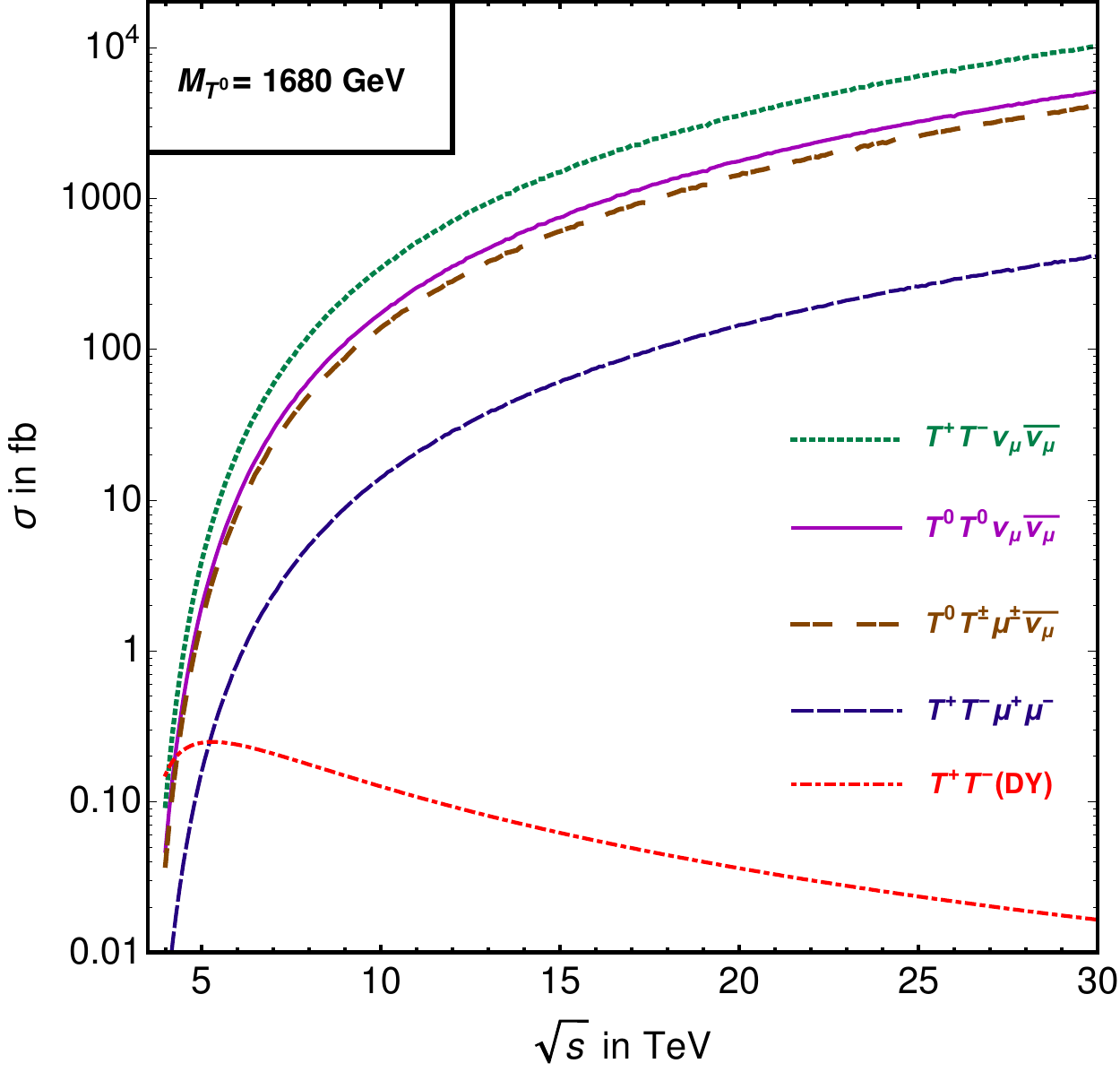}} \quad
	\subfigure[]{\includegraphics[width=0.48\linewidth]{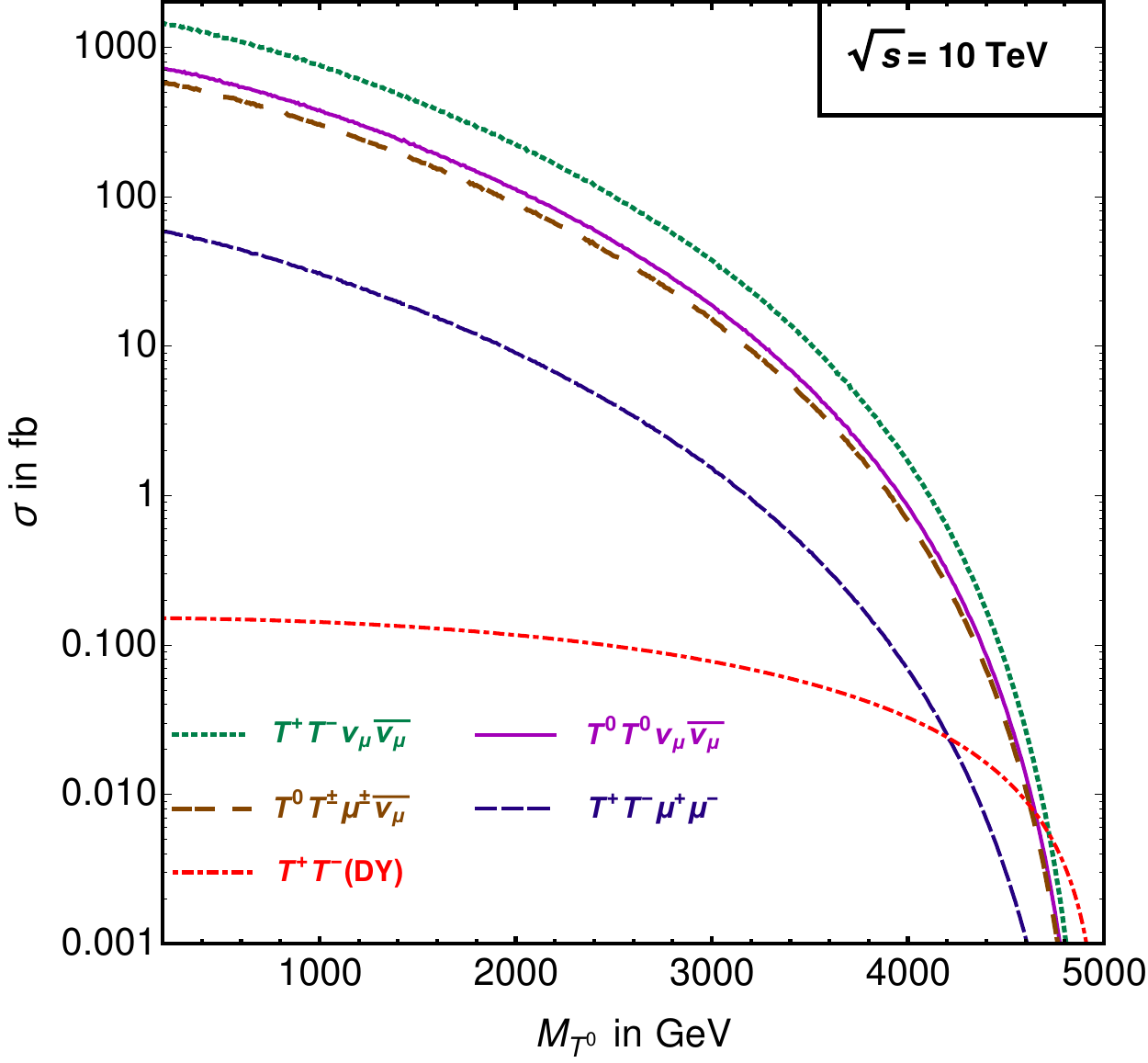}}
	\caption{ Various production cross-sections of triplet scalars as a function of  the MuC c.m.~energy $\sqrt{s}$ in the left panel (a),
	and as a function of $\mtz$  in the right panel (b). 
	For the left panel (a), we consider BP2 with $\mtz=1.68\tev$. In the right panel (b), 
	the c.m.~energy is fixed at $\sqrt{s}=10\tev$.}
	\label{fig:svcs}
\end{figure}

To determine the most efficient production channel for triplet scalars, 
we calculate the parton-level cross-sections of the DY and four VBF processes.
These calculations are performed using \texttt{MadGraph5\_aMC\@NLO} 
with the \texttt{.ufo} files generated by \texttt{SARAH}. 
\autoref{fig:svcs} displays these parton-level production cross-sections.
To observe the dependence of the cross-sections on the c.m.~energy,
we show the cross-sections as functions of $\sqrt{s}$ for BP2 with $\mtz = 1.68\tev$
in the left panel (a).
We observe that the cross-section for the DY process, marked by a red dotted line, 
peaks near $\sqrt{s}=5\tev$ before rapidly decreasing. 
In contrast, the VBF process cross-sections demonstrate a consistent increase with rising $\sqrt{s}$, 
showcasing the $\log^2({s}/{M_V^2})$ pattern.

Among the four VBF channels detailed in \autoref{eq:VBF:processes}, 
the $T^+ T^- \nu \bar{\nu}$ final state yields the highest cross-section, 
exceeding the DY cross-section at approximately $\sqrt{s} \approx 4.3\tev$. 
The process $\mu^+\mu^- \to T^0 T^0 \nu \bar{\nu}$, involving two charged currents from the initial muon beams, 
has the second highest cross-section. 
The third significant process, $\mu^+ \mu^- \to T^\pm T^0 \mu^\mp \nu$, 
is mediated by a charged current from one muon beam and a neutral current from the other. 
The VBF channel $\mu^+ \mu^- \to T^+ T^- \mu^+ \mu^-$, which involves two neutral currents, 
has the lowest cross-section,
attributed to $\cos^2\theta_W$ vertex suppressions.

In \autoref{fig:svcs}(b), we present the cross-sections  
as a function of the DM mass $\mtz$ at  $\sqrt{s}=10\tev$. 
As the mass increases, there is a noticeable decrease in the cross-sections for both the DY and VBF processes, 
attributed to phase space suppression effects. 
Despite this, the hierarchy in production cross-sections, as observed in \autoref{fig:svcs}(a), remains intact, up until about $\mtz \sim 4.2$ TeV. 
Consequently, the VBF channels emerge as robust options for producing heavy triplets, 
especially those channels involving one or two charged currents. 
Notably, even the $\mu^+ \mu^- \to T^+ T^- \mu^+ \mu^-$ process, 
which has the smallest cross-section among the four VBF processes, still outperforms the DY process at 
$\sqrt{s}=10\tev$, before the hierarchy flips at  $\mtz \sim 4.2$ TeV. 
The DY cross-section becomes dominant only at $\mtz \sim 4.8$ TeV (near the kinematical reach of the 10 TeV muon collider), compared to the three other VBF processes.

\begin{table}[h]
	\renewcommand{\arraystretch}{1.4}
	\centering
	\begin{tabular}{|c|c|c||c|c|c|c|c|}
		\hline
\multicolumn{3}{|c||}{Cases} &		\multicolumn{5}{c|}{Cross-section [fb] for $\mu^+ \mu^- \to$ final states} \\ \hline
 & $\mtz$ [TeV] & $\sqrt{s}$ [TeV] & \;$T^+ T^-$\; & $ T^0 T^{\pm} \mu^{\mp} \bar{\nu}_\mu$  & $ T^+ T^- \mu^+ \mu^-$ & $ T^0 T^0 \nu_\mu \bar{\nu}_\mu$ & $ T^+ T^- \nu_\mu \bar{\nu}_\mu$ \\ \hline
\multirow{2}{*}{BP1} & \multirow{2}{*}{1.21} & 6 & 0.32 &  31.8 & 3.2 & 39.5 & 79.4 \\
& & 10 &  0.13 & 244.8 & 24.6 & 304.4 & 608.3  \\ \hline
%
%
\multirow{2}{*}{BP2} & \multirow{2}{*}{1.68} & 6 &  0.23 & 8.1 & 0.8 & 10.2 & 20.1  \\
&  & 10 & 0.12  & 138.3 & 13.8 & 171.3 &  342.6 \\ \hline
\multirow{2}{*}{BP3} & \multirow{2}{*}{3.86} & 6 & -- & -- & -- & -- & -- \\
& & 10 &  0.03 & 1.2 & 0.12 & 1.5 & 3.0  \\
\hline	\end{tabular}
	\caption{Parton-level cross-sections of pair and associate production of $T^\pm$ and $T^0$ at a MuC with two different $\sqrt{s}$ values. }
	\label{tab:vbfcs}
\end{table}

In \autoref{tab:vbfcs}, we provide a detailed overview of the parton-level cross-sections 
	for the three benchmark points at a high-energy MuC, 
	specifically focusing on two c.m.~energies: $\sqrt{s}=6\tev$ and $10\tev$. The kinematical limit of the 6 TeV MuC allows the pair production of the scalars for only BP1 and BP2, whereas at the 10 TeV MuC, all three benchmarks can be produced with respectable cross-sections. The hierarchy in the VBF production cross-sections remains consistent, 
	resulting in the following order: 
	$T^+ T^- \nu \bar{\nu}$, $T^0 T^0 \nu \bar{\nu}$, $T^{\pm}T^0 \mu^{\mp} \nu$, 
	and $T^+ T^-\mu^+ \mu^-$.
	The comparison between the DY and VBF production cross-sections 
	varies depending on the benchmark points and the $\sqrt{s}$ values, 
	but the DY cross-sections are significantly lower than the VBF cross-sections across all benchmarks. Another notable observation is that the VBF cross-sections increase more significantly with rising $\sqrt{s}$ for heavier $\mtz$ masses. The enhancement factors are approximately 7.7 for BP1, and around 17.5 for BP2, highlighting the benefits of operating at higher c.m.~energies for larger triplet masses. Based on these observations, later in \autoref{sec:results}, we will present results at the 6 TeV MuC for BP1 and BP2 only, whereas at the 10 TeV MuC, all three BPs will be considered.


\subsection{Signatures of the ITM at a muon collider}\label{sec:signatures}

Let us examine the signatures of $T^0$ and $\tpm$ at the MuC. 
The DM candidate $T^0$, being stable and neutral, does not interact with the detector, creating a missing energy signature. 
Detecting the charged triplet scalars, which decay within the detector, 
also presents a challenge due to the small mass splitting of $\mtch-\mtz \simeq 166\mev$. 
The pion or charged lepton produced in the decay of $\tpm$ lacks the necessary energy 
for effective tracking or calorimeter detection, rendering $\tpm$ elusive.
As both $T^0$ and $\tpm$ contribute to missing energy signatures, 
we encounter significant SM backgrounds from $Z\to \nu\bar{\nu}$.

Fortunately, the intrinsic electric charge of $\tpm$ enables its detection within the innermost layers of the detector.
A DCT appears
as the proper decay length of about 5.7 cm for $\tpm$ yields no corresponding signals in the external layers.
However, DCT signatures alone are insufficient for effective signal triggering, necessitating an additional mechanism. 
One option is a hard mono-photon trigger, but it significantly reduces event rates. 

In the VBF processes of $T^\pm T^0 \mu^\mp \nu$ and $T^+ T^- \mu^+ \mu^-$, 
a notable feature is the presence of one or two high-energy, forward spectator muons, $\muf$. 
The current MuC detector design, 
focusing primarily on the central region with $|\eta|<2.5$ 
to reduce BIB through the use of tungsten nozzles, 
mostly fails to detect these forward spectator muons. 
This limitation has led to growing support for integrating a dedicated Forward muon detector into the MuC. 
The feasibility of this addition is supported by the fact that 
Forward muons in VBF processes typically exhibit momenta exceeding $p_{\muf} \geq 300\gev$, 
enabling them to penetrate the tungsten nozzles~\cite{Ruhdorfer:2023uea, Accettura:2023ked, Forslund:2023reu}. 
This capability distinctly sets them apart from the softer BIB particles, including muons, 
which generally possess less than 100 GeV of kinetic energy~\cite{Collamati:2021sbv}. 
Consequently, we propose the use of one or two Forward muons as triggers 
in our analysis of the DCT signals from the VBF processes: $T^\pm T^0 \mu^\mp \nu$ and $T^+ T^- \mu^+ \mu^-$. 
Furthermore, we underscore the importance of $\met$ as an instrumental tool in probing DM signals, 
particularly given that DCTs are mainly detectable in the central barrel region of the detector.
 
Building upon these observations, we suggest four distinct final states (FS) as follows:
\begin{align}
\label{eq:final:signatures}
\text{FS1:}& \quad  1\,\cdt + 1\, \muf + \met \,; \\ \nn
\text{FS2:}&   \quad   2\,\cdt + 2\, \muf +\met \,;  
\\ \nn
\text{FS3:}&   \quad   1\,\cdt + 2\, \muf +\met \,; 
\\ \nn
\text{FS4:}&   \quad   2\,\cdt + 1\, \muf +\met . 
\end{align}
It is noteworthy that FS1 primarily results from the $T^\pm T^0 \mu^\mp \nu$ process. 
However, it may also emerge from the $T^+ T^- \mu^+ \mu^-$ process 
when one Forward muon falls outside the rapidity coverage of the proposed Forward muon detector 
and one DCT signal goes undetected. 
On the other hand, FS2, FS3, and FS4 are exclusively associated with the $T^+ T^- \mu^+ \mu^-$ process.

\section{Characteristics of the triplet signals at a multi-TeV muon colliders}
\label{sec:characteristics:signal}
In order to refine our final states selections, 
we delve into the characteristics of triplet signals at the MuC,
focusing on three pivotal aspects:
(i) the large decay radius of $\tpm$; 
(ii) the considerable efficiency of DCTs; and (iii) the high efficiency of Forward muons. 
We will elucidate these features using detailed detector simulations, employing the following methodology. 
Event generation was accomplished using \texttt{MadGraph5\_aMC@NLO}~\cite{Alwall:2011uj} version 3.4.2, 
followed by showering with \py~\cite{Bierlich:2022pfr} version 8.3.07. 
Subsequently, \texttt{Delphes} version 3.5.0~\cite{deFavereau:2013fsa} facilitated a rapid detector simulation, 
utilizing the card of \texttt{delphes\_card\_MuonColliderDet.tcl}\footnote{The MuC detector Delphes card
can be found at  the GitHub repository:  https://github.com/ delphes/delphes/blob/master/cards/delphes\_card\_MuonColliderDet.tcl}.
It is noteworthy that the detection of Forward muons has been incorporated into this \texttt{Delphes} card.

\subsection{Brief review of the MuC tracker and the decay radius of $T^\pm$}

\begin{figure}[h]
	\centering
	\subfigure[]{\includegraphics[width=0.565\linewidth]{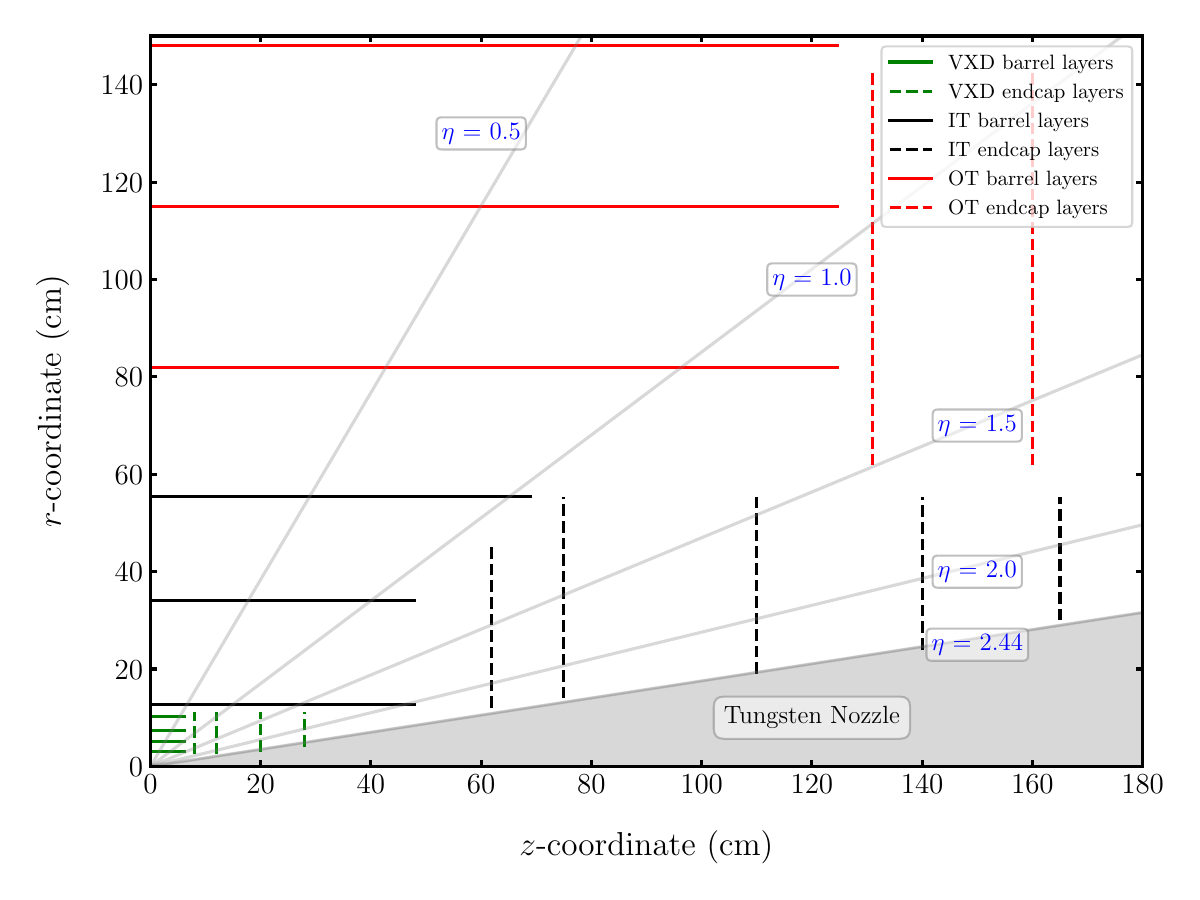}}
	\subfigure[]{\includegraphics[width=0.425\linewidth]{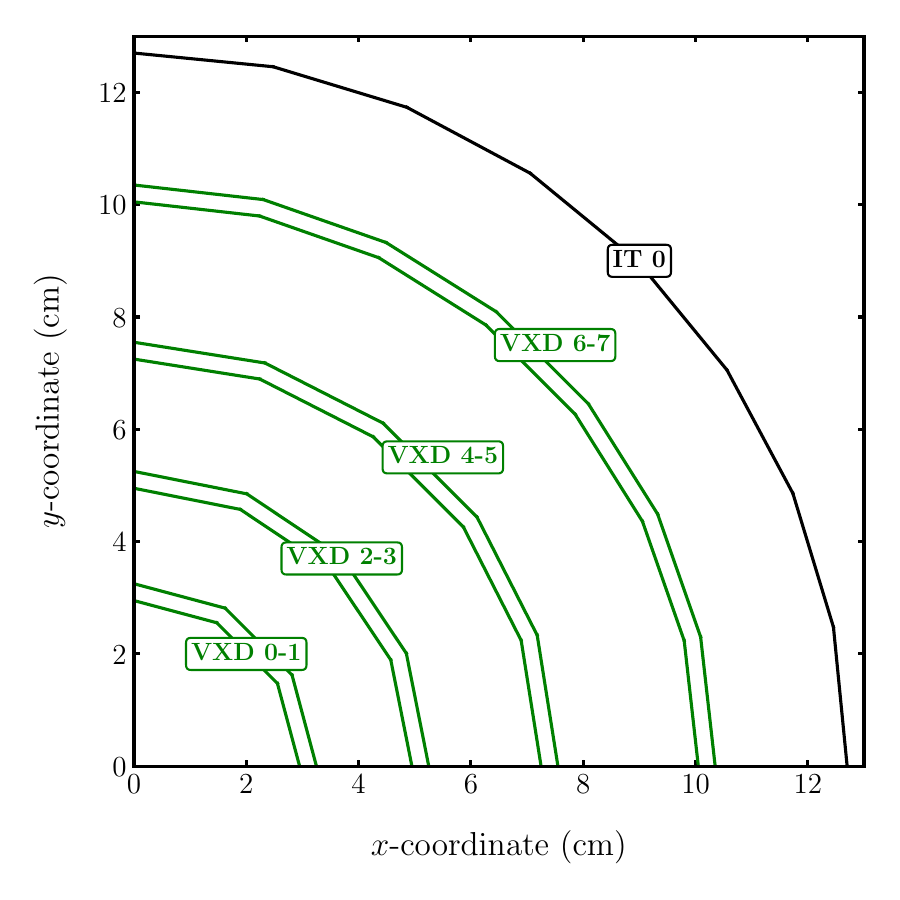}}
	\caption{(a) Schematic of the MuC tracker;
	 (b) detailed view of the radial plane focusing on the Vertex Detector (VXD) layers.
	 These figures are adapted for reader convenience from Refs.~\cite{Capdevilla:2021fmj,MuonCollider:2022ded}.}
	\label{fig:trkr}
\end{figure}

Disappearing charged tracks from $\tpm$ are pivotal for identifying ITM signatures. 
Analyzing these DCT signals demands a thorough understanding of the tracker configuration 
in the proposed MuC detector as outlined in Ref.~\cite{MuonCollider:2022ded}.
\autoref{fig:trkr} presents a schematic representation of the MuC tracker.  
As illustrated in \autoref{fig:trkr}(a), the tracker comprises three primary components: the Vertex Detector (VXD) depicted in green, the Inner Tracker (IT) in black, and the Outer Tracker (OT) in red. These components are segmented into barrel (represented by solid lines) and endcap (indicated by dashed lines) sections. The VXD barrel contains four double-layers positioned at specific radial distances from the interaction point, as shown in \autoref{fig:trkr}(b). Radially, the IT barrel ranges from 12.7 cm to 55.4 cm, and the OT spans from 81.9 cm to 148.1 cm. The endcap areas, comprising tracker disks oriented perpendicular to the beam axis~\cite{Collamati:2021sbv, MuonCollider:2022ded, Ally:2022rgk}, predominantly encounter a majority of the BIB hits.

Our investigation concentrates on the barrel region of the tracker, 
where the incidence of BIB hits is markedly lower compared to the endcap region~\cite{Capdevilla:2021fmj}. 
\autoref{fig:trkr}(b) zooms into the radial plane, 
focusing on the  double-layers of the VTD and the  initial single layer of the IT. 
For clear reference in subsequent discussions, 
the double layers in the VXD are designated as VXD0-1, VXD2-3, VXD4-5, and VXD6-7, 
while the IT layers are sequentially labeled as IT0, IT1, and so on.

Given the detailed design of the MuC tracker, 
we are able to determine the DCT reconstruction efficiency, 
which depends on the polar angle and decay length of $\tpm$. 
The decay length is expressed as $c\tau\beta\gamma$, 
where $\beta$ is the velocity component $\abs{\vec{p}}/{E}$ and $\gamma$ is the relativistic factor ${1}/{\sqrt{1-\beta^2}}$. 
However, for our analysis, we focus on the transverse decay length $c\tau\beta_T \gamma$ 
(with $\beta_T = {p_T}/{E}$), commonly referred to as the decay radius, 
since our DCT analysis is concentrated on the central barrel region of the detector.

\begin{figure}[h]
	\centering
	\includegraphics[width=0.7\linewidth]{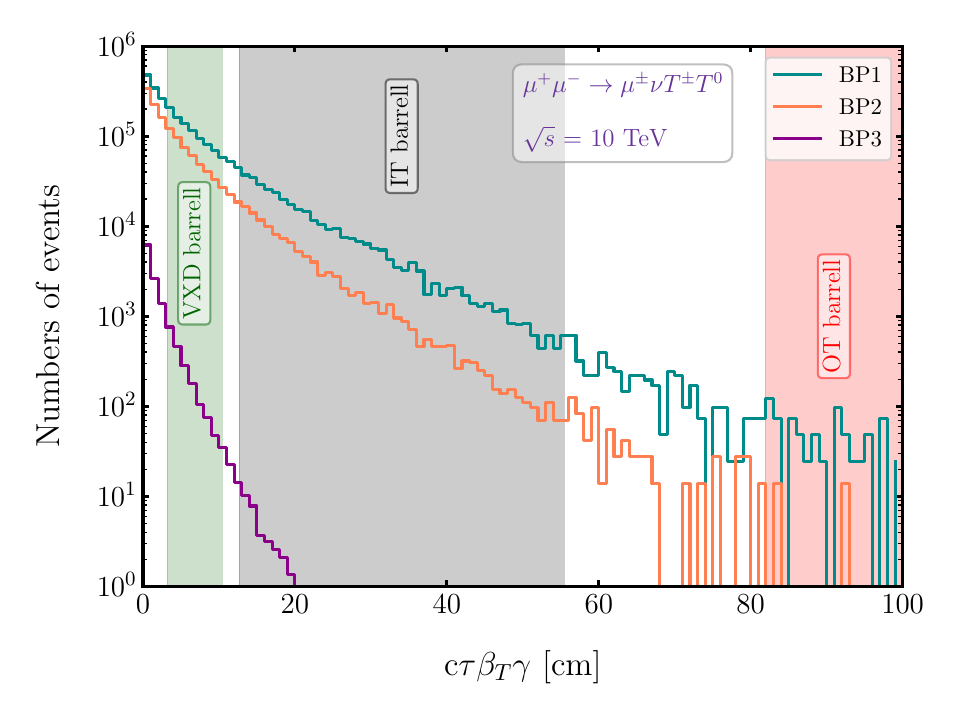}
	\caption{Event count distribution by the decay radius of $T^\pm$
	in the process of $\mu^+ \mu^- \to T^0 T^{\pm} \mu^{\mp} {\nu}$ 
	at a 10 TeV muon collider with $\lumtot=10\iab$. 
	Benchmark points are indicated as follows: BP1 ($\mtch=1.21\tev$) in turquoise, BP2 ($\mtch=1.68\tev$) in orange, and BP3 ($\mtch=3.86\tev$) in purple. The figure also delineates the radial extents of the barrel region trackers: Vertex Detector (VXD) in green, Inner Tracker (IT) in grey, and Outer Tracker (OT) in red. }
	\label{fig:dcrad}
\end{figure}

\autoref{fig:dcrad} illustrates the event count distributions 
relative to the decay radius of $T^\pm$ in the process $\mu^+ \mu^- \to T^0 T^{\pm} \mu^{\mp} {\nu}$ 
at the 10 TeV MuC with an integrated luminosity of $\lumtot=10\iab$. 
Our analysis includes three benchmark points: 
BP1 with $\mtch=1.21\tev$ (turquoise), BP2 with $\mtch=1.68\tev$ (orange), 
and BP3 with $\mtch=3.86\tev$ (purple). The radial extents of the three tracker barrel regions are also indicated: the VXD in green, the IT in grey, and the OT in red.

A consistent trend observed across all benchmark points is 
the rapid decrease in signal events as the decay radius of $\tpm$ increases. 
This pattern is attributed to reduced boost effects associated with the heavy $\mtch$. 
Fortunately, a considerable proportion of events exhibit $\tpm$ decaying within the VXD and IT regions. Such occurrences are vital for generating DCTs  that exhibit high levels of reconstruction efficiency.

\subsection{Disappearing charged tracks}

This subsection is dedicated to the reconstruction efficiency of DCT signals at the MuC. 
The current \texttt{Delphes} card for the MuC does not have the capability 
to track heavy unstable charged BSM particles, 
although it can reconstruct the tracks of SM particles using the \texttt{EFlowTrack} module. 
To bridge this gap, Ref.~\cite{Capdevilla:2021fmj} has calculated 
model-independent DCT reconstruction efficiencies using a \textsc{Geant}4 simulation. 
Our methodology integrates these results in two steps: first, by establishing criteria for selecting DCT candidates, and second, by applying the efficiency as a weight to each signal event.

For selecting DCT candidates, two conditions for $\tpm$ are set: 
a decay radius $\drad$ in the range of $5.1\cm < \drad < 148.1\cm$, 
and a polar angle $\theta$ within $0.7 < \theta < 2.44$. 
The lower $\drad$ threshold of 5.1 cm ensures that 
$\tpm$ tracks hit at least the first two double-layers of the VXD, 
i.e., up to VXD2-3. 
The upper limit for $\drad$ is set to guarantee that
$\tpm$ tracks end before reaching the  final layer of the OT at 148.1 cm.
This is crucial for reducing SM backgrounds by excluding calorimeter hits, 
akin to DCT search strategies at the LHC~\cite{CMS:2020atg,Capdevilla:2021fmj}. 
The $\theta$ restriction aims to minimize BIB hits in endcap regions beyond $|\eta|>2.44$. 
Further refinement to a more central region, $\theta \in [0.7,2.44]$ (or equivalently $|\eta| \leq 1$), is applied, as the DCT reconstruction efficiency significantly diminishes outside this range.

The second step involves applying the DCT efficiency $\pdct$ in Fig.11 of  Ref.~\cite{Capdevilla:2021fmj}
as a weighting factor to each signal event count.
This process requires an understanding of its dependency on both the decay radius and polar angle of $\tpm$. 
For $\drad$, the  efficiency  is high when the charged track extends into the VXD's outer layers without crossing IT0, peaking in the VXD6-7.
Between IT0 and the OT's final layer at 148.1 cm, the efficiency, though lower, remains non-zero, 
particularly within $\theta\in[0.7,1.1]$. 
We conservatively assume that the DCT reconstruction efficiency in this region is comparable to that of tracks reaching IT0. 
This assumption is supported by consistent BIB hit counts across all three tracker layers~\cite{MuonCollider:2022ded, Ally:2022rgk, Collamati:2021sbv}. 
Additionally, the efficiency is notably sensitive to $\theta$, 
with the highest efficiency observed within $\theta \in [1.05,1.22]$. 
This relationship, depicted as $\pdctf$, will be incorporated into the calculation of the expected number of signal events exhibiting a DCT signal.

\begin{figure}[h]
	\centering
	\includegraphics[width=\linewidth]{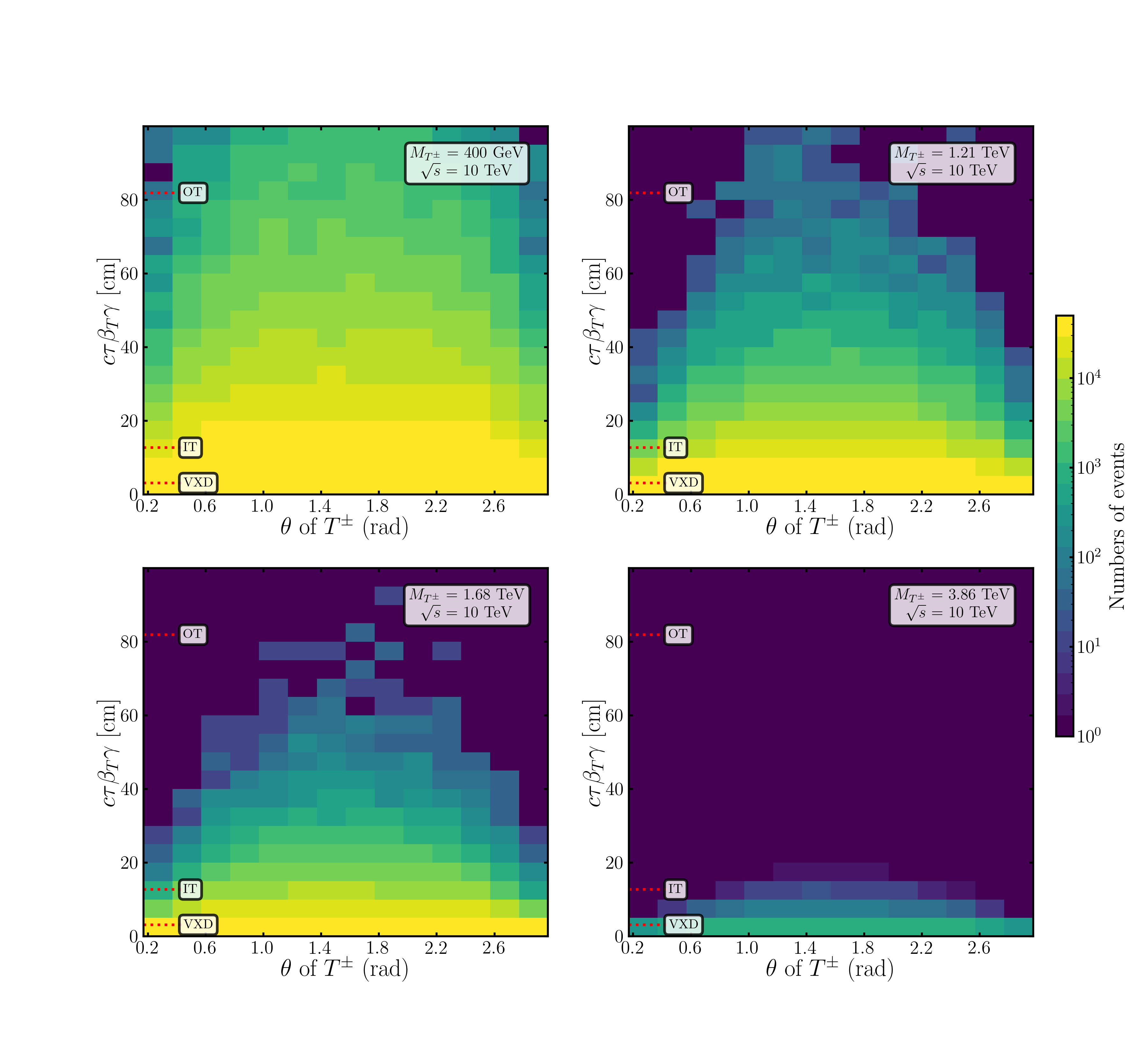}
	\caption{Two-dimensional histograms of normalized event counts 
	in $(\theta,c \tau\beta_T\gm)$ plane at a 10 TeV muon collider with $\lumtot=10\iab$. The figure showcases three benchmark $M_{T^\pm}$ values: 1.21 TeV
	in the upper right, 1.68 TeV in the lower left, and 3.86 TeV in the lower right panel.
	For comparison, the upper-left panel presents a reference scenario with a lower $T^\pm$ mass of 400 GeV.}
	\label{fig:tp2d}
\end{figure}

To assess the reconstructability of $\tpm$ as a DCT signal, 
we present in \autoref{fig:tp2d} the two-dimensional distribution of event counts 
in the $(\theta,\drad)$ plane. 
This analysis  is relevant to the signal $\mu^+ \mu^- \to T^0 T^{\pm} \mu^{\mp} \nu$ 
at a 10 TeV MuC with an integrated luminosity of $\lumtot=10\iab$. 
In addition to three benchmark scenarios (BP1, BP2, BP3),
we also include results for a reference case with $\mtch=400\gev$, 
providing insights into the impact of $\mtch$ on $\pdct$. 
Each panel notably features three red dashed lines, marking the starting radii for VXD, IT, and OT.
A critical observation from \autoref{fig:tp2d} is 
that a significant portion of our signal events lies within the range favoring DCT reconstruction. 
Additionally, we observe that signals from heavier $\tpm$ particles tend to be more centrally concentrated, 
a trend consistent with the findings in \autoref{fig:dcrad}. Specifically, for the 3.86 TeV mass of the charged scalar in BP3, the available boost is small enough to keep the tracks concentrated within $\theta$ values of [0.6, 2.6], mostly within the VXD.
These results suggest that the ITM with heavy triplet scalars exhibits a high potential 
for discovery at the MuC.

\subsection{Forward muons}

A key characteristic of the final states in \autoref{eq:final:signatures} is the presence of one or two spectator muons. These muons are expected to appear predominantly in the high pseudorapidity region. Although they can occasionally be detected in the main detector through statistical fluctuations, the probability of this occurring is notably low.
Therefore, dedicated Forward muon detection facilities are essential 
for the effective triggering of our VBF processes,
an option proactively considered by the Muon Collider R\&D group~\cite{Accettura:2023ked}.

The \texttt{Delphes} MuC card includes a Forward muon module,
although it is primarily optimized for the main detector setup, covering a pseudorapidity range of 
$\abs{\eta} \leq 2.44$.\footnote{The current nozzle configuration is tailored for BIB 
at a 3 TeV MuC~\cite{MuonCollider:2022glg}. 
At higher c.m.~energies, BIB hits are expected to occur more forward, 
justifying the use of $\abs{\eta} \leq 2.44$ for our analysis at $\sqrt{s}=6,10\tev$ as a conservative assumption.} 
This module spans the region $2.5 \leq \eta_{\mu} \leq 8.0$, achieving an optimistic efficiency of 90\%. 
In our simulations, we adopt a more conservative pseudorapidity range of 
$2.5 \leq \eta_{\mu} \leq 7.0$ for the Forward muons~\cite{Ruhdorfer:2023uea}.

\begin{figure}[h]
	\centering
	\subfigure[]{\includegraphics[width=0.49\linewidth]{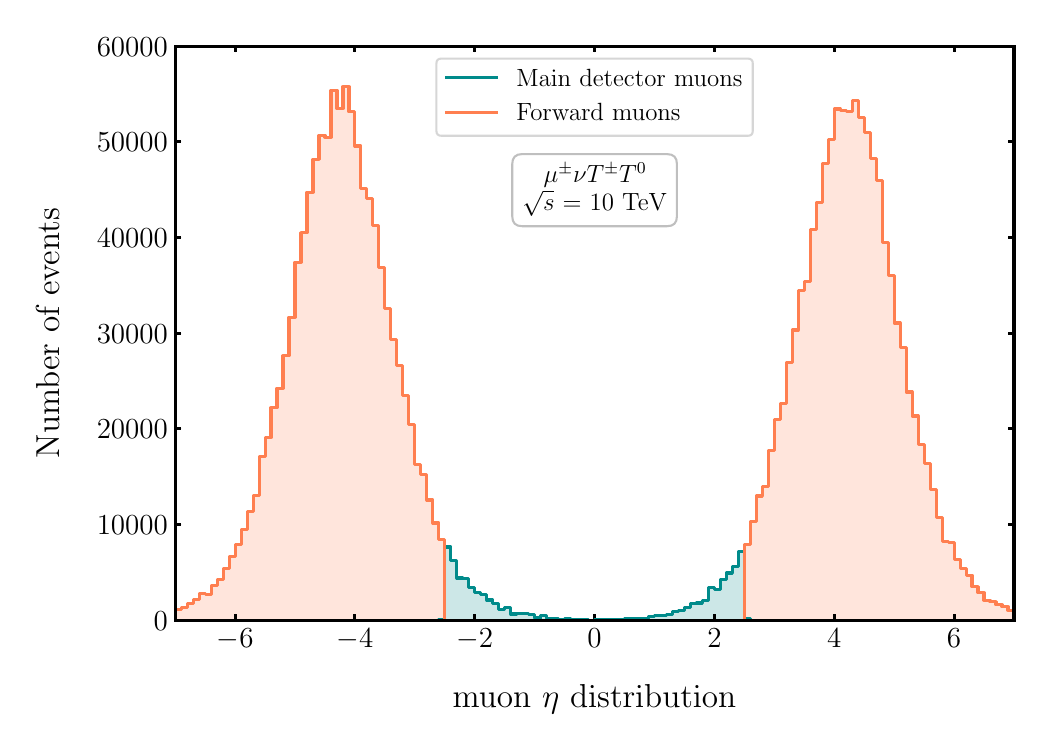}}
	\subfigure[]{\includegraphics[width=0.49\linewidth]{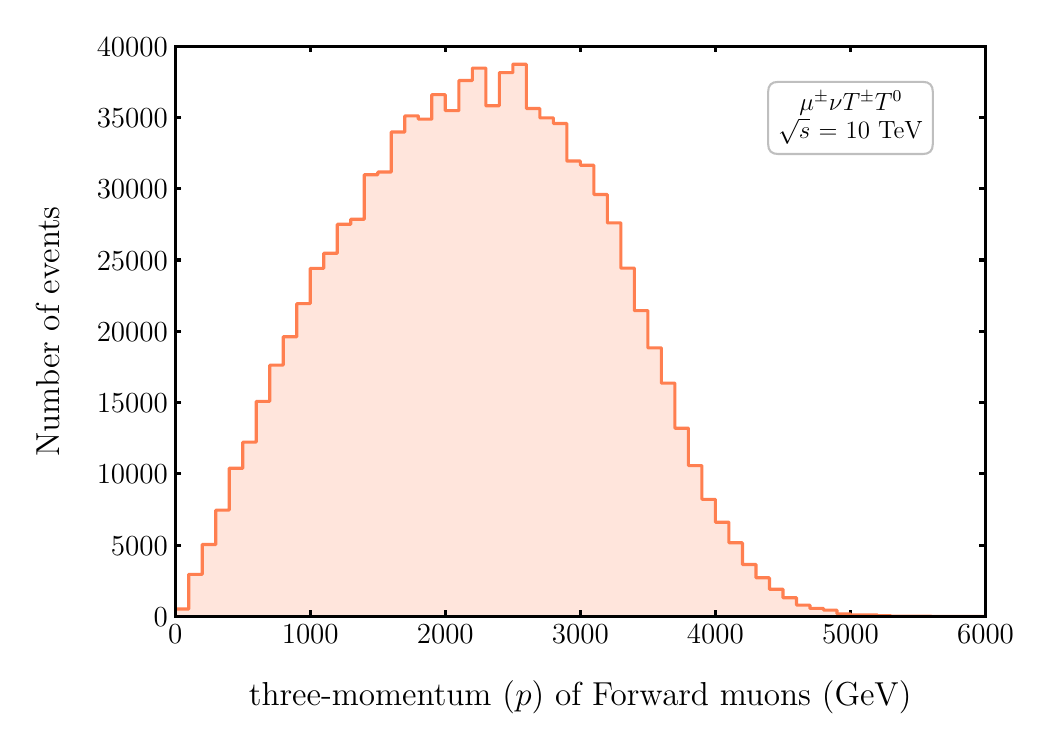}}
	\caption{ 
	(a) The $\eta$ distribution of muons produced 
	in the process $\mu^+ \mu^- \to T^0 T^{\pm} \mu^{\mp} {\nu}$ at a 10 TeV MuC with $\lumtot=10\iab$. 
	Forward muons are highlighted in orange, while muons detected by the main detector are shown in blue; 
	(b) The three-momentum distribution of the same Forward muons at the 10 TeV MuC.
}
	\label{fig:fm}
\end{figure}

To demonstrate the crucial role of Forward muon facilities in detecting the spectator muons of our signal, 
we display in \autoref{fig:fm}(a) the pseudorapidity distribution for muons produced 
in the $\mu^+ \mu^- \to T^0 T^{\pm} \mu^{\mp} {\nu}$ process at a 10 TeV MuC with $\lumtot=10\iab$. 
In this figure, Forward muons are represented in orange, 
contrasting with those detected by the main detector, shown in blue. 
The distribution indicates that 
the majority of muons in our signal predominantly occupy the $\abs{\eta} > 2.5$ region, 
with only a minor fraction being detected in the main detector.

An essential requirement for detecting the spectator muons by the Forward muon detector is 
their ability to penetrate the tungsten nozzle, which necessitates a sufficiently high energy
of the Forward muons. 
\autoref{fig:fm}(b) presents the three-momentum distribution of these spectator muons, 
confirming that they are indeed highly energetic, predominantly peaking around 3 TeV.  
This high energy profile significantly differentiates them from BIB particles, including muons, 
which generally possess much lower kinetic energy, typically less than 100 GeV. 
To accurately identify our signal, therefore,
we apply a threshold of $p_{\muf} \geq 300$ GeV, 
along with a pseudorapidity range of $2.5 \leq \abs{\eta_\mu} \leq 7.0$ for the Forward muons.

\subsection{Infeasibility of the charged pion signal}\label{sec:cpion}

Before delving into our analysis results, 
it is important to demonstrate the infeasibility of detecting charged pions originating from the decay $\tpm\to \pi^\pm T^0$. 
While the long-lived nature of $\tpm$ might suggest that the displaced charged pions could be detectable against SM backgrounds, in practice, these pions are not observable.
The small mass splitting between $T^\pm$ and $T^0$, approximately 166 MeV,
results in charged pions that lack the necessary momentum for reconstruction. 
Detecting these soft pions is particularly challenging in the MuC environment, overwhelmed by soft BIB pions.

\begin{figure}[h]
	\centering
	\includegraphics[width=0.7\linewidth]{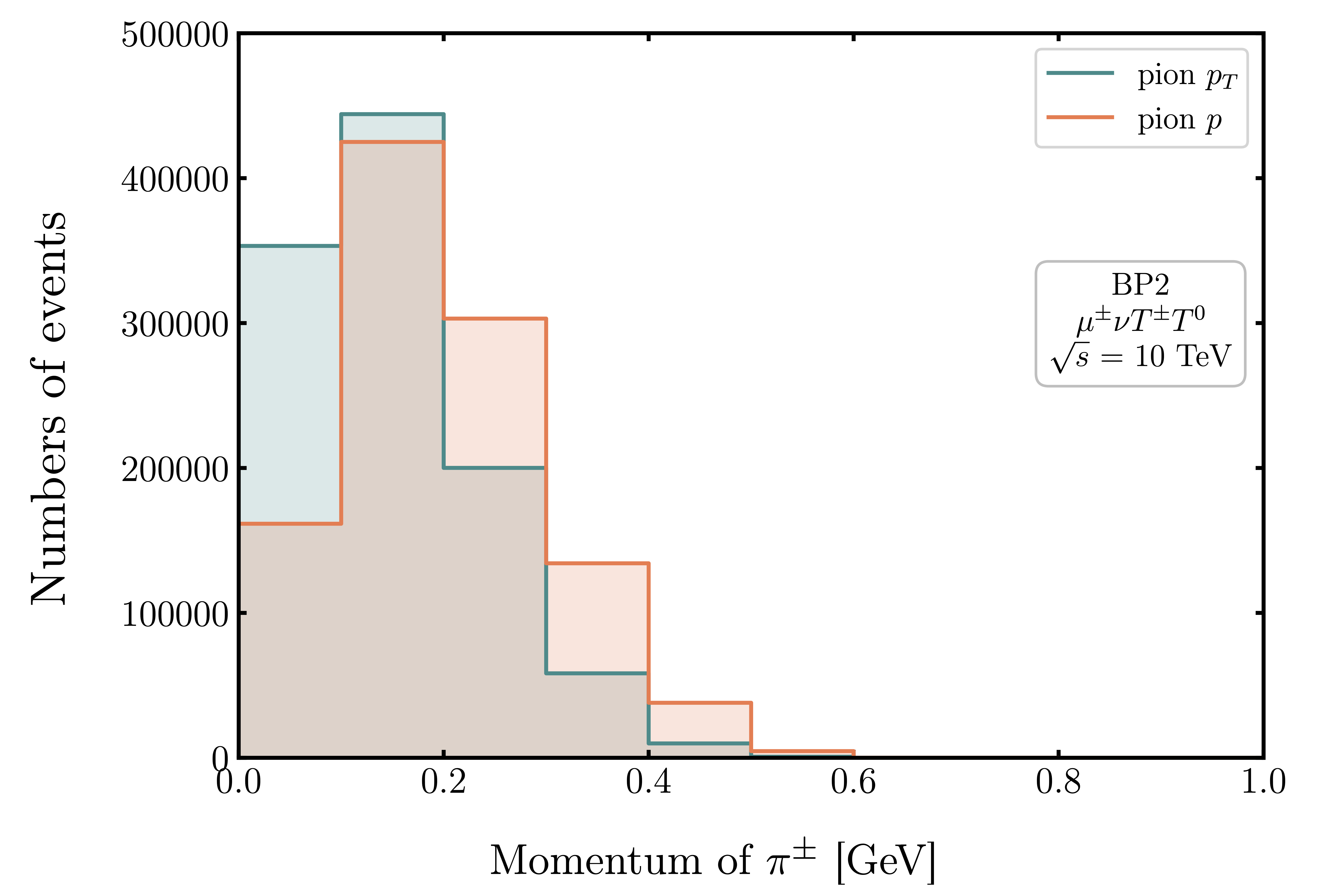}
	\caption{Distributions of  three-momentum (orange) and $p_T$ (blue) of the displaced charged pion for BP2, from the process $\mu^+ \mu^- \to T^0 T^{\pm} \mu^{\mp} {\nu}$ at a 10 TeV MuC with $\lumtot=10\iab$.}
	\label{fig:pimom}
\end{figure}

The idea that high collision energy at a 10 TeV MuC could provide enough boost 
to these displaced charged pions for detection is another aspect to consider.  
This possibility remains largely unexplored in the literature. 
We present in \autoref{fig:pimom} the momentum distribution of displaced charged pions 
from $\mu^+ \mu^- \to T^0 T^{\pm} \mu^{\mp} {\nu}$, followed by $\tpm\to\pi^\pm T^0$, 
at a 10 TeV MuC with $\lumtot=10\iab$. 
Our analysis reveals that, even at this collision energy level, 
the displaced pions do not attain enough boost to meet the track reconstruction threshold 
of $p_T \geq 0.5$ GeV. 
Consequently, they remain unreconstructed, indistinguishable 
among the multitude of soft BIB pions.

Looking forward, a 30 TeV MuC with extremely forward BIBs might provide a chance 
to detect some energetic displaced pion tracks. 
However, moving to a 30 TeV MuC involves a substantial jump in c.m.~energy and 
necessitates a longer timescale for development and analysis. 
Therefore, we defer the exploration of these scenarios to future studies.

\section{Results on the signal event counts}
\label{sec:results}

In the previous section,
we established criteria for DCT signals and Forward muons, summarized as: 
\begin{align}
\label{eq:def:track:fmuon}
\text{DCT}: & \quad 0.7 < \theta_{\text{tr}} < 2.44, \quad 
5.1 \cm < \trad< 148.1 \cm, \quad p_T^{\text{tr}} \geq 300 \gev, 
\\ \nn
\muf: & \quad p_{\muf} \geq 300\gev, ~~\quad 2.5\leq \abs{\eta_{\muf}} \leq 7.0,
\end{align}
where $\trad$ is the transverse length of the charged track.

Based on these criteria, our focus shifts to the signal-to-background analysis for 
$\mmu\to T^\pm T^0 \mu^\mp \nu$ and $\mmu\to T^+ T^- \mu^+ \mu^-$
through the four final states in \autoref{eq:final:signatures}.
We first generated events using \texttt{MadGraph5\_aMC@NLO}, 
followed by showering in \py and detector simulation with \texttt{Delphes} using  \texttt{delphes\_card\_MuonColliderDet.tcl}. 
For jet clustering, the inclusive Valencia algorithm was employed, 
with a jet radius parameter of $R = 0.5$ and a minimum momentum cut of $p_T^j \geq 20$ GeV. 
Charged leptons within the calorimeter coverage region 
were tagged with a cut of $p_T^\ell \geq 20$ GeV.

\begin{table}[t]
	\centering
	\renewcommand{\arraystretch}{1.3}
	\begin{tabular}{|c|c|c|c|c|}
		\toprule
		~~Final state & ~~VBF of $\mmu\to$~~ & \multicolumn{3}{c|}{Selections} \\ \hline
		\multirow{2}{*}{FS1} & $T^0 \tpm \mu^\mp \nu$, & \multirow{2}{*}{~~$N_\text{DCT}=1$~~} &
		\multirow{2}{*}{~~$N_{\muf}=1$~~} & \multirow{5}{*}{\shortstack[l]{~~$\met<10\gev$, \\[5pt] ~~$N_{\ell^\pm}=N_j=0$}} \\
		 & $T^+ T^- \mu^+\mu^-$ & & & \\ \cline{1-4}
		 FS2 & $T^+ T^- \mu^+\mu^-$ & $N_{\rm DCT}=2$ &
		$N_{\muf}=2$ & \\ \cline{1-4}
		FS3 & $T^+ T^- \mu^+\mu^-$ & $N_{\rm DCT}=1$ &
		$N_{\muf}=2$ & \\ \cline{1-4}
		FS4 & $T^+ T^- \mu^+\mu^-$ & $N_{\rm DCT}=2$ &
		$N_{\muf}=1$ & \\ \bottomrule
			\end{tabular}
	\caption{Summary of the selection cuts for the four final states of two signal processes, 
	$\mmu\to T^\pm T^0 \mu^\mp \nu$ and $\mmu\to T^+ T^- \mu^+ \mu^-$. 
	Here, $N_i$ represents the number of the object $i$, and $\met$ stands for the missing transverse energy.}
	\label{tab:summary:cuts}
\end{table}

In general, signals that involve disappearing tracks at the MuC can be relatively free from SM background processes, especially after vetoing out any calorimeter hits (jets and charged leptons). Demanding a hard Forward muon can effectively rule out non-VBF SM backgrounds such as $\mu^+ \mu^- \to VV$, or the invisible process $\mu^+ \mu^- \to \nu_\mu \bar{\nu}_\mu$, which can potentially mimic a DCT signal with some combinations of spurious detector hits. Certain VBF SM processes,
such as $\mu^+ \mu^- \to \mu^\pm \nu W^\mp Z$ and $\mu^+ \mu^- \to\mu^+ \mu^- \nu_\mu \bar{\nu_\mu}$, 
still possess large cross-sections and could potentially contaminate our signal.
To effectively suppress SM backgrounds to negligible levels, we employ four key discriminators:
(i) presence of one or two Forward muons with $p_T^{\muf}>300\gev$;
(ii) a veto on events with any jets or leptons; 
(iii) the requirement of exactly one or two DCTs with $p_T^{\rm tr}>300\gev$;
(iv) soft $\met$ below 10 GeV.
These discriminators were meticulously tailored to each final state,
as summarized in \autoref{tab:summary:cuts}.
In this table,  we present the number of Forward muons and DCT signals 
along with common requirements on $\met$ and lepton/jet multiplicity.

The proposed discriminators are highly efficient, nearly eliminating all SM backgrounds.
Detailed information is provided in \autoref{appendix:SM:backgrounds}. The only backgrounds that can mimic the signal are from BIB hits that can get reconstructed as fake tracks, which will be discussed later in \autoref{sec:projection}.
Consequently, the number of expected signal events, $N_{\rm sig}$, 
becomes a critical factor in assessing the potential of the MuC to probe ITM signals. 
We calculate $N_{\rm sig}$ by integrating the DCT reconstruction efficiency $\pdct$ as follows:
\begin{equation}
N_{\rm sig} = \pdct \times \frac{n_{\rm cut}}{n_{\rm gen}} \sigma_\tot \lumtot,
\end{equation}
where $n_{\rm gen}$ is the total number of events generated at the \texttt{MadGraph} level, 
$n_{\rm cut}$ is the count of events meeting the criteria in \autoref{tab:summary:cuts}, 
and $\sigma_\tot$ is the total cross-section for the respective signal process. 
Two MuC configurations are considered:
6 TeV  MuC with $\lumtot=4\iab$ and  10 TeV MuC with $\lumtot=10\iab$.

We now present our results for FS1, a final state characterized by exactly one DCT and one Forward muon. 
FS1 emerges from two distinct signal processes: 
$\mu^+ \mu^- \to T^0 T^{\pm} \mu^{\mp} {\nu}$ and $\mu^+ \mu^- \to T^+ T^- \mu^+\mu^-$. 
The notable promise of this final state is largely attributed 
to the high production cross-section of $\mu^+ \mu^- \to T^0 T^{\pm} \mu^{\mp} {\nu}$, 
positioning it as a key focus in our analysis.

\begin{table}[h]
	\centering
	\renewcommand{\arraystretch}{1.3}
	\begin{tabular}{|c|c|c|c|c|c|}
		\hline
		\multirow{2}{*}{$\sqrt{s}$} & \multirow{2}{*}{$\lumtot$} & \multirow{2}{*}{\makecell{VBF process \\  $\mu^+ \mu^- \to$ products}} & \multicolumn{3}{c|}{Number of events for FS1} \\
		\cline{3-6}
		& & & \hspace*{0.5cm} BP1 \hspace*{0.5cm} & \hspace*{0.5cm} BP2 \hspace*{0.5cm} & \hspace*{0.5cm} BP3 \hspace*{0.5cm} \\
		\hline\hline
		\multirow{3}{*}{$6\tev$} & \multirow{3}{*}{$4\iab$} & $T^0 T^{\pm} \mu^{\mp} {\nu}$ &12598.78 &1865.75&-- \\
		\cline{3-6}
		& & $T^+ T^- \mu^+\mu^-$ &153.86&26.26&--\\
		\cline{3-6}
		& & Total &12752.64 &1892.01& --\\
		\hline\hline
		\multirow{3}{*}{$10\tev$} & \multirow{3}{*}{$10\iab$} &$T^0 T^{\pm} \mu^{\mp} {\nu}$ & 402391.71& 177061.29& 246.80\\
		\cline{3-6}
		& & $T^+ T^- \mu^+\mu^-$ &4231.26&2102.84&4.32\\
		\cline{3-6}
		& & Total & 406622.97 & 179164.13& 251.12 \\
		\hline		
	\end{tabular}
	\caption{Number of signal events for the FS1  at a 6 TeV and a 10 TeV MuC, 
	with an integrated luminosity of 4 ab$^{-1}$ and 10 ab$^{-1}$, respectively.}
	\label{tab:fs1}
\end{table}

\autoref{tab:fs1} shows the expected number of signal events for FS1 at both 6 TeV and 10 TeV MuC, 
accounting for contributions from both signal processes across three benchmark points. 
Although the process $\mu^+ \mu^- \to T^+ T^- \mu^+\mu^-$ also contributes a substantial number of events, 
these values are smaller by almost two orders of magnitude 
compared to $\mu^+ \mu^- \to T^0 T^{\pm} \mu^{\mp} {\nu}$. 
This discrepancy arises from its lower cross-section and a small probability of missing one Forward muon. 

At the 6 TeV MuC, a significant number of signal events are predicted for BP1 and BP2, but no events for BP3 with its production kinematically disallowed. At the 10 TeV MuC, these numbers increase drastically in accordance with the cross-section, particularly for $\mu^+ \mu^- \to T^0 T^{\pm} \mu^{\mp} \nu$,  and we also obtain $\sim250$ events for BP3 as well.
The lower event count for BP3 is attributed to the reduced cross-section (as shown in \autoref{tab:vbfcs}) 
and the insufficient boost for the charged tracks to reach the detector region.

\begin{table}[h]
	\centering
	\renewcommand{\arraystretch}{1.2}
	\begin{tabular}{|c|c|c|c|c|c|}
		\hline
		\multirow{2}{*}{$\sqrt{s}$}  & \multirow{2}{*}{$\lumtot$} & \multirow{2}{*}{\makecell{VBF process \\ $\mu^+ \mu^- \to$ products}} & \multicolumn{3}{c|}{\makecell{Number of events for FS1$'$ \\ 
		$N_{\rm DCT}=N_\muc=1,~ \met<10\gev,~ N_{j,e}=0$ }} \\
		\cline{4-6}
		&& & \hspace*{0.5cm} BP1 \hspace*{0.5cm} & \hspace*{0.5cm} BP2 \hspace*{0.5cm} & \hspace*{0.5cm} BP3 \hspace*{0.5cm} \\
		\hline\hline
		6 TeV & $4\iab$ & $T^0 T^{\pm} \mu^{\mp} {\nu}$ &115.82&19.79&--\\
		\hline
		10 TeV & $10\iab$& $T^0 T^{\pm} \mu^{\mp} {\nu}$ &987.53&503.18&1.80\\
		\hline		
	\end{tabular}
	\caption{The number of signal events for FS1$'$ featuring a central muon 
	but without tagging a Forward muon at both 6 TeV and 10 TeV MuC, with integrated luminosities of 4 ab$^{-1}$ and 10 ab$^{-1}$, respectively. 
	Here, $N_\muc$ represents the number of hard central muons. }
	\label{tab:fsnofm}
\end{table}

Prior to discussing the results for FS2, 
it is important to highlight the significant role of tagging Forward muons in probing the ITM signals. 
To illustrate this, we consider an alternative final state to FS1, named FS1$'$. 
The primary distinction between FS1$'$ and FS1 lies in the type of muon required: 
FS1$'$ necessitates a central muon, 
characterized by $\abs{\eta_\muc} \leq 2.44$, 
while FS1 demands a Forward muon. 
The other selection conditions remain the same.
The number of signal events for FS1$'$, originating 
from the main signal process $\mu^+ \mu^- \to T^0 T^{\pm} \mu^{\mp} {\nu}$, is presented in \autoref{tab:fsnofm}.

It is apparent that the event count for FS1$'$ 
is significantly lower than that for FS1, by almost three orders of magnitude. 
Notably, at a 6 TeV MuC, BP2 yields only about 20 events. 
While the event numbers at a 10 TeV MuC are still relatively high, 
FS1$'$ may encounter substantial non-VBF SM backgrounds~\cite{Han:2020uak}, 
leading to a much lower discovery potential. 

Similar analyses about disappearing track signals in the context of wino, higgsino, or other fermionic models~\cite{Capdevilla:2021fmj, Han:2020uak} prefer to use a hard photon as a trigger instead of forward muons. This is primarily driven by the fact that, for fermions, the DY pair production seems to dominate over the VBF production, especially considering no Forward muon detectors. In contrast, for the scalar triplet case such as ours, we have already established in \autoref{sec:prod} that the VBF production rate almost always dominate, over a very large mass range. Nonetheless, we also consider a case where a final state of one DCT from the  $\mu^+ \mu^- \to T^0 T^{\pm} \mu^{\mp} {\nu}$ process is triggered via a hard photon inside the main detector coverage of $\abs{\eta} \leq 2.44$, with $p_T^\gamma \geq 25$ GeV. Removing requirements for any high-energy muon as a trigger, we present the event counts for the 1 DCT + 1$\gamma$ final state, denoted as FS1$''$, in the following \autoref{tab:fs1gamma}.

\begin{table}[h]
	\centering
	\renewcommand{\arraystretch}{1.2}
	\begin{tabular}{|c|c|c|c|c|c|}
		\hline
		\multirow{2}{*}{$\sqrt{s}$}  & \multirow{2}{*}{$\lumtot$} & \multirow{2}{*}{\makecell{VBF process \\ $\mu^+ \mu^- \to$ products}} & \multicolumn{3}{c|}{\makecell{Number of events for FS1$''$ \\ 
				$N_{\rm DCT}=N_\gamma=1,~ \met<10\gev,~ N_{j,e}=0$ }} \\
		\cline{4-6}
		&& & \hspace*{0.5cm} BP1 \hspace*{0.5cm} & \hspace*{0.5cm} BP2 \hspace*{0.5cm} & \hspace*{0.5cm} BP3 \hspace*{0.5cm} \\
		\hline\hline
		6 TeV & $4\iab$ & $T^0 T^{\pm} \mu^{\mp} {\nu} + \gamma$ &59.87&11.07&--\\
		\hline
		10 TeV & $10\iab$& $T^0 T^{\pm} \mu^{\mp} {\nu} + \gamma$ &672.79&340.04&1.05\\
		\hline		
	\end{tabular}
	\caption{The number of signal events for FS1$''$ with one DCT and one hard photon with $p_T^\gamma \geq 25$ at both 6 TeV and 10 TeV MuC, with integrated luminosities of 4 ab$^{-1}$ and 10 ab$^{-1}$, respectively. 
		Here, $N_\gamma$ represents the number of hard photons. }
	\label{tab:fs1gamma}
\end{table}

From \autoref{tab:fs1gamma}, it is evident that the event counts with a photon trigger instead of a central muon trigger decrease even further, mainly as a consequence of the phase space-suppressed cross-section that comes with the demand for a hard photon. This further exemplifies the enhanced sensitivity of the MuC for the ITM with the inclusion of a Forward muon detector to correctly tag the VBF process.

In additional advantage of utilizing Forward muons to tag VBF processes lies in the potential rejection of BIB events. Demanding a triggering particle such as a photon or a muon in the main detector makes the final state more susceptible to contamination by BIB (and other SM backgrounds). On the other hand, a soft muon from the BIB penetrating through the tungsten nozzles and reaching the Forward muon detectors is drastically less feasible. Hence, triggering VBF processes with Forward muons not only enhances the sensitivity for the triplet scalars, but also allows potentially less BIB events for a much cleaner signal. It is also important to note that, for scenarios where the DY cross-section dominates, such as the ones discussed in Refs.~\cite{Capdevilla:2021fmj, Han:2020uak}, the photon trigger will obviously work better. This choice of trigger is hence highly model-dependent, and a one-to-one comparison with different production modes for different models is beyond the scope of this work. With this discussion in mind, we will present the event counts for the FS2-4 exclusively with Forward muon tagging.

\begin{table}[h]
	\centering
	\renewcommand{\arraystretch}{1.2}
	\begin{tabular}{|c|c|c|c|c|c|}
		\hline
		\multirow{2}{*}{$\sqrt{s}$}  & \multirow{2}{*}{$\lumtot$} & \multirow{2}{*}{\makecell{VBF process \\ $\mu^+ \mu^- \to$ products}} & \multicolumn{3}{c|}{Number of events for FS2} \\
		\cline{4-6}
		& & & \hspace*{0.5cm} BP1 \hspace*{0.5cm} & \hspace*{0.5cm} BP2 \hspace*{0.5cm} & \hspace*{0.5cm} BP3 \hspace*{0.5cm} \\
		\hline\hline
		6 TeV & $4\iab$  & $T^+ T^- \mu^+\mu^-$ &206.29&18.71&--\\
		\hline
		10 TeV & $10\iab$  & $T^+ T^- \mu^+\mu^-$ &9979.87&3620.01&1.00\\
		\hline
		
	\end{tabular}
	\caption{Number of signal events for the FS2 at a 6 TeV and a 10 TeV muon collider, with an integrated luminosity of 4 ab$^{-1}$ and 10 ab$^{-1}$, respectively.}
	\label{tab:fs2}
\end{table}

We now turn our attention to FS2, characterized by two DCTs and two Forward muons. 
This final state exclusively arises from the $\mu^+ \mu^- \to T^+ T^- \mu^+\mu^-$ process.
The expected number of signal events for FS2 at both a 6 TeV and a 10 TeV MuC, 
with integrated luminosities of 4 ab$^{-1}$ and 10 ab$^{-1}$ respectively, are detailed in \autoref{tab:fs2}. 

When comparing the signal event counts for FS2 with those for FS1 from the same process, $\mu^+ \mu^- \to T^+ T^- \mu^+ \mu^-$, some interesting insights emerge.
Intuitively, one might expect the $T^+ T^- \mu^+ \mu^-$ process to yield more events for FS2 than for FS1.
However, this is not the case at the 6 TeV MuC.  
Specifically, for BP2, the event counts for FS2 are lower than those for FS1.
Even though BP1 shows a higher event count for FS2, the increase is only marginal. 
This outcome results from the requirement of two reconstructed DCTs, 
leading to a compounded suppression from two factors of $\pdct$. 
The limited boosting at the 6 TeV MuC leads to decreased reconstruction efficiency, thereby diminishing the overall event count.

Conversely, at a 10 TeV MuC, FS2 records more events than FS1 for the $T^+ T^- \mu^+\mu^-$ process. 
The number of signal events is substantial for BP1 and BP2. 
This increase is attributed to the enhanced boost available at the higher energy level, 
allowing more disappearing tracks to reach regions with higher efficiency. However, BP3 only yields $\sim1$ event, with heavily suppressed cross-section, and not enough boost to reach high-efficiency regions of the tracker.

\begin{table}[h]
	\centering
	\renewcommand{\arraystretch}{1.2}
	\begin{tabular}{|c|c|c|c|c|c|}
		\hline
		\multirow{2}{*}{$\sqrt{s}$}  & \multirow{2}{*}{$\lumtot$}  & \multirow{2}{*}{\makecell{VBF process \\ $\mu^+ \mu^- \to$ products}} & \multicolumn{3}{c|}{Number of events for FS3} \\
		\cline{4-6}
		& & & \hspace*{0.5cm} BP1 \hspace*{0.5cm} & \hspace*{0.5cm} BP2 \hspace*{0.5cm} & \hspace*{0.5cm} BP3 \hspace*{0.5cm} \\
		\hline\hline
		6 TeV & $4\iab$ &  $T^+ T^- \mu^+\mu^-$ &1330.21&211.22&--\\
		\hline
		10 TeV & $10\iab$  &  $T^+ T^- \mu^+\mu^-$ &37556.77&18783.92&31.33\\
		\hline		
	\end{tabular}
	\caption{Number of signal events for the FS3 at a 6 TeV and a 10 TeV muon collider, with an integrated luminosity of 4 ab$^{-1}$ and 10 ab$^{-1}$, respectively.}
	\label{tab:fs3}
\end{table}

We now examine FS3, characterized by one DCT and two Forward muons. 
This final state arises exclusively from the $\mu^+ \mu^- \to T^+ T^- \mu^+\mu^-$ process, 
with one of the two $\tpm$ particles not being reconstructed as a DCT signal at the detector level. 
The expected numbers of signal events for FS3 at both a 6 TeV and a 10 TeV MuC, 
with integrated luminosities of 4 ab$^{-1}$ and 10 ab$^{-1}$ respectively, are shown in \autoref{tab:fs3}.

Comparing FS3 with FS2, it becomes evident that FS3 consistently shows a higher number of events. 
This difference is attributable to the requirement of only a single DCT for FS3, 
which allows for the preservation of more signal events 
due to the application of $\pdct$ only once.

When comparing the results at the 6 TeV MuC with those at the 10 TeV MuC, 
a clear pattern emerges: the 10 TeV MuC consistently yields a significantly higher number of signal events compared to the 6 TeV MuC. 
Remarkably, the enhancement factor in the number of signal events exceeds that seen 
at the cross-section level in \autoref{tab:vbfcs}. 
This discrepancy is primarily due to the boost required for DCT reconstruction. 
While the 6 TeV MuC yields limited boosting, hindering effective DCT reconstruction, 
the 10 TeV MuC achieves the necessary boost, 
resulting in more effective DCT detection and a subsequent increase in signal events. In this final state, the heavy triplet of BP3 also yields $\sim31$ events at the 10 TeV MuC, despite the low cross-section and insufficient boost.

\begin{table}[h]
	\centering
	\renewcommand{\arraystretch}{1.2}
	\begin{tabular}{|c|c|c|c|c|c|}
		\hline
		\multirow{2}{*}{$\sqrt{s}$}  & \multirow{2}{*}{$\lumtot$} & \multirow{2}{*}{\makecell{VBF process \\ $\mu^+ \mu^- \to$ products}} & \multicolumn{3}{c|}{Number of events for FS4} \\
		\cline{4-6}
		& & &  \hspace*{0.5cm} BP1 \hspace*{0.5cm} & \hspace*{0.5cm} BP2 \hspace*{0.5cm} & \hspace*{0.5cm} BP3 \hspace*{0.5cm} \\
		\hline\hline
		6 TeV & $4\iab$ & $T^+ T^- \mu^+\mu^-$ &24.46&2.72&--\\
		\hline
		10 TeV & $10\iab$ & $T^+ T^- \mu^+\mu^-$ &1084.74&429.05&0.17\\
		\hline
		
	\end{tabular}
	\caption{Number of signal events for the FS4 at a 6 TeV and a 10 TeV muon collider, with an integrated luminosity of 4 ab$^{-1}$ and 10 ab$^{-1}$, respectively.}
	\label{tab:fs4}
\end{table}

Finally, we analyze FS4, which consists of two DCTs and one Forward muon. 
Originating from the $\mu^+ \mu^- \to T^+ T^- \mu^+\mu^-$ process, 
FS4 represents a scenario where one of the two Forward muons is not detected by the Forward detector, 
while both DCTs are successfully reconstructed. 
The expected number of events for FS4 at a 6 TeV and 10 TeV MuC, 
with integrated luminosities of 4 ab$^{-1}$ and 10 ab$^{-1}$ respectively, 
is presented in \autoref{tab:fs4}.

FS4 records the lowest number of events among the four final states. 
This is primarily due to the relatively small cross-section of the $T^+ T^- \mu^+\mu^-$ process
and the low probability of missing a hard Forward muon, given its expected detection efficiency of about 90\%. 
At the 6 TeV MuC, fewer than 25 events are observed for BP1, while the event count for BP2 is negligible.  
In contrast, at the 10 TeV MuC, 
a substantial number of events are observed for both BP1 and BP2, while BP3 remains unseen.

\section{Discovery projections} \label{sec:projection}

In this section, we conduct an in-depth analysis of the discovery potential of our model at two different MuC energies, 6 TeV and 10 TeV, taking into account the target luminosity. 
A critical factor in determining this potential is the signal significance, 
which depends heavily on the number of SM background events. 
As detailed in \autoref{appendix:SM:backgrounds}, the final states consisting of DCTs and hard Forward muons
are not contaminated by the direct SM background processes. 
However, the sheer volume of BIB hits can result in a significant number of events with fake tracks 
that persist even after applying all selection criteria for the suggested final states. In Ref.~\cite{Capdevilla:2021fmj}, the 10 TeV MuC results are evaluated with the BIB overlay from a 1.5 TeV MuC study, with which $\sim$200 fake track background events are obtained for the photon-triggered single DCT search with a total luminosity of $10\iab$, considering stub tracks only in the VXD barrel region. At higher collider energies, the BIB is expected to become more and more forward, and is expected to reduce significantly at a 10 TeV MuC. However, we consider this 200 background events as one of the scenarios in which we evaluate the reach of our final state searches. Taking into account the fact that the Forward muon tagging can suppress the events with fake BIB tracks even further, and considering the additional suppression at higher energies, we add another scenario to our discovery projection, with 10 background events.\footnote{In \cite{Capdevilla:2021fmj}, it is also found that for the two-track final state, the background counts are less than one. However, to maintain uniformity, we evaluate the reaches for both single- and double-DCT with a lowest of 10 background events. } Furthermore, we also take into account a very conservative scenario with 600 background events from fake tracks, considering the distribution of BIB hits in the IT and OT barrel regions~\cite{MuonCollider:2022ded}.


For each final state and collider energy, we explore the reach of the MuC for the three aforementioned background scenarios, scanning over a range of $\mtz$ values, starting from 500 GeV, until the kinematical limit of pair production at the 6 TeV and 10 TeV MuC. For each mass, we present the required luminosity for a $5\sigma$ discovery, considering each of the three background numbers, at both the MuC energies. We utilize the \texttt{pyhf} package~\cite{lukas_heinrich_2021_5009162, pyhf_joss} to evaluate the significance for each final state at each mass, with a likelihood-based analysis. For the statistical model, we use the \texttt{pyhf.simplemodels.uncorrelated\_backgrounds()} model, which takes as inputs the signal event counts, the background event counts, as well as a systematic uncertainty on the background, which for this analysis is considered to be 20\%. Based on the inputs, it constructs Poisson likelihoods for background-only and signal+background scenarios, with the uncertainty  by modeling it as a Gaussian probability distribution function. The hypothesis testing is then performed with \texttt{pyhf.infer.hypotest()} function, which returns $p$-values for each of the signal + background combination corresponding to the luminosity.\footnote{Further details of the \texttt{pyhf} usage can be found at \hyperlink{https://pyhf.readthedocs.io/en/v0.7.6/index.html}{https://pyhf.readthedocs.io/en/v0.7.6/index.html}} A $p$-value of $2.87\times10^{-7}$ corresponds to a 5$\sigma$ signal significance. We scale the event counts over a range of luminosities, until we obtain the $5\sigma$ significance for each mass.

\begin{figure}[h]
	\centering
	\subfigure[]{\includegraphics[width=0.49\linewidth]{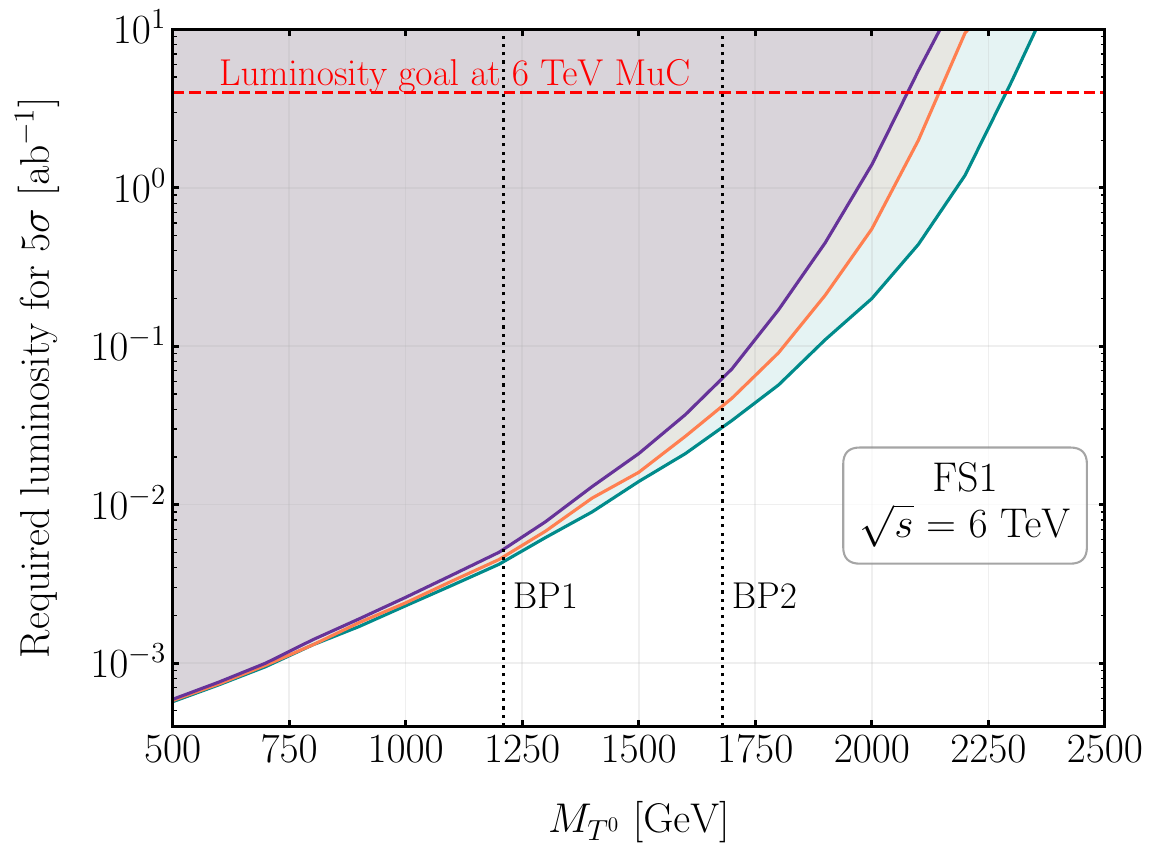}}
	\subfigure[]{\includegraphics[width=0.49\linewidth]{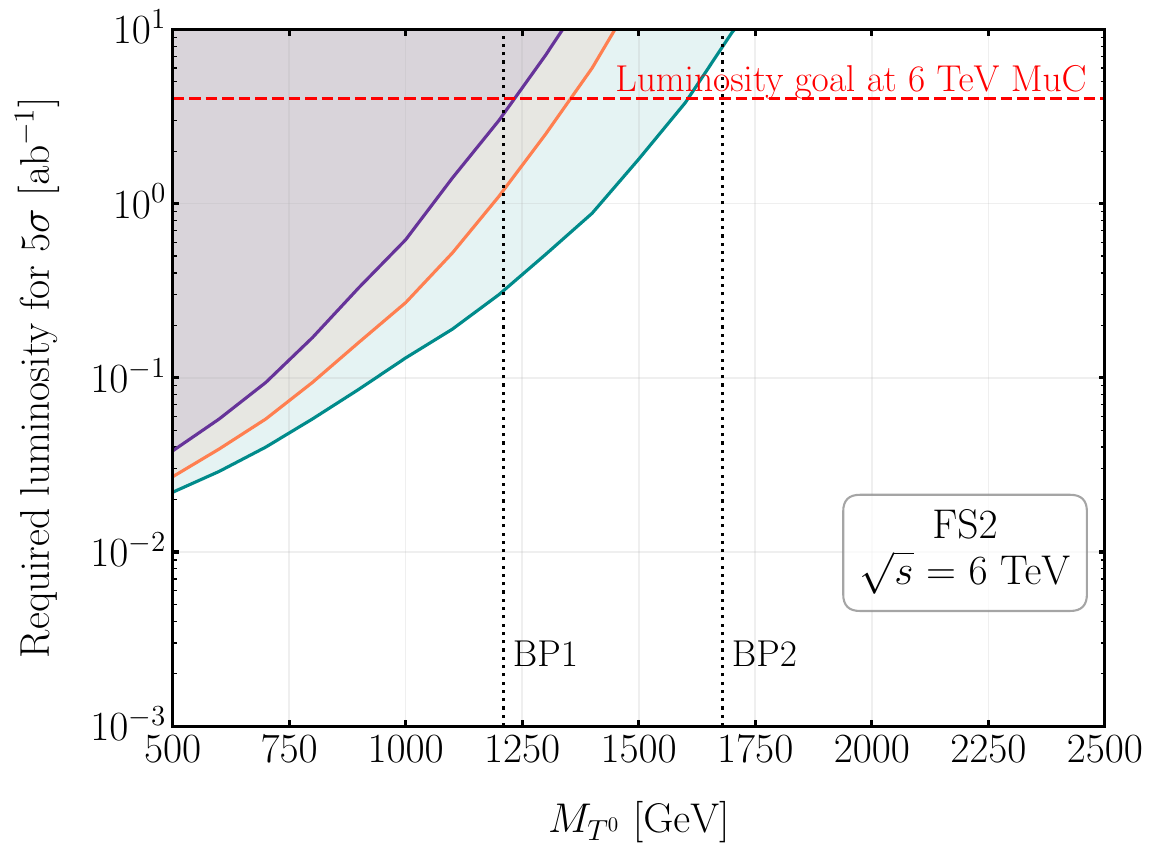}}
	\subfigure[]{\includegraphics[width=0.49\linewidth]{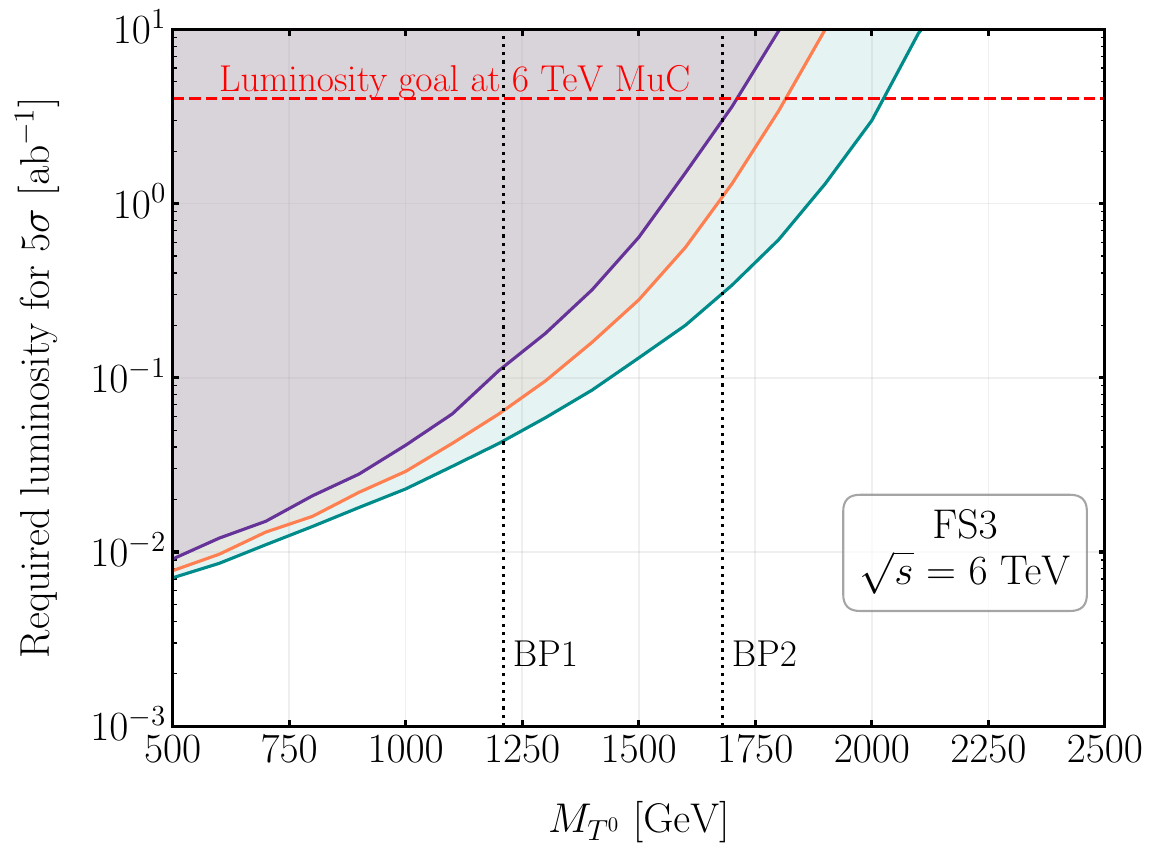}}
	\subfigure[]{\includegraphics[width=0.49\linewidth]{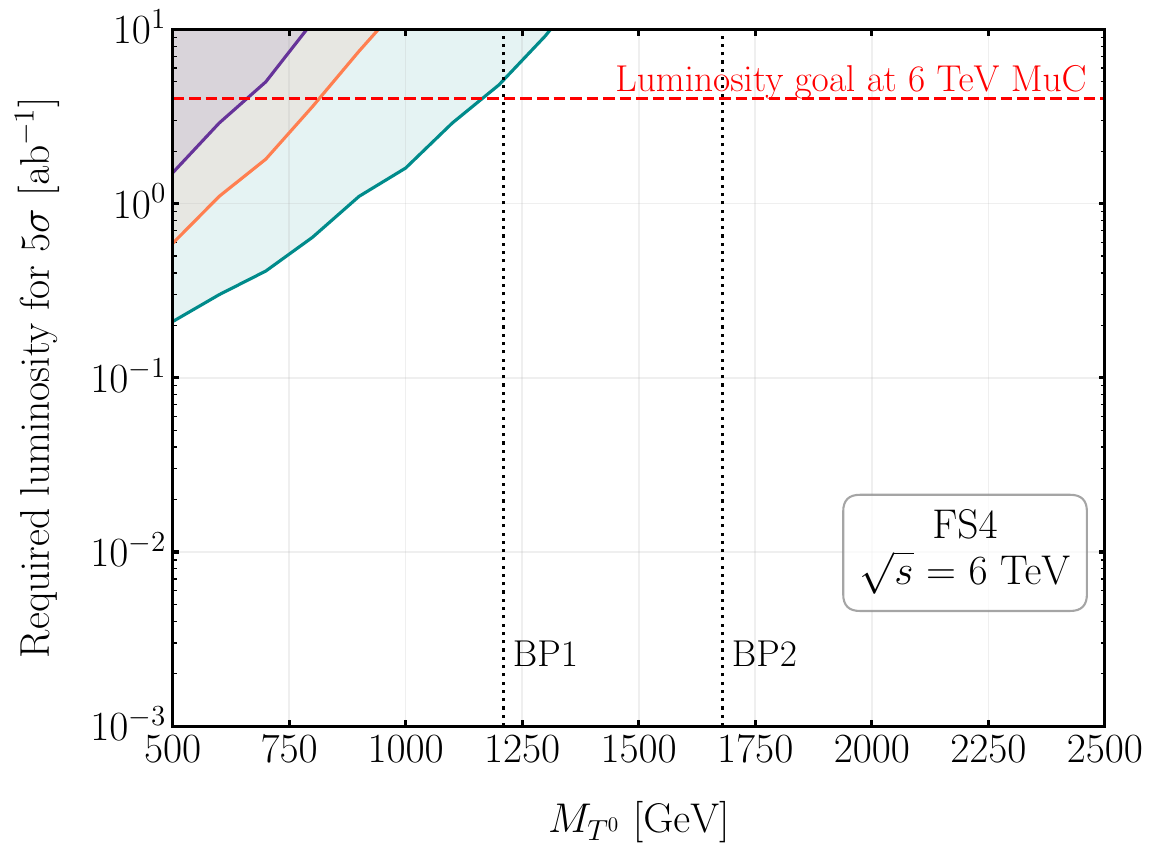}}
	\subfigure{\includegraphics[width=0.7\linewidth]{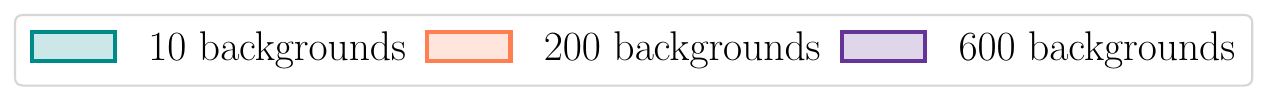}}
	\caption{Reach plots for triplet scalar masses and the required luminosity for 5$\sigma$ discovery, at the 6 TeV MuC, with event counts for (a) FS1, (b) FS2, (c) FS3, and (d) FS4. The green, orange, and purple colours correspond to scenarios with 10, 200, and 600 background events respectively, with 20\% systematic uncertainty. The 4 \abi luminosity goal is marked in the red dashed line, and the collider benchmark points are shown in black dotted vertical lines.
	}
	\label{fig:6sig}
\end{figure}

 Utilizing the analytical framework discussed above, in \autoref{fig:6sig} we present the reach plots for FS1-4 at the 6 TeV MuC, with the DM mass on the $x$-axis, and the required luminosity for $5\sigma$ significance in the $y$-axis. In each plot, the green, orange, and purple colors correspond to scenarios with 10, 200, and 600 background events respectively. The luminosity goal of 4 \abi for the 6 TeV MuC is marked with the red dashed line. The black dotted vertical lines correspond to the two benchmark points for which we have presented event counts in \autoref{sec:results}.

From \autoref{fig:6sig}(a), it is evident that FS1 is the most advantageous final state, 
requiring the least luminosity to achieve a 5$\sigma$ significance for all three backgrounds over the largest range of triplet scalar mass. Within the target luminosity of 4 \abi at the 6 TeV MuC, the ITM scalar with masses up to $\sim$2.3 TeV can be successfully probed for 10 background events. For 200 and 600 backgrounds, the upper limit for mass reduces to $\sim$2.1 TeV and $\sim$2.2 TeV, respectively.  For BP1, a luminosity of $\mathcal{O}(1)$ \fbi is enough to achieve the discovery, whereas for BP2, one can obtain $5\sigma$ significance with $\mathcal{O}(10)$ \fbi of luminosity.

FS3, depicted in \autoref{fig:6sig}(c), emerges as the next most favorable final state. For 10, 200, and 600 backgrounds, the upper limits on the triplet scalar mass that can be probed within the 4\abi luminosity goal are $\sim$2.1 TeV, $\sim$ 1.8 TeV, and $\sim$ 1.7 TeV, respectively. Here, BP1 can be probed with 40-100 \fbi of luminosity, and the heavier BP2 still remains discoverable at all three background scenarios, with a maximum luminosity requirement of 3 \abi.

In contrast, FS2 and FS4, as shown in \autoref{fig:6sig}(b) and (d) respectively, 
exhibit very limited significances. 
Through FS2, 
BP1 can still be probed for all three background considerations below the 4 \abi luminosity goal,
but BP2 cannot be probed with the target luminosity of the 6 TeV MuC with 10 backgrounds.
FS4 has further suppressed significance,  for which the maximum mass reach, even with 10 backgrounds, is found to be $\sim$ 1.15 TeV, which is below the BP1 mass value.


\begin{figure}[h]
	\centering
	\subfigure[]{\includegraphics[width=0.49\linewidth]{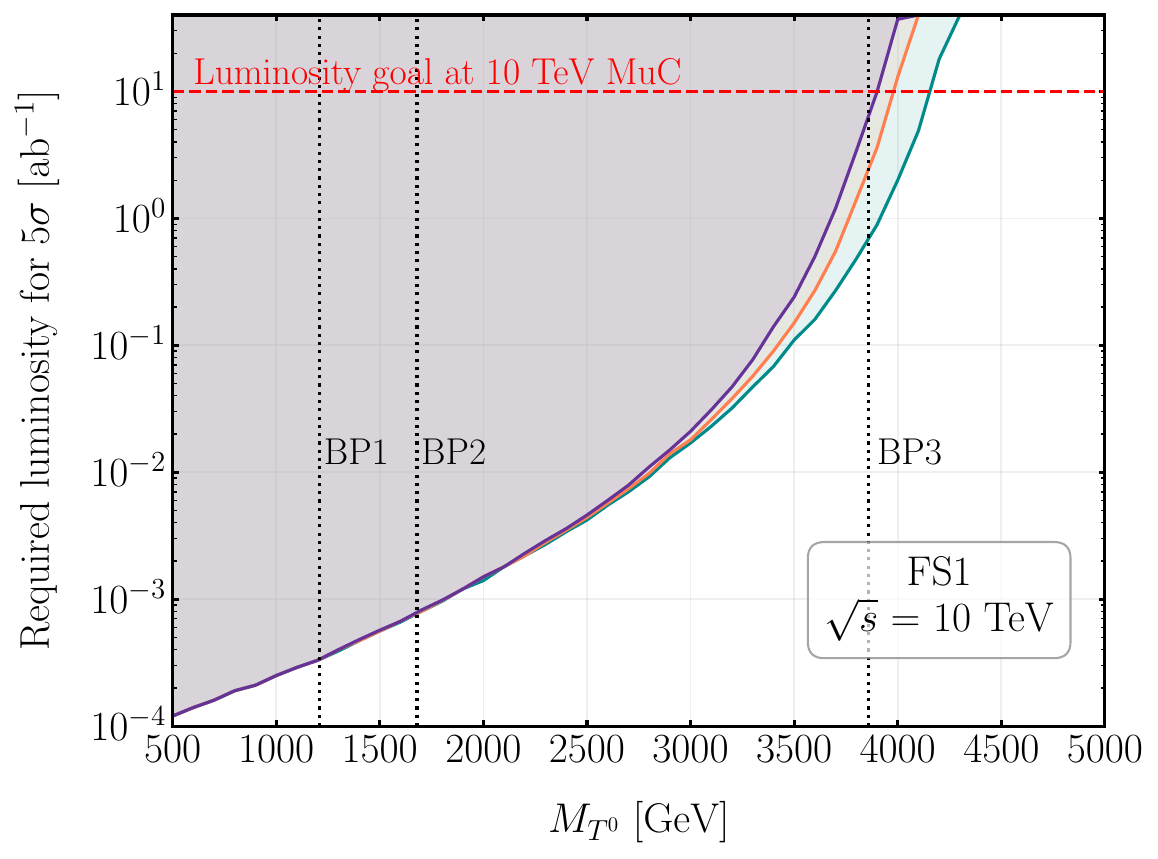}}
	\subfigure[]{\includegraphics[width=0.49\linewidth]{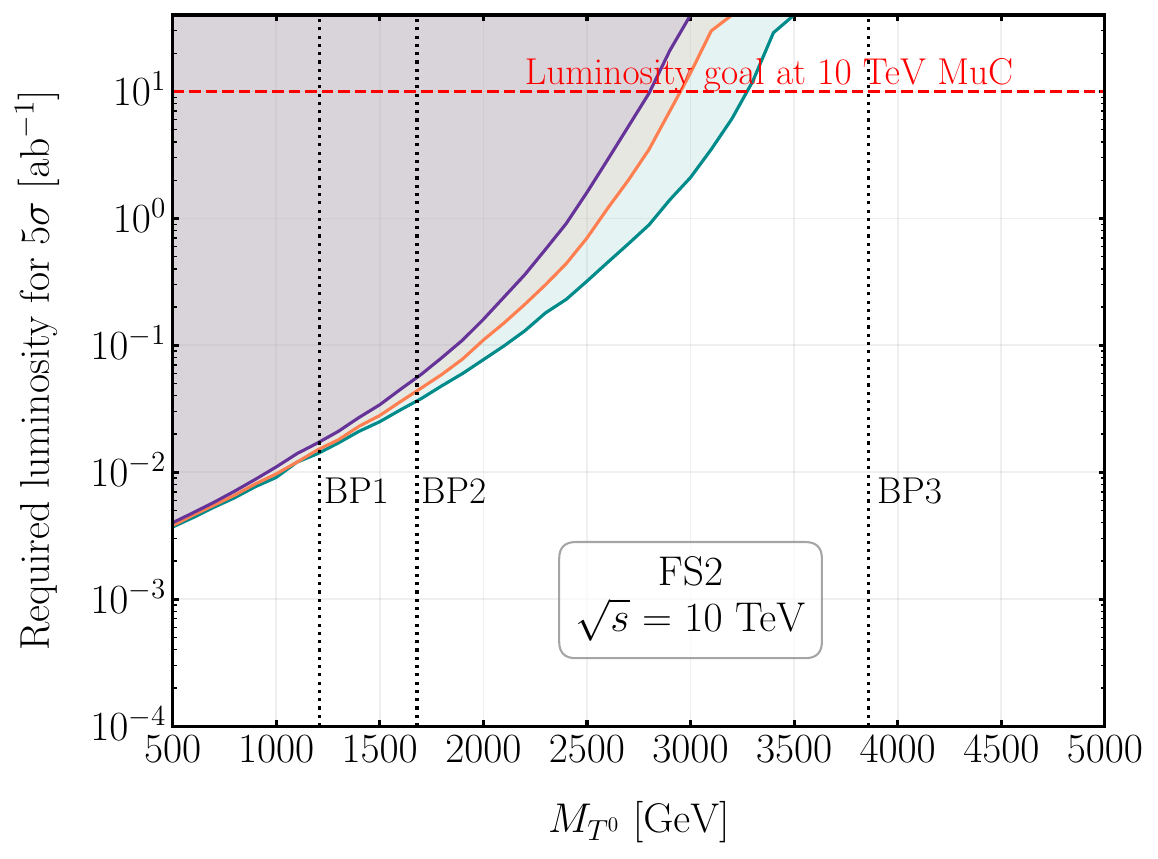}}
	\subfigure[]{\includegraphics[width=0.49\linewidth]{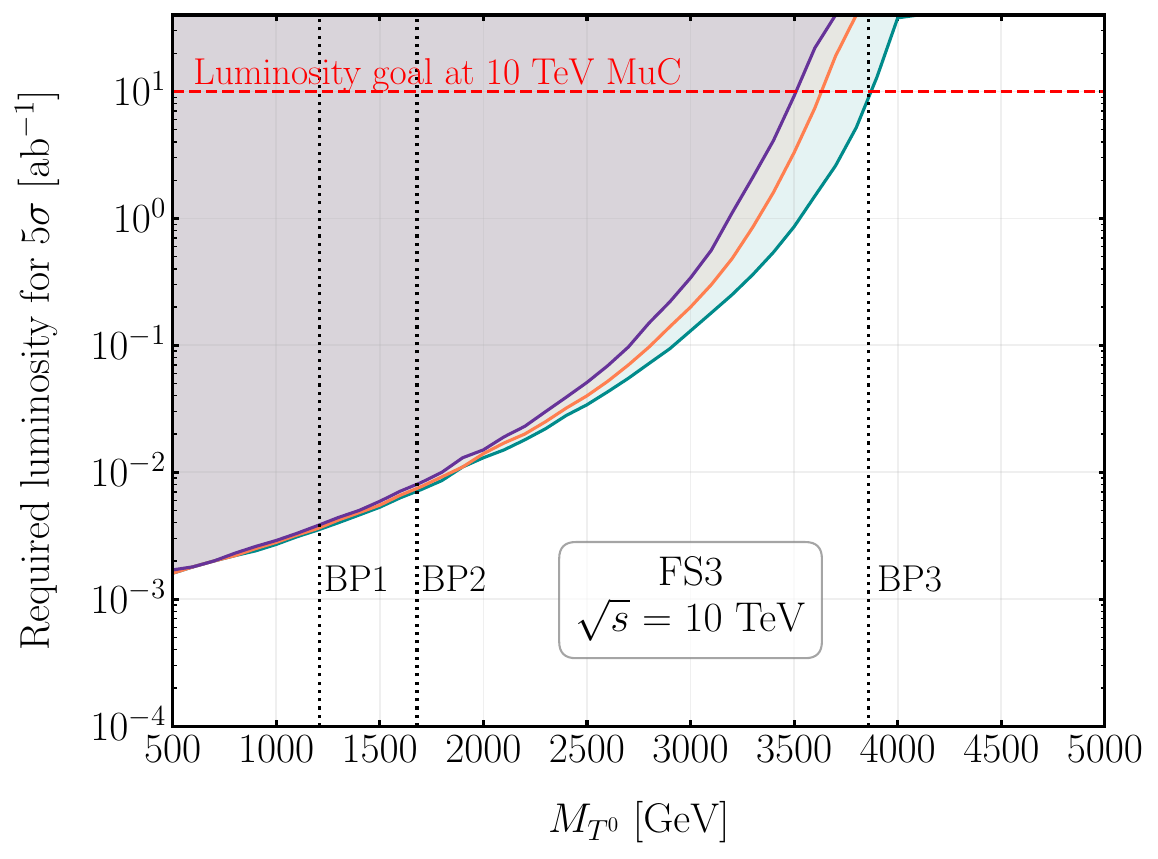}}
	\subfigure[]{\includegraphics[width=0.49\linewidth]{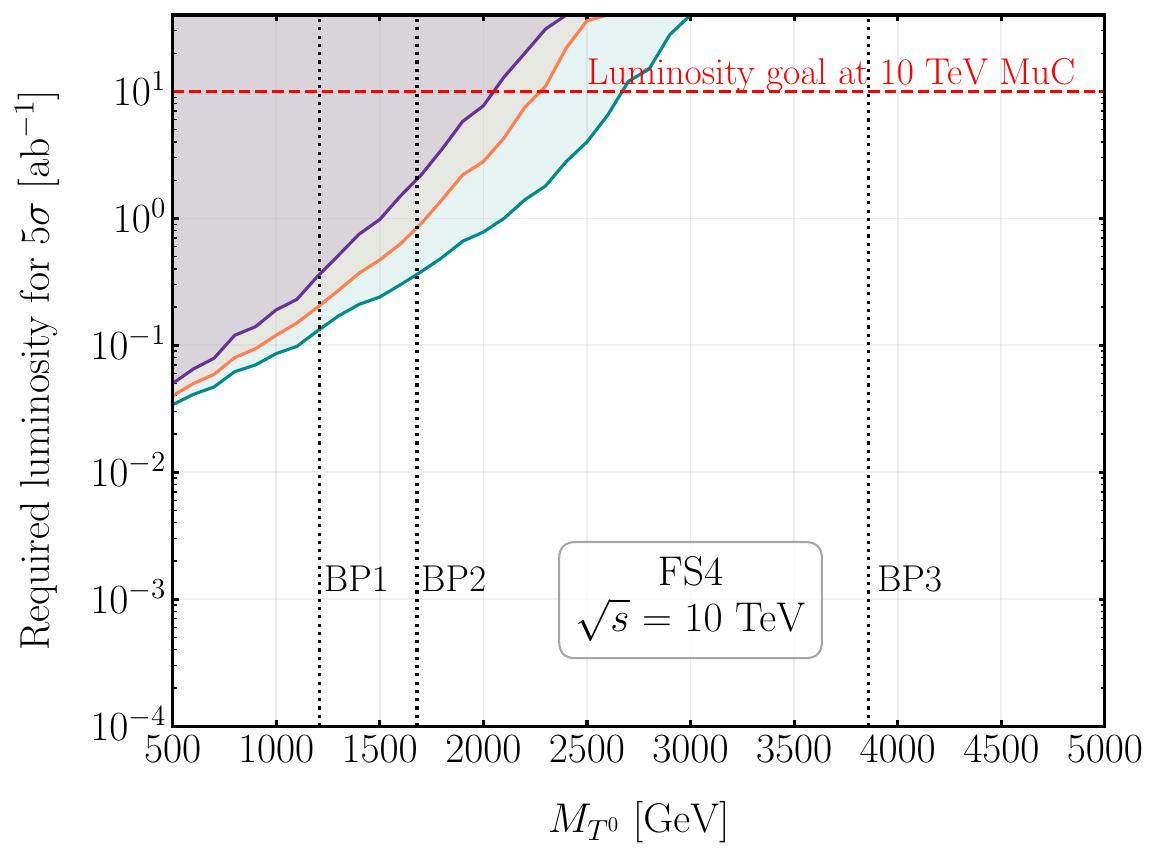}}
	\subfigure{\includegraphics[width=0.7\linewidth]{./plots/reachlegends-new.pdf}}
	\caption{Reach plots for triplet scalar masses and the required luminosity for 5$\sigma$ discovery, same as \autoref{fig:6sig}, but at the 10 TeV MuC.
	}
	\label{fig:10sig}
\end{figure}

Shifting our focus to the 10 TeV MuC, \autoref{fig:10sig} presents 
the same reach plots, with an increased mass range corresponding to the kinematical limit of the higher-energy collider.
The enhanced cross-section and improved tracking efficiency at the 10 TeV MuC enable 
all three benchmark points to achieve a $5\sigma$ discovery 
in the final states FS1 and FS3. Specifically, in FS1, the high cross-section of the VBF process $\mu^+ \mu^- \to T^0 T^{\pm} \mu^{\mp} {\nu}$ plays a crucial role. 
Coupled with the higher c.m.~energy and an increased number of events meeting the DCT reconstruction criteria, \autoref{fig:10sig}(a) shows that FS1 allows achieving a $5\sigma$ significance, within the 10 \abi target luminosity, for triplet scalar masses up to $\sim$4.2 TeV for 10 background events, which reduces to $\sim$4 TeV and $\sim$3.9 TeV for 200 and 600 backgrounds, respectively. Both BP1 and BP2 can be probed with less than 1 \fbi of luminosity via FS1, and the very massive BP3 with $\mtz = 3.86$ TeV also achieves $5\sigma$ discovery within a range between 0.7 \abi and 7 \abi luminosity, considering the different background counts. FS3 again remains the next favorable channel, with which
both BP1 and BP2 can be discovered with less than 10 \fbi luminosity for all three backgrounds.
FS3 also accommodates the  BP3 mass value of $\mtz =3.86$ TeV as the maximum reach limit with 10 \abi luminosity, considering 10 background events. However BP3 does not reach the required significance for 200 and 600 backgrounds. While FS2 and FS4 both provide enough events to discover BP1 and BP2 before reaching the luminosity goal, the maximum mass that they can reach are limited to $\sim 3.3$ TeV and $\sim 2.7$ TeV, respectively.


\begin{figure}[h]
	\centering
	\subfigure[]{\includegraphics[width=0.49\linewidth]{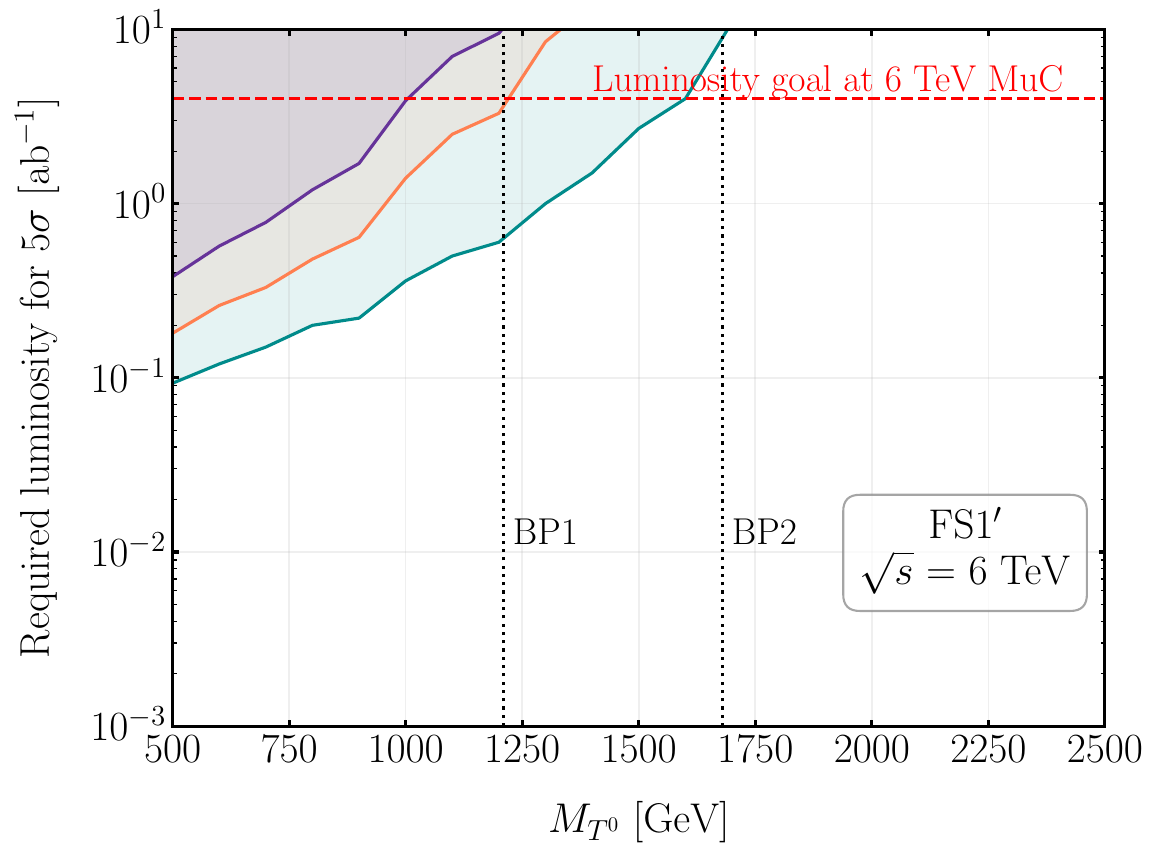}}
	\subfigure[]{\includegraphics[width=0.49\linewidth]{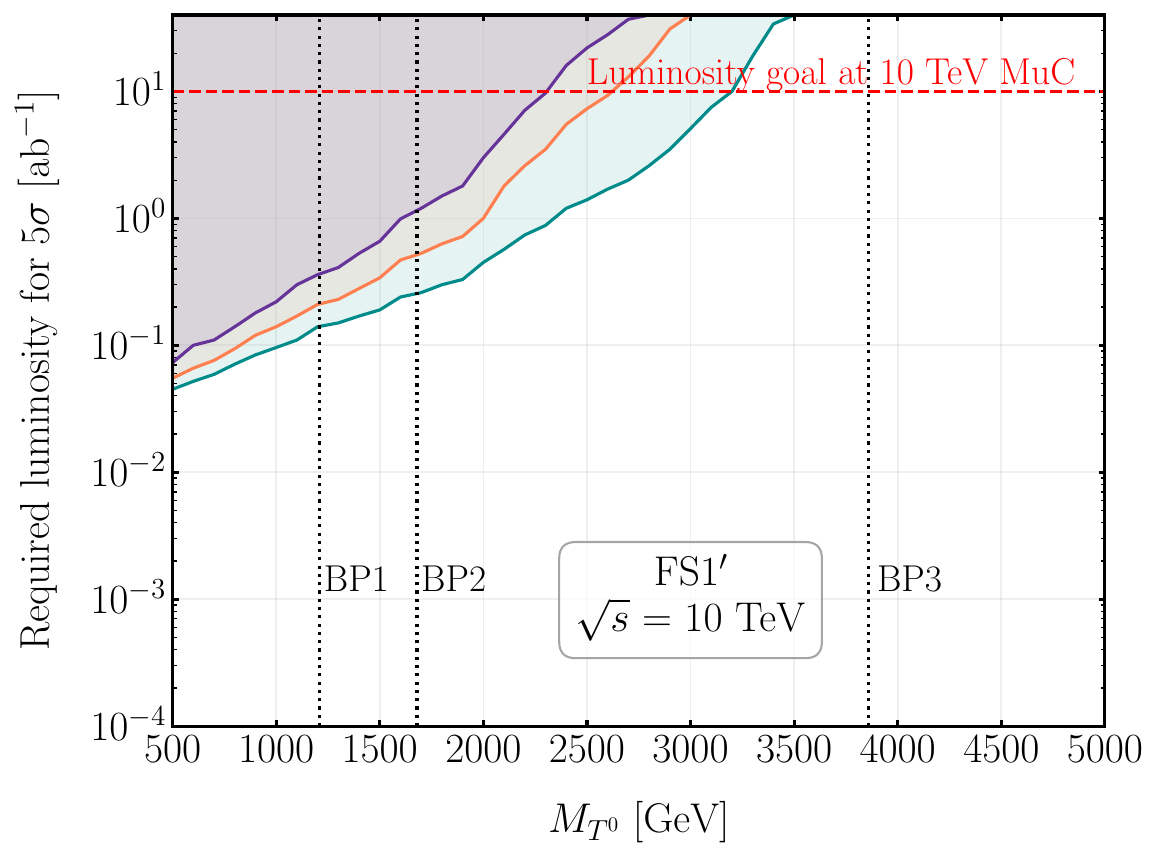}}
	\subfigure{\includegraphics[width=0.7\linewidth]{./plots/reachlegends-new.pdf}}
	\caption{Reach plots for triplet scalar masses and the required luminosity for 5$\sigma$ discovery, using events from FS1$'$, where one DCT is demanded with one muon in the main detector, and no Forward muons tagged, at MuC energies of (a) 6 TeV and (b) 10 TeV. The colour codes are same as \autoref{fig:6sig} and \autoref{fig:10sig}. 
		}
	\label{fig:nfmsig}
\end{figure}

Lastly, we discuss FS1$'$, an alternative to the FS1 scenario. 
FS1$'$ includes one DCT and a central muon, but distinctively omits tagging Forward muons.
The same luminosity versus significance plots for FS1$'$ are shown in \autoref{fig:nfmsig}. 
This alternative final state significantly differs from the original FS1, 
leading to a notable decrease in detection sensitivity, as we have already established in \autoref{sec:results}.
At the 6 TeV MuC, only the 10 and 200 background scenarios ensure 5$\sigma$ discovery of BP1 before reaching 4 \abi, with the maximum possible reach terminating at $\sim$ 1.6 TeV. While both BP1 and BP2 can be probed for all three backgrounds at the 10 TeV MuC from FS1$'$, the required luminosity is almost two orders of magnitude higher compared to those for FS1 with a Forward muon tagged. BP3 still remains unreachable here, with a maximum of $\sim$ 3.2 TeV triplet scalar mass being feasibly probed for the 10 background scenario.

A comparison of the collider reach for these final states can be drawn with the results from Ref.~\cite{Bottaro:2021snn}.
In their Figure 12, the 5$\sigma$ reach for a scalar triplet Minimal Dark Matter is provided for a 14 TeV MuC with 20 \abi luminosity, leading to the conclusion that the single DCT final state with a photon trigger provides 5$\sigma$ reach only till $\mtz \sim 2.5$ TeV.
In comparison, the FS1 of our analysis, with one Forward muon being tagged, allows the 10 TeV MuC to reach $\mtz \sim 4$ TeV, with just 10 \abi of luminosity. Coming to the double-track searches, the Figure 12 of Ref.~\cite{Bottaro:2021snn} also predicts a 5$\sigma$ reach up to $\sim4.1$ TeV at the 14 TeV MuC with 20 \abi luminosity. In their analysis, much milder cuts on the DCT $p_T$ are used, and the significance is evaluated with a zero-background hypothesis, corresponding to the observation of four events for the 5$\sigma$ discovery. A comparable final state for us is FS2 with two DCTs and two Forward muons, which provides a maximum mass reach of $\sim$3.4 TeV at the 10 TeV MuC with 10 \abi luminosity, considering 10 background events. The zero-background hypothesis in this case  (which is not included in the analysis and plots) can enhance this reach up to $ \mtz \sim 3.7$ TeV. Moving to a 14 TeV MuC and 20 \fbi luminosity, the order-of-magnitude jump in the VBF cross-section can ensure a reach of $\mtz \sim 5$ TeV for FS2, which is almost a GeV higher than what the photon-triggered final state offers in Ref. \cite{Bottaro:2021snn}.  These findings once again underscore the importance of dedicated Forward muon detectors, to fully harness the potential of VBF productions at a multi-TeV muon collider.

\section{Conclusions}\label{sec:conclusions}

In this study, we have investigated the potential of a multi-TeV Muon Collider (MuC) in probing the Inert Triplet Model (ITM), which introduces a scalar triplet field with zero hypercharge ($Y=0$). The ITM, featuring a neutral triplet $T^0$ and charged triplets $T^\pm$, emerges as an intriguing Beyond the Standard Model scenario, offering a viable Dark Matter (DM) candidate $T^0$, thanks to an incorporated $Z_2$ symmetry. 

Our evaluation against a range of constraints, including theoretical, collider, and DM experimental considerations, 
has identified three benchmark points with heavy triplet scalar masses of 1.21 TeV, 1.68 TeV, and 3.86 TeV.  
A notable characteristic of the ITM, distinguishing it from fermionic DM models, 
is the existence of a $TTVV$ four-point vertex. 
This feature enables efficient pair production via Vector Boson Fusion (VBF) processes, 
positioning the MuC as an ideal venue for ITM exploration, especially given the amplified VBF cross-sections at high collision energies. However, the extremely narrow mass gap between $T^\pm$ and $T^0$ complicates the detection of $T^\pm$. The dominant decay mode $T^\pm \to T^0 \pi^\pm$ produces pions that are too soft for detection amidst the beam-induced background (BIB) at the MuC.  
To address this, we have advocated the use of Disappearing Charged Tracks (DCTs) from $T^\pm$. 
Additionally,  we propose leveraging Forward muons, characterized by their high momenta in high $\eta$ regions, as innovative triggers. 
By identifying four final states, comprising combinations of one or two DCTs and one or two Forward muons, 
we have calculated event counts for the three benchmark points at MuC center-of-mass  energies of 6 TeV and 10 TeV, with integrated luminosities of 4 ab$^{-1}$ and 10 ab$^{-1}$, respectively.

While SM backgrounds can be effectively suppressed, 
BIBs at the MuC pose challenges, potentially leading to fake tracks that contaminate the signal.
Addressing this, we consider scenarios with 10, 200, and 600 background events and 
present integrated luminosity projections for a $5\sigma$ discovery over a range of the triplet scalar mass, at both the 6 TeV and 10 TeV MuC energies. 
The most promising outcome is observed in the final state with one DCT and one Forward muon. 
Remarkably, less than 100 \fbi luminosity at the 6 TeV MuC is sufficient 
for a $5\sigma$ discovery of BP1 and BP2 with triplet scalar masses of 1.21 TeV and 1.68 TeV, with a maximum possible mass reach of $\sim 2.3$ TeV with the 4 \abi luminosity goal.
At the 10 TeV MuC, the heavier scalar with a mass of 3.86 TeV can also be probed within the 10 \abi luminosity goal, with a maximum reach of $\mtz \sim 4.2$ TeV. These projections showcase the enhanced sensitivity of the MuC for the ITM, when one incorporates dedicated Forward muon detectors to successfully harness the potential of VBF processes.

In conclusion, our study highlights the significant  potential of a multi-TeV MuC 
in exploring heavy triplet scalars in the ITM via VBF pair production channels. 
The MuC emerges not just as a powerful tool for probing intricate aspects of particle physics, but as a crucial instrument in unveiling previously unknown facets of BSM.

\acknowledgments

PB wants to thank SERB's MTR/2020/000668 grant for support during project. SP acknowledges the Council of Scientific and Industrial Research (CSIR), India for funding his research (File no: 09/1001(0082)/2020-EMR-I). SP also thanks Rodolfo Capdevilla, Nathaniel Craig, Simone Pagan Griso, Lorenzo Sestini, and Massimo Casarsa for useful inputs, facilitated by the Muon4Future workshop. CS would like to thank the MoE, Government of India for supporting her research via SRF. CS also acknowledges Barbara Mele for valuable discussions during her visit at the INFN Sezione di Roma. 
The work of JS is supported by the National Research Foundation of Korea, Grant No.~NRF-2022R1A2C1007583.
PB and JS are also supported by the Visiting Professorship at Korea Institute for Advanced Study.

\appendix

\section{SM Background rejection}
\label{appendix:SM:backgrounds}

In this appendix, we demonstrate the effectiveness of four discriminators in significantly reducing the backgrounds from direct SM processes, as mentioned briefly in \autoref{sec:results}. We focus our discussion on the signal process $\mmu\to T^\pm T^0 \mu^\mp \nu$ for BP2. 
The discriminators are as follows:
(i) the presence of one hard Forward muon with $p_T^{\muf}>300\gev$;
(ii) a veto on any jets or leptons in the event;
(iii) the requirement of exactly one disappearing charged track (DCT) with $p_T^{\rm tr}>300\gev$;
(iv) soft missing transverse energy (MET) below 10 GeV.

The first discriminator, focusing on a hard Forward muon, 
is essential in distinguishing our signals from the majority of non-VBF or neutral-current-VBF SM background events. 
An example of such backgrounds is the invisible process $\mu^+ \mu^- \to \nu_\mu \bar{\nu_\mu}$,
which could mimic nearly-invisible disappearing track signals~\cite{Capdevilla:2021fmj}. 
Implementing the hard Forward muon criterion effectively nullifies this background.

The second discriminator,  vetoing out any calorimeter hits (jets and charged leptons),
leverages the unique nature of our signal,
which is  the absence of hadronic or leptonic activities in the main detector region. 
This method efficiently eliminates most non-VBF SM processes, such as $\mu^+ \mu^- \to VV$. 
However, certain VBF SM processes, 
such as $\mu^+ \mu^- \to \mu^\pm \nu W^\mp Z$ and $\mu^+ \mu^- \to\mu^+ \mu^- \nu_\mu \bar{\nu_\mu}$, 
still possess large cross-sections and could potentially contaminate our signal.

\begin{figure}[h]
	\centering
	\subfigure[]{\includegraphics[width=0.49\linewidth]{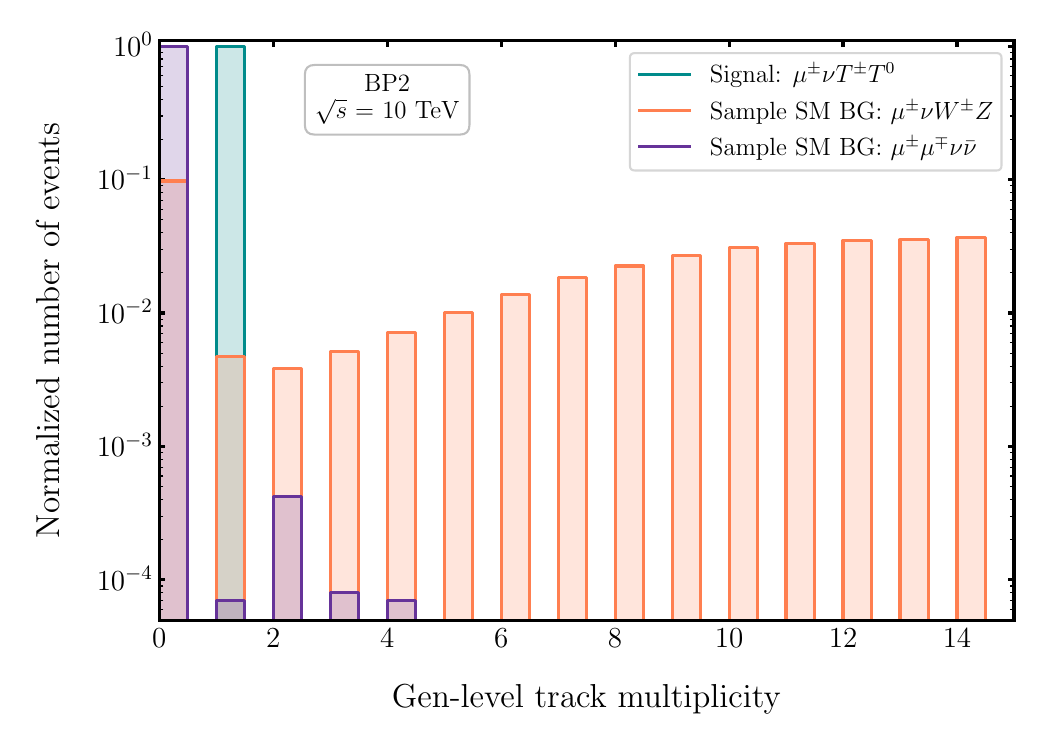}}
	\subfigure[]{\includegraphics[width=0.49\linewidth]{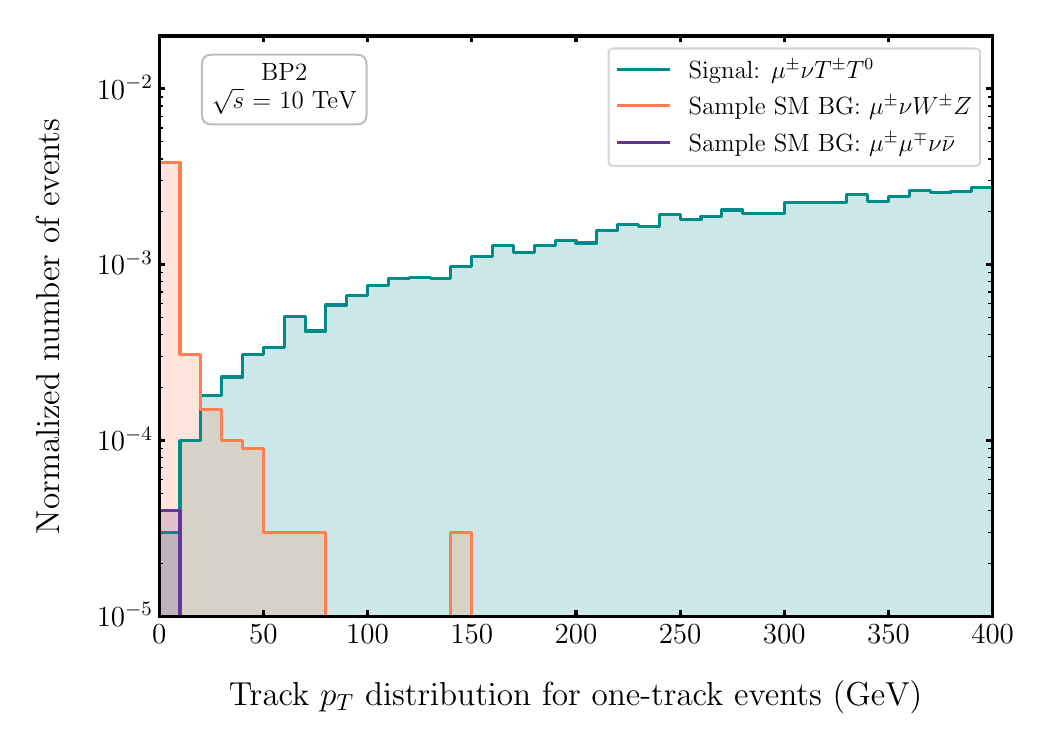}}	
	\caption{Panel (a) presents the distribution of charged track multiplicity and panel (b) showcases the $p_T$ distribution 
	of the track. This figure features the signal process $\mu^+ \mu^- \to T^0 T^{\pm} \mu^{\mp} {\nu}$ for BP2, in comparison with two main SM backgrounds ($\mu^+ \mu^- \to \mu^\pm \nu W^\mp Z$ and $\mu^+ \mu^- \to\mu^+ \mu^- \nu_\mu \bar{\nu_\mu}$) at the 10 TeV MuC.
}
	\label{fig:smbg}
\end{figure}

The third discriminator, which requires one DCT with high $p_T$, plays a vital role
in distinguishing the signal from background events. 
In \autoref{fig:smbg}(a), we illustrate the DCT multiplicity distributions
for both the signal and SM backgrounds at a 10 TeV MuC. 
The backgrounds include $\mu^+ \mu^- \to \mu^\pm \nu W^\mp Z$ and $\mu^+ \mu^- \to\mu^+ \mu^- \nu_\mu \bar{\nu_\mu}$.
The signal events primarily consist of just one track, corresponding to a single $T^\pm$.
However, a notable number of background events also display a single DCT signal, 
though they represent only a minor fraction of the total backgrounds. 

In addressing this, the transverse momentum of the single track becomes an essential discriminator. 
\autoref{fig:smbg}(b) presents the $p_T$ distributions of DCTs for both signal and SM backgrounds, each with precisely one track. 
The greater mass of $T^\pm$ leads to a notably higher $p_T$ for its track, 
in contrast to the generally softer SM background tracks, which mostly remain below $150\ \text{GeV}$.  
Therefore, applying a cut of $p_T^{\text{tr}} \geq 300\ \text{GeV}$ removes nearly all SM backgrounds with a single track.

\begin{figure}[h]
	\centering
\includegraphics[width=0.5\linewidth]{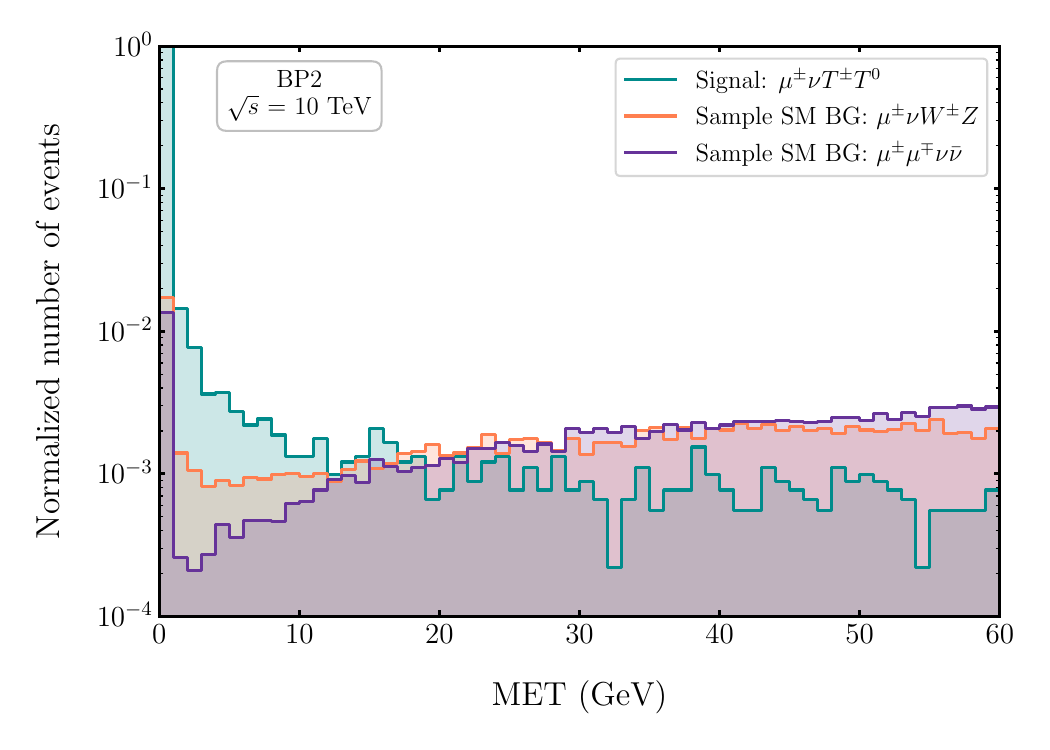}	
	\caption{Distribution of the missing transverse energy (MET) of the BP2 signal process $\mu^+ \mu^- \to T^0 T^{\pm} \mu^{\mp} {\nu}$ against two sample SM backgrounds at the 10 TeV MuC.}
	\label{fig:smbg:met}
\end{figure}

The final discriminator, requiring $\met$ below 10 GeV, arises from the observation that in our signal process, 
the two heavy triplet scalars are almost back-to-back, resulting in relatively low MET. 
\autoref{fig:smbg:met} shows the $\met$ distributions for the BP2 signal process $\mu^+ \mu^- \to T^0 T^{\pm} \mu^{\mp} {\nu}$, 
alongside two representative SM backgrounds at the 10 TeV MuC.  
Despite the signal involving two heavy particles that escape detection, 
a significant cancellation effect results in most events having MET below 10 GeV. 
This contrasts with the SM backgrounds, which typically exhibit much higher $\met$ values. 
Therefore, we implement a stringent cut of $\met < 10\gev$ for all our signal processes.

\bibliographystyle{JHEPMod}
\bibliography{2024ITMatMuonC}


\end{document}